\documentclass[11pt,a4paper]{article} 

\pdfoutput=1
\usepackage{jheppub}
\usepackage[T1]{fontenc}
\usepackage[utf8]{inputenc}
\usepackage{mathbbol}
\usepackage{mathrsfs}

\usepackage{bm}
\usepackage{booktabs}
\usepackage{colortbl}
\usepackage{enumerate}
\usepackage{latexsym}
\usepackage{lscape}
\usepackage{multirow}
\usepackage{multicol}
\usepackage{psfrag}
\usepackage{cancel}
\usepackage{shuffle}

\usepackage{acro}
\acsetup{first-style=short}
\DeclareAcronym{qft}{
	short = QFT,
	long  = quantum field theory,
	class = abbrev
}
\DeclareAcronym{qcd}{
	short = QCD,
	long  = quantum chromodynamics,
	class = abbrev
}
\DeclareAcronym{2d}{
	short = 2D,
	long  = two-dimensional,
	class = abbrev
}
\DeclareAcronym{bfkl}{
	short = BFKL,
	long  = Balitsky--Fadeev--Kuraev--Lipatov,
	class = abbrev
}
\DeclareAcronym{jimwlk}{
	short = Balitsky-JIMWLK,
	long  = Balitsky--Jalilian-Marian--Iancu--McLerran--\\Weigert--Leonidov--Kovner,
	class = abbrev
}
\DeclareAcronym{ll}{
	short = LL,
	long  = leading logarithm(ic accuracy),
	class = abbrev
}
\DeclareAcronym{nll}{
	short = NLL,
	long  = next-to-leading logarithm(ic accuracy),
	class = abbrev
}
\DeclareAcronym{nnll}{
	short = NNLL,
	long  = next-to-next-to-leading logarithm(ic accuracy),
	class = abbrev
}
\DeclareAcronym{hpl}{
	short = HPL,
	long  = harmonic polylogarithm,
	class = abbrev
}
\DeclareAcronym{hpls}{
	short = HPLs,
	long  = harmonic polylogarithms,
	class = abbrev
}
\DeclareAcronym{svhpl}{
	short = SVHPL,
	long  = single-valued harmonic polylogarithm,
	class = abbrev
}
\DeclareAcronym{svhpls}{
	short = SVHPLs,
	long  = single-valued harmonic polylogarithms,
	class = abbrev
}
\DeclareAcronym{mzv}{
	short = MZV,
	long  = multiple zeta value,
	class = abbrev
}
\DeclareAcronym{mzvs}{
	short = MZVs,
	long  = multiple zeta values,
	class = abbrev
}
\DeclareAcronym{oeis}{
	short = OEIS,
	long  = The On-Line Encyclopedia of Integer Sequences,
	class = abbrev
}

\def\beq{\begin{equation}}
	\def\eeq{\end{equation}}
\def\beqa{\begin{eqnarray}}
	\def\eeqa{\end{eqnarray}}

\def\be{\begin{equation}}
	\def\ee{\end{equation}}
\def\bea{\begin{eqnarray}}
	\def\eea{\end{eqnarray}}

\newcommand{\bal}{\begin{align}}
	\newcommand{\eal}{\end{align}}

\newcommand{\bei}{\begin{itemize}}
	\newcommand{\eei}{\end{itemize}}

\def\nn{\nonumber}

\newcommand\Eqn[1]     {Eq.~\eqref{#1}}

\newcommand\eqn[1]     {eq.~\eqref{#1}}
\newcommand\eqns[2]    {eqs.~\eqref{#1} and~\eqref{#2}}
\newcommand\eqnss[2]   {eqs.~\eqref{#1}--\eqref{#2}}

\newcommand{\pfrac}[2]{\left(\frac{#1}{#2}\right)}
\newcommand{\mratio}[2]{\left(\frac{#1^2}{#2^2}\right)}

\newcommand{\eps}{\epsilon}

\newcommand{\as}{\alpha_s}

\newcommand{\wb}{\bar{w}}
\newcommand{\zb}{\bar{z}}

\newcommand{\dd}{\mathrm{d}}

\newcommand{\der}[1]{\frac{\dd}{\dd {#1}}}
\newcommand{\ddeps}{\dd^{2-2\eps}}
\newcommand{\ddz}{\der{z}}
\newcommand{\ddw}{\der{w}}

\newcommand{\Dd}{\mathrm{D}}

\newcommand{\Dk}{[\Dd k]}
\newcommand{\Dkp}{[\Dd k']}

\newcommand{\T}{\mathbf{T}}
\newcommand{\Ts}{\T_s^2} 
\newcommand{\Tt}{\T_t^2} 
\newcommand{\Tu}{\T_u^2} 
\newcommand{\Tsu}{\T_{s-u}^2}
\newcommand{\Ca}{C_A}
\newcommand{\Cone}{(2\Ca-\Tt)}
\newcommand{\Ctwo}{(\Ca-\Tt)}

\newcommand{\CA}{C_{A}}

\newcommand{\disc}[1]{\mathrm{disc_{#1}}}
\newcommand{\reg}{\mathrm{reg}}
\newcommand{\con}{\mathrm{con}}

\newcommand{\RE}{{\rm Re}}
\newcommand{\IM}{{\rm Im}}

\newcommand{\MM}{\mathcal{M}}
\newcommand{\M}{\mathcal{M}}
\newcommand{\Mtree}{\M^{\mathrm{(tree)}}}

\newcommand{\Mreduced}{\hat{\mathcal{M}}}
\newcommand{\Mh}{\hat{\M}}

\newcommand{\Mhnll}{\Mh_{\mathrm{NLL}}}
\newcommand{\Mhnllp}{\Mhnll^{(+)}}
\newcommand{\Mhnllpl}[1]{\Mhnll^{(+,{#1})}}

\newcommand{\Hhard}{\mathcal{H}} 

\newcommand{\Ha}{\hat{H}}
\newcommand{\Hi}{\Ha_{\mathrm{i}}}

\newcommand{\Hm}{\Ha_{\mathrm{m}}}
\newcommand{\Has}{\Ha_{\mathrm{s}}}

\newcommand{\Htd}{\Ha_{\mathrm{2d}}}
\newcommand{\Hitd}{\Ha_{\mathrm{2d,i}}}
\newcommand{\Hitdone}{\Ha_{\mathrm{2d,i}_1}}
\newcommand{\Hitdtwo}{\Ha_{\mathrm{2d,i}_2}}
\newcommand{\Hitdn}{\Ha_{\mathrm{2d,i}_n}}
\newcommand{\Hmtd}{\Ha_{\mathrm{2d,m}}}

\newcommand{\El}{\mathcal{L}}
\newcommand{\Ef}{\mathcal{F}}
\newcommand{\Li}{\mathrm{Li}}
\newcommand{\ct}{\text{(contact terms)}}

\newcommand{\Bn}[1]{B_{#1}}
\newcommand{\bn}[1]{B_{#1}}
\newcommand{\hbn}[1]{\hat B_{#1}} 
\newcommand{\bzero}{B_0} 
\newcommand{\Bh}[1]{\hat{B}_{#1}}

\newcommand{\gE}{\gamma_{\mathrm{E}}}

\newcommand{\Wt}[1]{{\Omega}^{(#1)}}
\newcommand{\Wts}[1]{{\Omega}_s^{(#1)}}

\newcommand{\Wh}{\Wf_{\mathrm{h}}}

\newcommand{\Wf}{\Omega}
\newcommand{\Wfl}[1]{\Wf^{({#1})}}
\newcommand{\Ws}{\Wf_{\mathrm{s}}}
\newcommand{\Wsl}[1]{\Wfl{#1}_{\mathrm{s}}}
\newcommand{\Wtd}{\Wf_{\mathrm{2d}}}
\newcommand{\Wtdl}[1]{\Wfl{#1}_{\mathrm{2d}}}

\newcommand{\ord}{\mathcal{O}}

\allowdisplaybreaks

\title{Two-parton scattering amplitudes in the Regge limit to high loop orders}

\author[a]{Simon Caron-Huot,}
\author[b]{Einan Gardi,}
\author[b]{Joscha Reichel,}
\author[c]{Leonardo Vernazza}

\affiliation[a]{Department of Physics, 
	McGill University, 3600 rue University, 
	Montr\'eal, QC Canada H3A 2T8}
\affiliation[b]{Higgs Centre for Theoretical Physics, 
	School of Physics and Astronomy, \\
	The University of Edinburgh, Edinburgh EH9 3FD, Scotland, UK}
\affiliation[c]{Dipartimento di Fisica Teorica, 
	Universit\`a di Torino \\
	and INFN, Sezione di Torino, Via P. Giuria 1, 
	I-10125 Torino, Italy}

\emailAdd{schuot@physics.mcgill.ca}
\emailAdd{Einan.Gardi@ed.ac.uk}
\emailAdd{joscha.reichel@ed.ac.uk}
\emailAdd{leonardo.vernazza@to.infn.it}

\abstract{We study two-to-two parton scattering amplitudes 
in the high-energy limit of perturbative QCD by iteratively 
solving the BFKL equation.  This allows us to predict the 
imaginary part of the amplitude to leading-logarithmic 
order for arbitrary $t$-channel colour exchange.
The corrections we compute correspond to ladder diagrams 
with any number of rungs formed between two Reggeized 
gluons. Our approach exploits a separation of the 
two-Reggeon wavefunction, performed directly in momentum 
space, between a soft region and a generic (hard) region. 
The former component of the wavefunction leads to infrared 
divergences in the amplitude and is therefore computed in 
dimensional regularization; the latter is computed directly 
in two transverse dimensions and is expressed in terms of 
single-valued harmonic polylogarithms of uniform weight.
By combining the two we determine exactly both 
infrared-divergent and finite contributions to the 
two-to-two scattering amplitude order-by-order in 
perturbation theory. We study the result numerically 
to 13 loops and find that finite corrections to the 
amplitude have a finite radius of convergence which 
depends on the colour representation of the $t$-channel 
exchange.}


\begin{document} 

\begin{flushright} 
\vspace*{-25pt}
\end{flushright}

\maketitle
\flushbottom



\section{Introduction} \label{intro}

\begin{figure}[htb]
	\centering
	\includegraphics{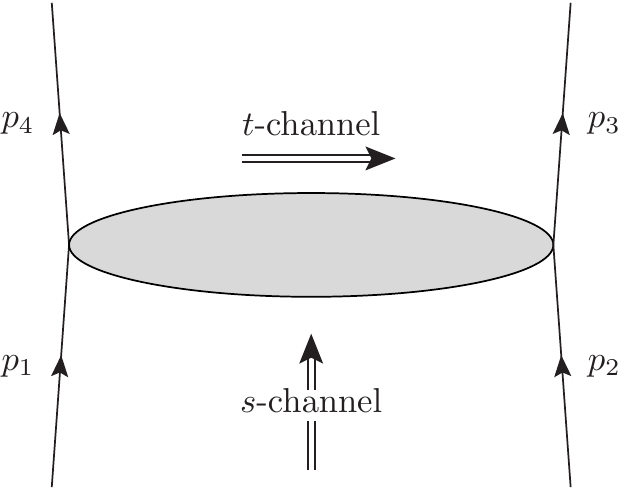}
	\caption{The $t$-channel exchange dominating the high-energy limit, 
		$s\gg -t>0$. The figure also defines our conventions for momenta assignment 
		and Mandelstam invariants. We shall assume that particles 2 and 3 (1 and 4) are 
		of the same type and have the same helicity.}
	\label{setup_fig}
\end{figure}

The study of QCD scattering in the Regge limit has been 
an active area of research for over half a century, e.g.~\cite{Kuraev:1977fs,Balitsky:1978ic,Lipatov:1985uk,Mueller:1993rr,Mueller:1994jq,Brower:2006ea,Moult:2017xpp}. 
While the general problem of high-energy scattering is 
non-perturbative, in the regime where the exchanged 
momentum $-t$ is high enough,  i.e. 
$s\gg-t\gg\Lambda_{\rm QCD}^2$ (see figure~\ref{setup_fig}), 
perturbation theory offers systematic tools to analyse 
this limit. Central to this is the Balitsky-Fadin-Kuraev-Lipatov 
(BFKL) evolution equation~\cite{Kuraev:1977fs,Balitsky:1978ic}, 
which provides a systematic theoretical framework to resum 
high-energy (or rapidity) logarithms, $\ln (s/(-t))$, to all orders 
in perturbation theory. This approach was used extensively to 
study a range of physical phenomena including the small-$x$ 
behaviour of deep-inelastic structure functions and parton 
densities, and jet production with large rapidity gaps. 
Furthermore, non-linear generalisations of BFKL, 
known as the Balitsky-JIMWLK 
equation~\cite{Balitsky:1995ub,Balitsky:1998kc,Kovchegov:1999yj,JalilianMarian:1996xn,JalilianMarian:1997gr,Iancu:2001ad}, 
are today a main tool in the theoretical description of 
dense states of nuclear matter, notably in the context 
of heavy-ion collisions.

While many applications of rapidity evolution equations 
to phenomenology require the scattering particles to be 
colour-singlet objects, in the present paper we are 
concerned with the more theoretical problem of 
understanding  \emph{partonic} scattering amplitudes 
in the high-energy limit, similarly to refs.~\cite{Sotiropoulos:1993rd,Korchemsky:1993hr,Korchemskaya:1996je,Korchemskaya:1994qp,DelDuca:2001gu,DelDuca:2013ara,DelDuca:2014cya,Bret:2011xm,DelDuca:2011ae,Caron-Huot:2013fea,Caron-Huot:2017fxr,Caron-Huot:2017zfo}. 
This is part of a more general programme of understanding 
the structure of gauge-theory amplitudes and the underlying 
physical and mathematical principles governing this structure. 
The basic observation is that gauge dynamics drastically 
simplifies in the high-energy limit, which renders the 
amplitudes computable to all orders in perturbation 
theory, to a given logarithmic accuracy.

The present paper continues our recent study~\cite{Caron-Huot:2013fea,Caron-Huot:2017fxr,Caron-Huot:2017zfo} 
of $2\to 2$ partonic amplitudes ($qq\to qq$, $gg\to gg$, 
$qg\to qg$) in QCD and related gauge theories. 
A key ingredient in these studies is provided once again by 
rapidity evolution equations, BFKL and its generalisations, 
which are used to compute high-energy logarithms in these 
amplitudes order-by-order in perturbation theory. 

Scattering amplitudes of quarks and gluons are dominated 
at high energies by the $t$-channel exchange (figure~\ref{setup_fig}) 
of effective degrees of freedom called \emph{Reggeized 
gluons}.  $2\to 2$ amplitudes are conveniently decomposed 
into \emph{odd} and \emph{even} signature characterising 
their symmetry properties under $s\leftrightarrow u$ 
interchange, or crossing symmetry:
\begin{equation}\label{Odd-Even-Amp-Def}
	{\cal M}^{(\pm)}(s,t) = \tfrac12\Big( {\cal M}(s,t) \pm {\cal M}(-s-t,t) \Big)\,,
\end{equation}
where odd (even) amplitudes ${\cal M}^{(-)}$ (${\cal M}^{(+)}$) 
are governed by the exchange of an odd (even) number of 
Reggeized gluons. Furthermore, as shown in 
ref.~\cite{Caron-Huot:2017fxr}, these have respectively 
\emph{real} and \emph{imaginary} coefficients, when 
expressed in terms of the natural signature-even 
combination of logarithms,
\begin{equation}
	\label{L-def}
	\frac12\left(\log\frac{-s-i0}{-t}+\log\frac{-u-i0}{-t}\right)
	\simeq \log\left|\frac{s}{t}\right| -i\frac{\pi}{2} \equiv L\,.
\end{equation}

The real part of the amplitude, ${\cal M}^{(-)}$, is governed, 
at leading logarithmic (LL) accuracy, by the exchange of a 
single Reggeized gluon in the $t$ channel. To this accuracy, 
high-energy logarithms admit a simple exponentiation pattern, 
namely 
\begin{equation}
	\label{Mreal}
	{\cal M}^{(-)}_{\rm LL} = (s/(-t))^{\alpha_g(t)} \times {\cal M}^{\rm tree}
\end{equation}
where the exponent is the \emph{gluon Regge trajectory} 
(corresponding to a Regge pole in the complex angular 
momentum plane), $\alpha_g(t)=\frac{\alpha_s}{\pi} C_A 
\alpha_g^{(1)}(t)+{\cal O}(\alpha_s^2)$, whose leading order 
coefficient $\alpha_g^{(1)}(t)$ is infrared singular, 
$\alpha_g^{(1)}(t)\sim \frac{1}{2\epsilon}$ in dimensional 
regularization with $d=4-2\epsilon$ (see eq.~(\ref{alphag1}) 
below). Infrared singularities are well-known to exponentiate, 
independently of the high-energy limit.
Importantly, however, eq.~(\ref{Mreal}) illustrates the fact that 
the exponentiation high-energy logarithms must be compatible 
with that of infrared singularities, which is a nontrivial constraint 
on both. This observation and its extension to higher logarithmic 
accuracy underpins a long line of investigation in refs.~\cite{Sotiropoulos:1993rd,Korchemsky:1993hr,Korchemskaya:1996je,Korchemskaya:1994qp,DelDuca:2001gu,DelDuca:2013ara,DelDuca:2014cya,Bret:2011xm,DelDuca:2011ae,Caron-Huot:2013fea,Caron-Huot:2017fxr,Caron-Huot:2017zfo}. 

The key property of the Reggeized gluon being signature-odd 
greatly constrains the structure of higher-order corrections.
For the real part of  amplitude, the simple exponentiation 
pattern generated by a single Reggeized gluon is preserved 
at the next-to-leading logarithmic (NLL) accuracy, except that 
it requires ${\cal O}(\alpha_s^2)$ corrections to the trajectory 
and also  the introduction of ($s$-independent) impact factors.
This simple picture only breaks down when three Reggeized 
gluons can be exchanged,
which first occurs at NNLL accuracy and leads to Regge cuts. 
This contribution was computed in ref.~\cite{Caron-Huot:2017fxr} 
through three-loops, by constructing an iterative solution of the 
non-linear Balitsky-JIMWLK equation which tested the mixing 
between one and three Reggeized gluons.

In this paper we focus on the imaginary part of the amplitude, 
${\cal M}^{(+)}$, extending our work~\cite{Caron-Huot:2017zfo}.
Here the leading tower of logarithms, in which we are interested, 
is generated by the exchange of \emph{two} Reggeized gluons, 
starting with a non-logarithmic term at one loop:
\begin{equation} 
	\label{MevenOneloop}
	{\cal M}^{(+)}_{\rm NLL}\simeq  i\pi 
	\left[\frac{1}{2\epsilon} \frac{\alpha_s}{\pi}
	+{\cal O}\left(\alpha_s^{2} L\right)\right] 
	{\mathbf T}^2_{s-u}  {\cal M}^{\rm tree}\,.
\end{equation}
Here we suppressed subleading terms in $\epsilon$ as well 
as multiloop corrections, which take the form $\alpha_s^{\ell} 
L^{\ell-1}$ at $\ell$ loops; because the power of the energy 
logarithm $L$ is one less than that of the coupling, these 
are formally next-to-leading logarithms (NLL).   
In eq.~(\ref{MevenOneloop}) one may observe another 
salient feature of this tower of corrections, namely the 
colour structure, which is even under $s\leftrightarrow u$ 
interchange (${\cal M}^{\rm tree}$ is odd, and so is the 
operator ${\mathbf T}^2_{s-u}$ acting on it). The first term 
in the square brackets in (\ref{MevenOneloop}) is the exact 
result in the planar limit; we will be interested in the full 
series of corrections $\alpha_s^{\ell} L^{\ell-1}$, which are 
all subleading in the large $N_c$ limit (see the definitions 
of colour operators in eq. (\ref{TtTsTu}) below). 

All higher-order corrections,  ${\cal O}(\alpha_s^{\ell}L^{\ell-1})$, 
in (\ref{MevenOneloop}) can be described by the well-known 
ladder graphs, where each additional loop constitute an 
additional rung in the ladder (see figure~\ref{fig:simplebfkl} 
below). 
\begin{figure}[!ht]
	\centering
	\includegraphics{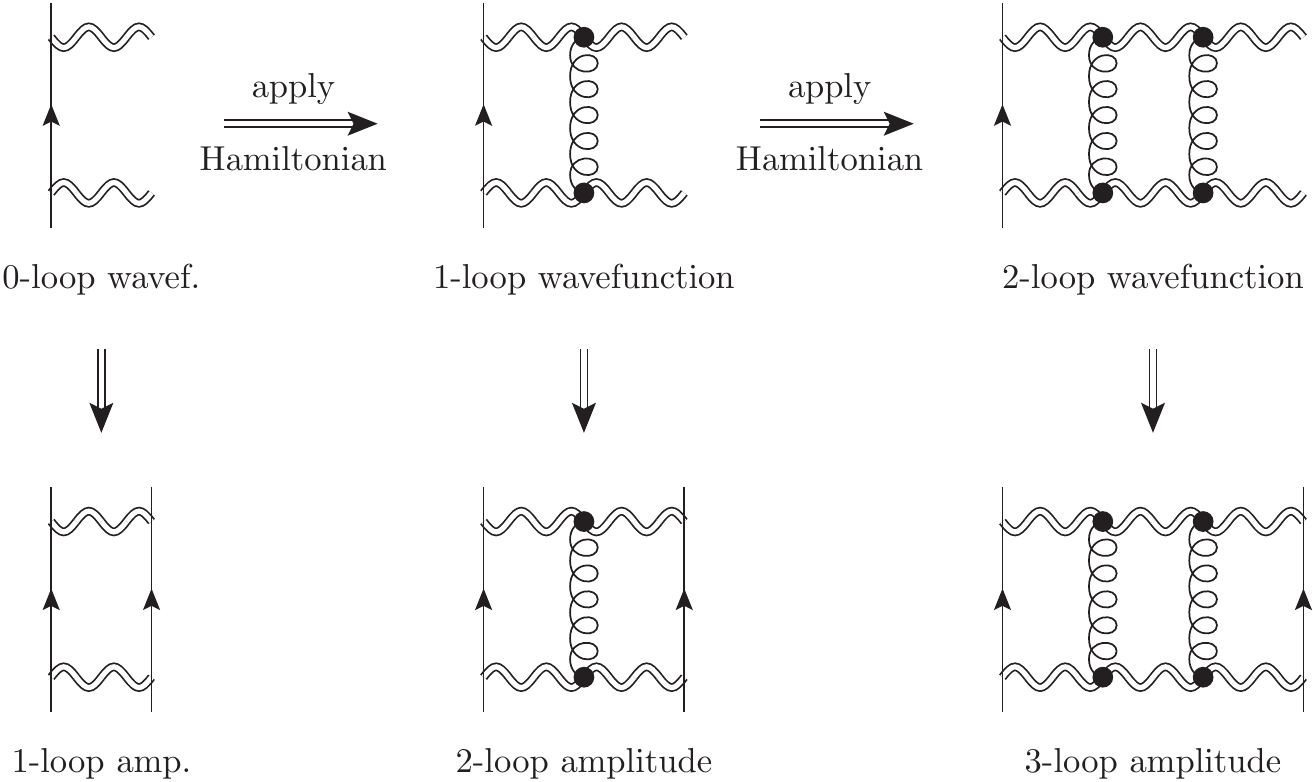}
	\caption{Sketch of \ac{bfkl} evolution generating ladder graphs 
	in the imaginary part of the amplitude. Considering initially emission 
	from the projectile side only, the 0-loop wavefunction (top left) describes 
	a state involving two reggeized gluons. The Reggeized gluons are both 
	off-shell and are characterized by their transverse momenta $k$ and 
	$p-k$. Each application of the BFKL Hamiltonian (the top row) 
	generates an additional rung in the ladder. Upon integrating the 
	$(\ell-1)$-loop wavefunction with the target one obtains the 
	$\ell$-loop amplitude (bottom row).}
	\label{fig:simplebfkl}
\end{figure}
Being the leading contributions to the imaginary part of 
the amplitude, they are particularly important, and clearly 
at high energies, where $\alpha_s L\sim {\cal O}(1)$, one 
should aim at an all-order calculation.
These corrections, however, do not feature a simple 
exponentiation pattern as in eq.~(\ref{Mreal}); they give 
rise to a Regge cut rather than a pole. We shall study 
these corrections using an iterative solution of the BFKL 
equation, continuing the work of 
ref.~\cite{Caron-Huot:2013fea,Caron-Huot:2017fxr,Caron-Huot:2017zfo}.
In \cite{Caron-Huot:2013fea} higher-order terms in 
eq.~(\ref{MevenOneloop}) were computed through 
four loops -- the first order where finite contributions 
appear (see eqs. (28-29) in~\cite{Caron-Huot:2017fxr}). 
Subsequently, in ref.~\cite{Caron-Huot:2017zfo} 
infrared-singular contributions were computed in 
dimensional regularization to all orders.
The purpose of the present paper is to extend the 
calculation to finite contributions, and in particular,
to obtain the infrared-renormalized amplitude, or 
hard function, which we expect (together with the 
soft anomalous dimension) to control any 
infrared-safe cross section.

We are interested in the exact perturbative solution 
of the BFKL equation for any colour exchange, that is, 
not restricted to the planar limit.  While the BFKL 
Hamiltonian was famously diagonalized by its authors
in the case of color-singlet exchange, the solution is 
not known in the general case. Adding to the complexity 
is the fact that amplitudes are infrared singular, forcing 
us to work in dimensional regularization. While it is not 
known how to diagonalise the BFKL Hamiltonian in these 
circumstances, we are able to solve the problem by using 
two complementary approaches, the first by taking the 
soft approximation while maintaining dimensional 
regularization, and the second by considering general 
(hard) kinematics in strictly two transverse dimensions. 
Let us briefly describe each of these approaches.  

The first approach is a computation of the wavefunction 
describing the emission of two Reggeons at $(\ell-1)$ loops, 
and the corresponding $\ell$-loop $2\to 2$  amplitude, in the 
\emph{soft approximation}, where one of the two Reggeized 
gluons carries transverse momentum $k^2$ which is 
significantly smaller than the total momentum transfer by 
the pair, $-t = p^2$,  i.e. the limit characterized by a double 
hierarchy of scales $k^2\ll p^2 \ll s$.
This is the limit used in ref.~\cite{Caron-Huot:2017zfo} to 
determine all infrared-singular contributions the amplitude. 
This was achieved using the simple observation that the 
wavefunction is itself finite to all orders in perturbation 
theory and that BFKL evolution closes within this 
approximation. All the singularities of the amplitude at 
any given loop order are in turn produced in the final 
integration over the wavefunction (corresponding to the 
transition from the top to the bottom row in figure~\ref{fig:simplebfkl}). 
In the present paper, building upon the computation of 
the wavefunction in~\cite{Caron-Huot:2017zfo} we 
introduce a symmetrized solution accounting simultaneously 
for the two soft limits, $k^2\ll p^2$ and $(p-k)^2\ll  p^2$, which 
amounts to an elegant separation between soft and hard 
contributions to the wavefunction and amplitude. Within 
this approximation we are able to write down a resummed 
analytic expression for the amplitude, including its finite 
contributions.

The second approach, which we develop in the present 
paper, is based on starting with the BFKL equation in 
exactly two dimensions.
Without making any further approximation, we set up 
an iterative solution of the equation by identifying 
differential operators that commute with (parts of) the 
Hamiltonian up to a computable set of contact terms. 
Evolution induced by the Hamiltonian then becomes 
trivial within a class of iterated integrals dictated by 
the nature of the problem, these are the Single-Valued 
Harmonic Polylogarithms (SVHPLs), first systematically 
classified by Francis Brown in ref.~\cite{Brown:2004ugm} 
and then studied and applied in the context of motivic 
periods~\cite{Brown:2013gia} and Feynman 
integrals~\cite{Chavez:2012kn,Schnetz:2013hqa}. The 
relevance of this class of functions for gauge-theory 
amplitudes within the Regge limit~\cite{Pennington:2012zj,Dixon:2012yy,DelDuca:2013lma,Dixon:2014voa,DelDuca:2016lad,DelDuca:2018hrv} 
(and beyond~\cite{Almelid:2017qju,Dixon:2019lnw}) has 
been recognised in recent years, and it is important also 
in our current problem: the hard wavefunction, defined in 
strictly two dimensions, is fully expressible in terms of 
SVHPLs, and the corresponding contribution to the 
amplitude can in turn be written in terms of Single-Valued 
Multiple Zeta Values (SVMZVs). For the ladder graphs 
relevant here, each additional loop increases the 
transcendental weight by one unit. The resulting 
uniform-weight expressions in terms of single-valued 
functions are significantly simpler as compared to the 
corresponding ones in terms of ordinary polylogarithms 
and zeta values.
For the final integration over the wavefunction we 
develop two independent approaches, one relying on 
analytic continuation and integration over the discontinuities 
of the wavefunction away from the region were they are 
single-valued, and the other relying instead on a modified 
application of the evolution algorithm itself. The two yield 
identical results. By combining the hard contribution to the 
amplitude with the dimensional-regularized soft contribution 
we compute the full amplitude, in principle to any order, 
and in practice to thirteen loops.

The structure of the paper is as follows. In section~\ref{chap:bfkl}
we present the BFKL equation in dimensional regularisation, 
bring it to a form suitable for iterative solution and review the 
relation between the off-shell wavefunction and the two-to-two 
scattering amplitude. We also show how an iterative solution 
can be obtained for the first few orders  directly in dimensional 
regularization without resorting to any approximation, and 
explain why this approach does not practically extend to 
higher orders. In this context we compute the amplitude 
numerically through five loops, providing a valuable check 
for our subsequent calculations.
Next, in section~\ref{soft} we review the soft approximation 
developed in~\cite{Caron-Huot:2017zfo} and explain how 
infrared factorization, combined with the finiteness of the 
wavefunction, facilitate a systematic separation of the latter 
into `soft' and `hard' components, such that eventually, finite 
corrections to the infrared-renormalized scattering amplitude 
can be determined in full. To this end we introduce a symmetrized 
version of the soft wavefunction, which captures both soft limits, 
and then derive an analytic expression for the amplitude as a 
function of $\alpha_sL$, which resums both infrared-divergent 
and finite contributions to all loops, within the soft approximation.
In section~\ref{2d-bfkl} we turn to discuss the wavefunction in 
general (hard) kinematics. Working directly in two dimensions 
we introduce the relevant kinematic variables, analyse the action 
of the BFKL Hamiltonian and demonstrate that evolution 
generated by this Hamiltonian translates into an algorithmic 
procedure in the space of SVHPLs. 
Having determined the wavefunction order by order, we turn 
in section~\ref{amplitude} to compute the corresponding 
two-to-two scattering amplitude. In section~\ref{numerics} we 
perform a numerical study of the resulting wavefunctions  and 
amplitudes, and address the convergence of the perturbative 
expansion. Finally, in section~\ref{conclusion} we make some 
concluding comments and present an outlook for future 
investigation.


\section{\ac{bfkl} equation in dimensional regularisation and the $2\to 2$ amplitude}
\label{chap:bfkl} 

In the high-energy limit, scattering amplitudes are
conveniently described in terms of Wilson lines, 
which dress the external partons. The evaluation 
of vacuum expectation values of Wilson lines stretching
from minus to plus infinity leads to rapidity divergences, 
which needs to be renormalised. As a consequence, the 
renormalised amplitude obeys a rapidity evolution equation, 
which can be shown to correspond to the 
Balitsky-JIMWLK equation. In this paper we are 
interested to study the two-Reggeon exchange 
contribution to two-parton scattering amplitudes, 
for which the evolution equation reduces to the 
BFKL equation 
\cite{Caron-Huot:2013fea,Caron-Huot:2017fxr}. 
The scattering amplitude can be determined formally 
to any order in perturbation theory as an iterative solution
of the dimensionally-regularised BFKL equation.
This procedure was described in 
\cite{Caron-Huot:2017zfo}, to which we refer for 
further details. In this section we review the 
definitions necessary to set up the calculation. 

In the following we consider the two-reggeon
exchange contribution to $2 \to 2$ scattering
amplitudes. We can single out this contribution 
by introducing a reduced amplitude, in which 
the one-Reggeon exchange has been removed:
\beq
\label{Mreduced}
\Mh_{ij\to ij} \equiv e^{-\alpha_g(t) L \Tt} \M_{ij\to ij},
\eeq
where $L$ is the signature-even high-energy logarithm 
defined in eq.~(\ref{L-def}), $\Tt$ represents the total 
colour charge exchanged in the $t$ channel (see 
eq.~\eqref{TtTsTu} below) and $i,j$ are the species 
indices defining the two-parton scattering; in what 
follows we will drop these indices, unless explicitly 
needed. Finally, the function 
\begin{align}
	\alpha_g(t)=\frac{\alpha_s}{\pi} \alpha_g^{(1)}(t)+{\cal O}(\alpha_s^2)
\end{align}
is the \emph{gluon Regge trajectory} introduced already 
in \eqn{Mreal}, where  the leading-order coefficient in 
dimensional regularization with $d=4-2\epsilon$ is 
given by 
\begin{equation}
	\label{alphag1}
	\alpha_g^{(1)}(t)= \frac{B_0}{2\epsilon} 
	\left(\frac{-t}{\mu^2}\right)^{-\epsilon}
\end{equation}
where
\begin{equation} \label{B0}
	\Bn{0} \equiv \Bn{0}(\eps)=e^{\eps\gamma_{\rm E}}  
	\frac{\Gamma^2(1-\eps)\Gamma(1+\eps ) }{\Gamma(1-2 \eps )}
	= 1 - \frac12 \eps^2 \zeta_2 -\frac73 \eps^3\zeta_3 + O(\eps^4)
\end{equation}
belongs to a class of bubble integrals which will be defined below.

The two-Reggeon cut contributes only to the even amplitude 
defined in \eqn{Odd-Even-Amp-Def}, thus we focus only on this
component in the following. As discussed in \cite{Caron-Huot:2017zfo},
the reduced amplitude takes the form of an integral over the 
two-Reggeon wavefunction $\Wf(p,k)$, as follows:
\beq
\label{ReducedAmpNLL}
\Mhnllp \left(\frac{s}{-t}\right) = -i\pi \int \Dk \frac{p^2}{k^2(p-k)^2} 
\, \Wf(p,k) \, \Tsu \, \Mtree_{ij\to ij},
\eeq
where $p^2 = -t$. In \eqn{ReducedAmpNLL} the integration 
measure is
\beq
\label{measure}
\Dk \equiv \frac{\pi}{\Bn{0}} 
\left( \frac{\mu^2}{4\pi e^{-\gE}} \right)^{\eps} 
\frac{\ddeps k}{(2\pi)^{2-2\eps}},
\eeq
and $\Mtree_{ij\to ij}$ represent the tree amplitude, 
given by
\beq
\Mtree_{ij\to ij} = 4\pi \alpha_s \frac{2s}{t} (T_i^b)_{a_1 a_4} (T_j^b)_{a_2 a_3} 
\delta_{\lambda_1\lambda_4}\delta_{\lambda_2\lambda_3},
\eeq
where $\lambda_i$ for $i=1$ through $4$ are helicity indices. 
The colour operator $\Tsu$ in eq.~\eqref{ReducedAmpNLL} acts 
on $\Mtree_{ij\to ij}$ and it is defined in terms of the usual 
basis of quadratic Casimirs corresponding to colour flow through 
the three channels~\cite{Dokshitzer:2005ig,DelDuca:2011ae}:
\beq
\label{TtTsTu}
\Tsu \equiv \frac{\Ts-\Tu}{2} \qquad \rm{with} \qquad \left\{ \begin{array}{c}
	\T_s = \T_1+\T_2=-\T_3-\T_4, \\ 
	\T_u = \T_1+\T_3=-\T_2-\T_4, \\
	\T_t = \T_1+\T_4=-\T_2-\T_3,
\end{array} \right.
\eeq
where $\T_i$ is the colour-charge operator 
\cite{Catani:1998bh} associated with parton $i$.  

The BFKL equation~\cite{Kuraev:1977fs,Balitsky:1978ic} 
for the wavefunction $\Wf(p,k)$ in eq.~(\ref{ReducedAmpNLL}) 
takes the form
\begin{equation}
	\label{BFKL_evolution}
	\frac{d}{dL}\Omega(p,k)=
	\frac{\alpha_s B_0(\epsilon)}{\pi} \hat{H} \Omega(p,k)\,,
\end{equation}
where $L$ is the high-energy logarithm (\ref{L-def}) and where
the Hamiltonian takes the form \cite{Caron-Huot:2017zfo} 
\beq
\label{Hdef1}
\Ha = \Cone \, \Hi + \Ctwo \, \Hm,
\eeq
where two independent colour factors come along 
with two different operations:
\begin{subequations}
	\label{Hamil}
	\begin{align}
		\Hi \, \Psi(p,k) &= \int \Dkp \, f(p,k,k') \left[ \Psi(p,k') - \Psi(p,k) \right]  \label{Him}, \\
		\Hm \, \Psi(p,k) &= J(p,k) \, \Psi(p,k)\,   \label{eq:Hm}\,.
	\end{align}
\end{subequations}
The function $f(p,k,k')$ in eq.~(\ref{Him})
represents the \ac{bfkl} evolution kernel 
\beq
\label{bfkl-kernel}
f(p,k,k') \equiv \frac{k^2}{(k')^2 (k-k')^2} 
+ \frac{(p-k)^2}{(p-k')^2 (k-k')^2} 
- \frac{p^2}{(k')^2 (p- k')^2}\,,
\eeq
and $J(p,k)$ in eq.~(\ref{eq:Hm}) is defined by
\begin{align}  \label{Jp-def2}
	J(p,k) = \frac{1}{2\eps} + \int \Dkp \, f(p,k,k')
	= \frac{1}{2\eps} \left[2- \pfrac{p^2}{k^2}^{\eps} - \pfrac{p^2}{(p-k)^2}^{\eps} \right].
\end{align}

While it is unknown how to diagonalise this 
$d$-dimensional Hamiltonian, we may invoke 
a perturbative solution~\cite{Caron-Huot:2013fea,Caron-Huot:2017zfo}  
by expanding the wavefunction in the strong coupling constant:
\beqa
\label{OmegaEven}
\Wf(p,k) &=& \sum_{\ell=1}^{\infty} 
\left( \frac{\alpha_s}{\pi} \right)^{\ell} L^{\ell-1} 
\frac{\Bn{0}^{\ell}}{(\ell -1)!} \, \Wfl{\ell-1}(p,k), 
\eeqa
where we set the renormalisation scale equal 
to the momentum transfer, $\mu^2 = -t = p^2$. 

Substituting the expanded form of the wavefunction 
in (\ref{OmegaEven}) into the BFKL evolution equation 
(\ref{BFKL_evolution}) one deduces that 
\beq
\label{Hdef0}
\Wfl{\ell-1}(p,k) = \Ha \Wfl{\ell-2}(p,k),
\eeq
where $\Ha$ is the BFKL hamiltonian of eq.~(\ref{Hdef1}), 
that is, the wavefunction at any given order is found by 
repeated application of the \ac{bfkl} Hamiltonian, where 
the initial condition in our normalization is simply
\beq
\label{0th-wavefunction-tilde}
\Wfl{0}(p,k) = 1\,\,.
\eeq

Next, let us consider the on-shell $2\to 2$ amplitude. 
Substituting the expanded wavefunction (\ref{OmegaEven}) 
into (\ref{ReducedAmpNLL}) we readily obtain the following 
expansion 
\beqa
\label{MhatEven}
\Mhnllp \left( \frac{s}{-t} \right) &=& 
\sum_{\ell=1}^\infty \left( \frac{\as}{\pi} \right)^\ell 
L^{\ell -1} \, \Mhnllpl{\ell},
\eeqa
with 
\beq
\label{ReducedAmpNLL2}
\Mhnllpl{\ell} = -i\pi \frac{\Bn{0}^\ell}{(\ell-1)!} 
\int \Dk \frac{p^2}{k^2(p-k)^2} \Wfl{\ell-1}(p,k) \Tsu \Mtree\, .
\eeq
Namely, integrating over the $(\ell-1)$-th order contribution to 
the wavefunction yields the $\ell$-th order contribution to the 
amplitude.

\begin{figure}[ht]
	\centering
	\includegraphics{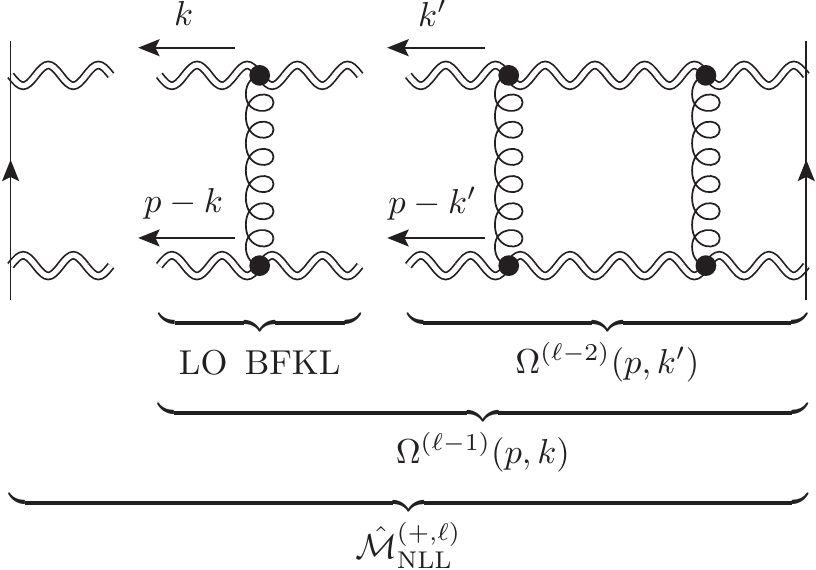}
	\caption{Graphical representation of the amplitude at \ac{nll} accuracy, as 
		obtained through \ac{bfkl} evolution. The addition of one rung corresponds 
		to applying once the leading-order \ac{bfkl} evolution on the wavefunction 
		of order $(\ell-2)$. This gives the wavefunction at order $(\ell-1)$, according 
		to eq.~\eqref{Hdef1}. Closing the ladder and integrating over the resulting 
		loop momentum gives the reduced amplitude, according 
		to eq.~\eqref{ReducedAmpNLL2}.}
	\label{fig:bfklwfamp}
\end{figure}

A graphical illustration of eq.~\eqref{ReducedAmpNLL2} is provided 
in figure \ref{fig:bfklwfamp}. As discussed in the introduction, because of 
\ac{bfkl} evolution, the amplitude at \ac{nll} accuracy can be represented 
as a ladder. At order $\ell$ it is obtained by closing the ladder and integrating 
the wavefunction of order $(\ell-1)$ over the resulting loop momentum, 
according to eq.~\eqref{ReducedAmpNLL2}. The wavefunction 
$\Wfl{\ell-1}(p,k)$ in turn is obtained by applying once the leading-order 
\ac{bfkl} evolution kernel to the wavefunction of order $(\ell-2)$. 
Graphically, this operation corresponds to adding one rung to 
the ladder. 

Inspecting \eqns{Him}{eq:Hm} we see that the
BFKL evolution consists of an integration and a multiplication
part. The effect of evolution is thus expressed formally in a 
compact form by introducing a class of functions
\begin{subequations}
	\begin{align}
		\label{J_im_general}
		\Wf_{\mathrm{i},w}(p,k) &\equiv \int \Dkp f(p,k,k') \left[ \Wf_{w}(p,k') - \Wf_{w}(p,k) \right],  \\
		\Wf_{\mathrm{m},w}(p,k) &\equiv J(p,k) \, \Wf_{w}(p,k) \label{eq:Wmdef}, 
	\end{align}
\end{subequations}
where $\Wf_{\varnothing}(p,k) \equiv 1$, and $w$ indicates a word 
made of indices ``i'' or ``m'', which stand for integration and multiplication, 
respectively, according to the action of the two Hamiltonian operators in 
eq.~\eqref{Him} and (\ref{eq:Hm}), respectively. In this notation the first 
four orders of the wavefunction read, for instance,
\begin{align}
	\Wfl{1}(p,k) &= \Ctwo \Wf_{\rm m},  \label{WavefunctionTwoLoops-b} \\
	\Wfl{2}(p,k) &= \Ctwo^2 \Wf_{\rm m,m} + \Cone \Ctwo \Wf_{\rm i,m},  \label{WavefunctionTwoLoops-b2} \\ \nn
	\Wfl{3}(p,k) &= \Ctwo^3 \Wf_{\rm m,m,m} + \Cone \Ctwo^2 \left( \Wf_{\rm i,m,m} + \Wf_{\rm m,i,m} \right) \\
	& +\,  \Cone^2 \Ctwo \Wf_{\rm i,i,m},  \label{eq:wf3loops}  \\ \nn
	\Wfl{4}(p,k) &= \Ctwo^4 \Wf_{\rm m,m,m,m} \\ \nn
	& +\, \Cone \Ctwo^3 \left( \Wf_{\rm m,m,i,m} + \Wf_{\rm m,i,m,m}  + \Wf_{\rm i,m,m,m}  \right) \\ \nn
	& +\,  \Cone^2 \Ctwo^2 \left( \Wf_{\rm m,i,i,m} + \Wf_{\rm i,m,i,m} + \Wf_{\rm i,i,m,m} \right) \\
	& +\,  \Cone^3 \Ctwo \Wf_{\rm i,i,i,m}. \label{eq:wf4loops}
\end{align}

Symmetries play an important role in determining the general 
structure of the wavefunction, and from a practical perspective
they can be useful to reduce the number of integrals that need 
to be evaluated at each loop order. The wavefunction is symmetric 
under swapping the two $t$-channel Reggeons, which can be 
understood from the graphical representation of the \ac{bfkl} 
evolution in figure~\ref{fig:bfklwfamp}. This 
implies
\beq\label{left-right-symmetry}
\Wfl{\ell}(p,k) = \Wfl{\ell}(p,p-k)\,,
\eeq
which can be easily verified by showing that the functions $f(p,k,k')$ 
in \eqref{bfkl-kernel}, $J(p,k)$ in \eqref{Jp-def2} and $\Wfl{0}(p,k)$ 
in \eqref{0th-wavefunction-tilde} obey the same symmetry.
This symmetry property will become handy in section \ref{soft}, 
making it possible to capture simultaneously both soft limits, 
$k^2\to 0$ and $(p-k)^2\to 0$. This, in turn, will be important 
for implementing a systematic separation between the soft 
and hard regimes, without needing an extra regulator.  

\begin{figure}[ht]
	\centering
	\includegraphics[width=0.72\textwidth]{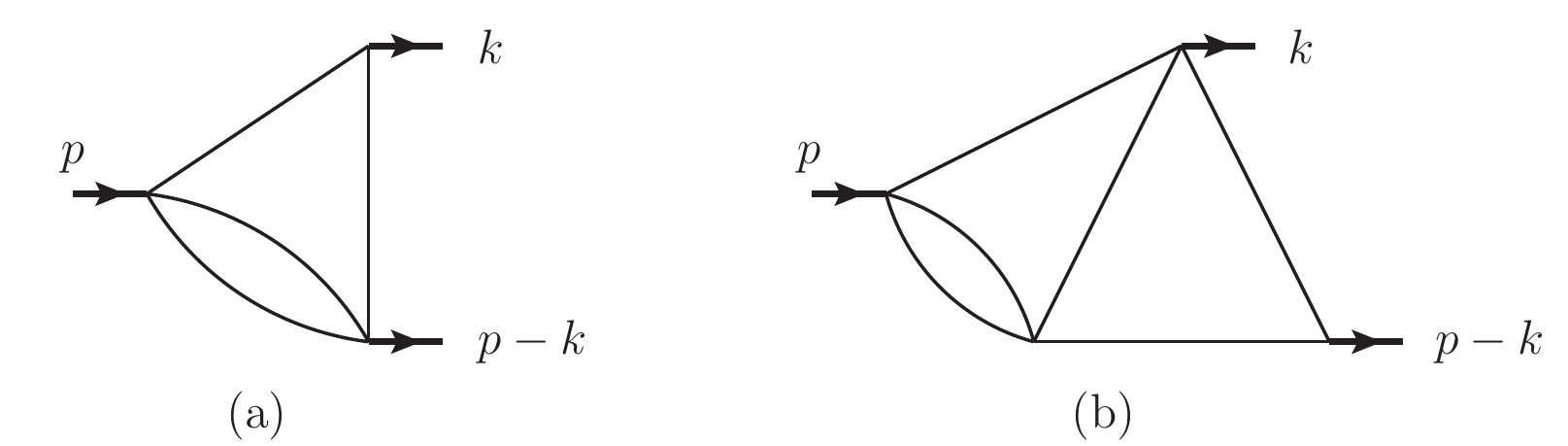}
	\caption{Three-mass triangle integrals with massless propagators, 
		which appear in the calculation  of the wavefunction at two and three loops. 
		These integrals contribute to the amplitude only starting respectively 
		at four and five loops, due to symmetry constraints, as discussed in the main 
		text. The bubble integral on one of the edges of the triangle clarifies the origin 
		of the propagator which is raised to power $\eps$ in eq.~\eqref{triang1}.}
	\label{fig:triang}
\end{figure}

Despite the simplifications allowed by symmetries, though, the
evaluation of the wavefunction in $2-2\eps$ transverse dimensions
without additional simplifications becomes quickly infeasible. 
For instance, already the wavefunctions with one or two 
integrations (one or two occurrences of the index ``i'') involve integrals 
of the type 
\begin{align}
	\label{triang1} \nn
	\Wf_{\rm i,m}  &\ni \int \Dkp \, 
	\frac{(p-k)^2}{(p-k')^2 (k-k')^2} \pfrac{p^2}{(k')^2}^{\eps}, \\ 
	\Wf_{\rm i,i,m}  &\ni  \int \Dkp  [Dk''] \, 
	\frac{k^2 (p-k'')^2}{(k'')^2  (p-k')^2 (k-k'')^2 (k'-k'')^2} \pfrac{p^2}{(k')^2}^{\eps},
\end{align}
which are represented respectively in figure~\ref{fig:triang} 
(a) and (b). Such integrals evaluates to Appell, and more 
in general Lauricella functions in dimensional regularisation. 
Given the lack of a systematic classification of these functions
in terms of iterated integrals, the evaluation of the wavefunction
beyond the third order is not practical.

\begin{figure}[ht]
	\centering
	\includegraphics[width=0.76\textwidth]{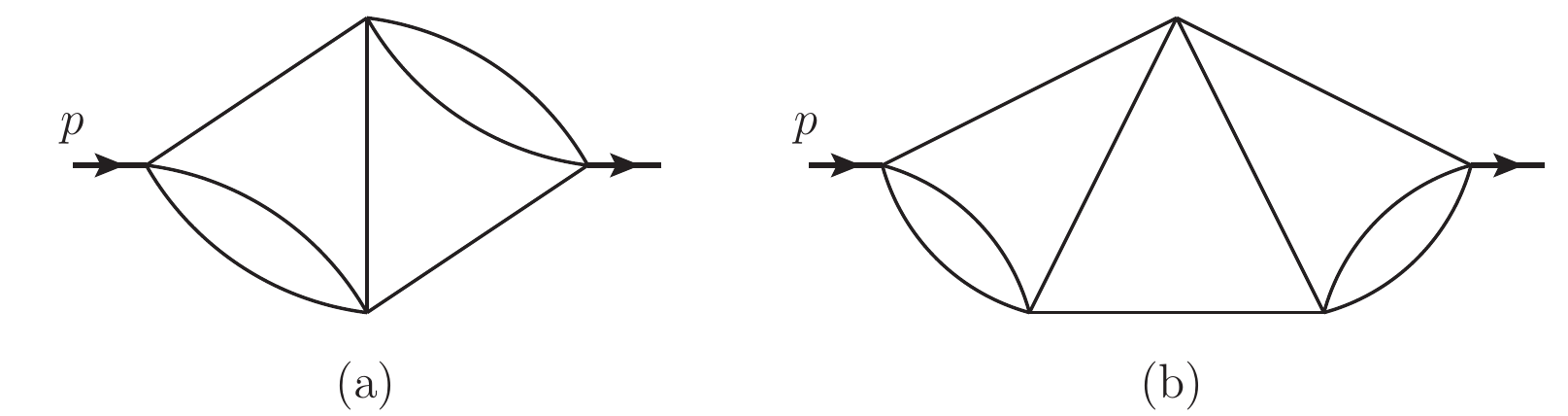}
	\caption{Example of a four- and five-loop integrals that enters the calculation 
		of the four- and five-loop amplitude respectively. The two bubbles may be integrated 
		out, turning it into a two- and three-loop integral with two propagators raised to 
		non-integer powers, \emph{cf.}\ eq.~\eqref{eq:5loopex}.}
	\label{fig:5loopex}
\end{figure}

The amplitude at order $\ell$ is obtained upon integrating the 
wavefunction of order $\ell-1$, as indicated in \eqn{ReducedAmpNLL2}. 
As in case of the wavefunction, symmetries turn out to be important for 
a simplification of the calculation and interpretation of the result. 
While the two Reggeons in the wavefunction can be \emph{defined} to 
originate from either the projectile \emph{or} target Wilson line --- which 
gives the corresponding ladder graphs a sense of direction --- this is no 
longer true at the level of the amplitude. Physically the two cases become 
indistinguishable, and we refer to this as the target-projectile symmetry. 
In general, this implies the relation~\cite{Caron-Huot:2017zfo}
\beq
\label{JiJmSymAllOrders}
\int \Dk \frac{p^2}{k^2 (p-k)^2} \Hi \Wf_{w}(p,k) = 
\int \Dk \frac{p^2}{k^2 (p-k)^2} \Wf_{\mathrm{i},w}(p,k) = 0.
\eeq
Furthermore, in the notation of eqs.~\eqref{J_im_general} and 
\eqref{eq:Wmdef} reversal of the rungs directly translates to the 
reversal of the indices of the wavefunction. The target-projectile 
symmetry thus guarantees the equality
\beq
\int \Dk \frac{p^2}{k^2 (p-k)^2} \Wf_{a_1,\dots,a_n}(p,k) 
= \int \Dk \frac{p^2}{k^2 (p-k)^2} \Wf_{a_n,\dots,a_1}(p,k).
\eeq
The symmetries discussed above can reduce the number of functions to be computed 
significantly, and make the calculation of the amplitude trivial up to three loops, since
it can be shown that the integration of the wavefunction involves only bubble integrals.
Furthermore, the calculation of the amplitude at four loops in dimensional regularisation 
is still feasible, as it involve bubble integrals and a single more involved kite-like integral, 
represented in figure \ref{fig:5loopex} (a). Up to four loops one obtains 
\cite{Caron-Huot:2017zfo}
\begin{align}
	\label{ReducedAmpNLL2-one-loop}
	\Mhnllpl{1} &= i\pi \frac{\Bn{0}}{2\eps} \Tsu \Mtree,  \\
	\label{ReducedAmpNLL2-two-loop} \nn
	\Mhnllpl{2} &= i\pi \frac{(\Bn{0})^2}{2} \left[ \frac{1}{(2\eps)^2} 
	+ \frac{9\zeta_3}{2}\eps + \frac{27\zeta_4}{4}\eps^2 
	+ \frac{63\zeta_5}{2}\eps^3 + \ord(\eps^4) \right]  \\
	&\hspace{55mm} \times \Ctwo \Tsu \Mtree, \\ \nn
	\label{ReducedAmpNLL2-three-loops}
	\Mhnllpl{3} &= i\pi \frac{\Bn{0}^3}{3!} \left[ \frac{1}{(2\eps)^3} 
	- \frac{11\zeta_3}{4} - \frac{33\zeta_4}{8}\eps 
	- \frac{357\zeta_5}{4}\eps^2 + \ord(\eps^3) \right] \\
	& \hspace{55mm} \times \Ctwo^2 \Tsu \Mtree, \\ \nn
	\label{ReducedAmpNLL2-four-loops}
	\Mhnllpl{4} &= i\pi \frac{\Bn{0}^4}{4!} 
	\bigg\{ \Ctwo^3 \left( \frac{1}{(2\eps)^4} 
	+ \frac{175\zeta_5}{2}\eps + \ord(\eps^2) \right) \\ 
	& \hspace{8mm} + \Ca \Ctwo^2 \left( -\frac{\zeta_3}{8\eps} 
	- \frac{3}{16}\zeta_4 - \frac{167\zeta_5}{8}\eps 
	+ \ord(\eps^2) \right) \bigg\} \Tsu \Mtree . 
\end{align}
A thorough discussion of the target-projectile symmetry, and its 
effect on the colour structure of the amplitude has been given 
in \cite{Caron-Huot:2017zfo}, to which we refer for further details. 
In this paper we are interested to evaluate the amplitude, including 
finite terms, to higher orders in the perturbative expansion. Despite
the symmetries discussed above, however, beyond four loops the 
iterated integrals appearing are all but easy with current methods. 

A simple and fast way to extend the study in 
ref.~\cite{Caron-Huot:2013fea,Caron-Huot:2017zfo} 
to higher loops is provided by numerical integration methods. 
In particular, we find sector decomposition as implemented in 
\texttt{pySecDec}/\texttt{SecDec} \cite{Carter:2010hi,Borowka:2017esm} 
to be suited to calculate the nested integrals that enter the five-loop 
amplitude. Provided a high numerical accuracy it is straightforward 
to extract from the results the rational coefficients of the  
zeta numbers appearing at this loop order. This procedure relies on the 
observed \emph{homogeneous transcendental weight} property of the 
$\ell$-loop amplitude: Assigning $o(\eps) = -1$, $o(\pi) = 1$ and 
$o(\zeta_n) = n$ one sees that the terms of the $\ell$-loop amplitude 
are uniformly of weight $o(\Mhnllpl{\ell} ) = \ell$. We can hence deduce 
which zeta numbers (or powers of $\pi$) may appear at any given 
order in $\eps$. 

Another observation facilitates this procedure at five loops; after 
dividing the $\ell$-loop amplitude by $B_0^\ell$ \eqref{B0} 
there are no occurrences of $\zeta_2 = \pi^2/6$ up to four loops, 
see e.g.\ the $\ord(\eps)$ terms of eq.~\eqref{ReducedAmpNLL2-four-loops}. 
If we assume this absence of $\zeta_2$ to be an actual property 
of the amplitude, the finite terms of the five-loop amplitude can 
only be proportional to one transcendental number, $\zeta_5$, 
whereas $\zeta_3 \zeta_2$ is excluded. At this point this approach 
may seem rather conjectural. However, over the course of the next 
two sections we develop methods that prove this assumption, 
and we shall briefly return to it at the end of section~\ref{sec:finiteamp}. 

To obtain the five-loop amplitude $\Mhnllpl{5}$ we integrate the 
four-loop wavefunction $\Wfl{4}(p,k)$ of~\eqref{eq:wf4loops} according 
to eq.~\eqref{ReducedAmpNLL2}. In doing so one is faced with a 
plethora of multi-loop integrals. Many of them correspond to bubble 
graphs and can be easily evaluated analytically.
Others vanish because of the symmetries discussed above. The 
remaining integrals can be computed numerically using \texttt{pySecDec}. 
One of the more difficult examples is shown in 
figure~\ref{fig:5loopex}. In the depicted case one can integrate out 
the two internal bubbles and is left with a three-loop integral with 
two of the propagators raised to non-integer powers:
\beq \label{eq:5loopex}
\text{figure~\ref{fig:5loopex} (b)} \sim \int 
\frac{\Dk \Dkp [\Dd k'']\, (p-k')^2}{
	(k^2)^\eps  (k')^2 ((k'')^2)^\eps (k-k')^2 
	(k'-k'')^2 (p-k)^2 (p-k'')^2}. 
\eeq
After combining all contributions (and reconstructing the zeta 
numbers in case of the numerical results) we find
\begin{multline}
	\Mhnllpl{5} = i\pi \frac{\Bn{0}^5}{5!} 
	\left\{ \Ctwo^4 \left( \frac{1}{32 \eps^5} - \frac{53 \zeta_5}{2} \right) \right. \\
	\left. + \Ca \Ctwo^3 \left( -\frac{\zeta_3}{16 \eps^2} - \frac{3 \zeta_4}{32 \eps} 
	+ \frac{253 \zeta_5}{16} \right) - \frac{5}{2} \Ca^2 \Ctwo^2 \zeta_5 \right\} 
	\Tsu \Mtree. \label{eq:m5num}
\end{multline}
This result will serve as a consistency check for our computation below.


\section{The soft approximation} \label{soft}

In section \ref{chap:bfkl} we have shown how 
the two-Reggeon contribution to the two-parton 
scattering amplitude is conveniently described 
in terms of the reduced amplitude $\hat {\cal M}$. 
The latter is defined in~\eqn{Mreduced} by 
(multiplicatively) removing the single-Reggeon 
effect from the full amplitude~${\cal M}$. This 
allowed us to use BFKL evolution to express the 
two-Reggeon contribution to $\hat {\cal M}$ in 
terms of iterated integrals. Beyond four loops 
these integrals become difficult to
evaluate exactly in $d = 4-2\eps$ dimensions, 
but as we are going to show now, this is also 
not necessary. 

Ultimately we are interested in extracting physical 
information about the scattering process, and 
dimensional regularization is used in the present 
context for the sole purpose of regularizing 
long-distance singularities\footnote{Note that 
ultraviolet renormalization is irrelevant for the 
signature-even amplitude at the logarithmic 
accuracy considered.}. Here infrared factorization 
come into play: the long-distance singularities of 
${\cal M}$ can be factorized, ${\cal M}={\bf Z} 
\Hhard$, where the ``infrared renormalization'' 
factor ${\bf Z}$ captures all divergences (which 
famously exponentiate in terms of the soft 
anomalous dimension, see e.g. \cite{Sterman:1995fz,Collins:1989gx,Korchemskaya:1994qp,Catani:1998bh,Aybat:2006mz,Sterman:2002qn,Gardi:2009qi,Becher:2009cu,Becher:2009qa,Almelid:2015jia,Almelid:2017qju}) 
while the infrared-renormalized amplitude $\Hhard$ -- 
sometimes referred to as the ``hard function'' -- is finite, 
and can  be evaluated in four space-time dimensions 
(or equivalently, two transverse dimensions).
To understand this from a physical perspective recall 
that physical quantities such as cross sections are 
finite: starting from the infrared-singular amplitude 
${\cal M}$, their calculation inevitably incorporates a 
mechanism of cancellation of the singularities involving 
soft real-gluon emission. Once this was implemented, 
the finite, physical result can only depend on 
four-dimensional quantities, namely the \emph{soft 
anomalous dimension} and the \emph{infrared-renormalized 
amplitude}~$\Hhard$.

In Ref.~\cite{Caron-Huot:2017zfo} we have shown 
that the soft anomalous dimension associated with 
the signature-even amplitude, or indeed the relevant 
infrared renormalization factor ${\bf Z}$, can be 
computed to all orders by evaluating the reduced 
amplitude $\hat\MM$ to $\ord(\eps^{-1})$.
Similarly, we are going to show now (section 
\ref{sec:IRfact}) that the infrared-renormalized 
amplitude $\Hhard$ (in four dimensions) can be 
completely determined from the reduced amplitude 
$\hat\MM$, evaluated at the same accuracy, i.e. to 
$\ord(\eps^0)$. This, along with the fact that the 
corresponding wavefunction $\Omega$ is finite, 
greatly simplifies the task of performing BFKL 
evolution to high loop orders, because it allows 
us to follow an ``expansion by region'' approach: 
in section~\ref{sec:soft_hard_split}  we split the 
wavefunction into soft and hard components, 
each of which is rendered computable using 
different considerations. The soft wavefunction -- 
giving rise to all the singularities in the amplitude -- 
can be computed analytically in dimensional 
regularization owing to the drastic simplification 
of BFKL evolution in this limit, while the hard 
wavefunction is only required in strictly two 
transverse dimensions, where BFKL evolution 
again simplifies (see section~\ref{2d-bfkl}).
These two wavefunction components will 
subsequently serve to compute the corresponding 
soft and hard contributions to the reduced amplitude 
$\hat\MM$ to the required order, $\ord(\eps^0)$. 
In section~\ref{sec:softwave} we review the main 
results of Ref.~\cite{Caron-Huot:2017zfo} regarding 
the all-order computation of the wavefunction within 
the soft approximation. We also introduce there a 
symmetrized soft wavefunction which captures both 
soft limits. This, in turn, is used in section~\ref{sec:softAmpl} 
to compute the corresponding $\ord(\eps^0)$ contributions 
to the reduced amplitude. Finally, in section~\ref{hardFdef} 
we make use of the results of sections \ref{sec:IRfact} and 
\ref{sec:softAmpl}  to evaluate the $\ord(\eps^0)$ soft 
contributions to the infrared-renormalized 
amplitude~$\Hhard$.

\subsection{Infrared factorisation 
in the high-energy limit\label{sec:IRfact}}

According to the infrared factorisation theorem (see e.g. \cite{Sterman:1995fz,Collins:1989gx,Korchemskaya:1994qp,Catani:1998bh,Aybat:2006mz,Sterman:2002qn,Gardi:2009qi,Becher:2009cu,Becher:2009qa,Almelid:2015jia,Almelid:2017qju}), 
infrared singularities of an amplitude $\MM$ are 
multiplicatively renormalised by a factor ${\bf Z}$,
\beq \label{IRfacteq}
\MM \left(\{p_i\},\mu, \as (\mu) \right) \, = \, 
{\bf Z} \left(\{p_i\},\mu, \as (\mu) \right) \Hhard \left(\{p_i\},\mu, \as (\mu) \right)\,,
\eeq
such that the infrared-renormalized amplitude ${\cal H}$ is finite
as $\eps\to 0$. We use a minimal subtraction scheme, 
where the renormalisation factor ${\bf Z}$ consists of pure poles. 
It is then given explicitly as the path-ordered exponential 
of the soft anomalous dimension:
\beq \label{RGsol}
{\bf Z} \left(\{p_i\},\mu, \as (\mu) \right) \, = \,  
{\cal P} \exp \left\{ -\int_0^{\mu} \frac{d\lambda}{\lambda}\,
{\bf \Gamma} \left(\{p_i\},\lambda, \as(\lambda) \right) \right\},
\eeq
where, to the accuracy needed in this paper, we can restrict to 
tree-level running coupling: $\as(\lambda) = \as(p)
\left(p^2/\lambda^2\right)^{\eps}$. 
Given that ${\bf Z}$ was determined in Ref.~\cite{Caron-Huot:2017zfo} 
to NLL accuracy in the high-energy logarithm, 
our goal here is to determine the infrared-renormalized 
amplitude~${\cal H}$ to the same accuracy. 
Thus we need to specialise \eqn{IRfacteq} to the 
high-energy limit. Recalling that in this limit the 
amplitude splits naturally into even and odd components 
under the $s \leftrightarrow u$ signature symmetry, we 
may focus directly on the even component (the odd 
component was analysed already in~\cite{Caron-Huot:2017fxr}): 
\be	\label{MtoHeven}
\MM^{(+)}_{\rm NLL}  =  {\bf Z}^{(-)}_{\rm NLL} \, \Hhard^{(-)}_{\rm LL} 
+ {\bf Z}^{(+)}_{\rm LL} \, \Hhard^{(+)}_{\rm NLL}.
\ee
Our final goal is to determine $\Hhard^{(+)}_{\rm NLL}$. 
Let us begin by inverting (\ref{MtoHeven}), i.e.
\be \label{getH}
\Hhard^{(+)}_{\rm NLL} 
= - \big({\bf Z}^{-1}\big)^{(+)}_{\rm LL} {\bf Z}^{(-)}_{\rm NLL} \, \Hhard^{(-)}_{\rm LL}
+\big({\bf Z}^{-1}\big)^{(+)}_{\rm LL} \MM^{(+)}_{\rm NLL}.
\ee
In \eqn{getH} both the leading- and next-to-leading 
logarithmic renormalisation factors are known: 
${\bf Z}^{(+)}_{\rm LL}$, and hence also 
$\big({\bf Z}^{-1}\big)^{(+)}_{\rm LL}$ is easily 
determined from the single-Reggeon exchange,
see \eqns{Mreal}{alphag1}:
\be\label{Zll}
{\bf Z}^{(+)}_{\rm LL} = e^{\frac{x}{2\eps} \T_t^2}\,\qquad\quad\Longrightarrow \qquad\quad 
\big({\bf Z}^{-1}\big)^{(+)}_{\rm LL}= e^{-\frac{x}{2\eps} \T_t^2}\,,
\ee
where we defined $x\equiv \frac{\alpha_s}{\pi} L$.
The factor ${\bf Z}^{(-)}_{\rm NLL}$ was
determined to all orders in perturbation theory in 
\cite{Caron-Huot:2017zfo}: comparing eqs.~(4.12),~(4.14) 
and~(4.17) there we express ${\bf Z}^{(-)}_{\rm NLL} $ 
as 
\be\label{Znll}
e^{-\frac{x}{2\eps} \T_t^2} \, 
{\bf Z}^{(-)}_{\rm NLL} 
= i\pi \, \frac{ e^{\frac{x}{2\eps} (\CA-\Tt)} - 1}{L(\CA-\Tt)}
\left( 1 - \frac{C_A}{C_A -\Tt} R(\eps) \right)^{-1}
\Tsu \Bigg|_{\rm poles},
\ee
where the function $R(\eps)$ reads
\bea \nn
R(\eps)  &=& 
\frac{\Gamma^{3}(1-\eps)\Gamma(1+\eps)}{\Gamma(1-2\eps)} -1 \\
\label{Rdef}
&=& 
-2\zeta_3 \, \eps^3 -3\zeta_4 \, \eps^4 -6\zeta_5 \eps^5
-\left(10 \zeta_6-2\zeta^2_3 \right) \eps^6 + {\cal O}(\eps^7),
\eea 
see also eq.~(3.16) of \cite{Caron-Huot:2017zfo}.
In \eqn{Znll} the factor $\exp[- (x \T_t^2)/(2\eps)]$ 
on the l.h.s.~is left there, because it corresponds directly 
to the factor $\big({\bf Z}^{-1}\big)^{(+)}_{\rm LL}$ appearing
in \eqn{getH} to the left of~${\bf Z}^{(-)}_{\rm NLL}$. 
Notice also that the $-1$ in the numerator of the first fraction 
on the r.h.s.~of \eqn{Znll} can actually be removed, given 
that we need to consider only the poles originating from 
\eqn{Znll}, and this term contributes only at 
$\ord(\eps^0)$, given that the second $\epsilon$-dependent 
factor is regular, i.e. $\big[ 1 - C_A/(C_A -\Tt) 
\, R(\eps) \big]^{-1} = 1 + \ord(\eps^3)$. 

\Eqn{getH} contains also the leading-logarithmic 
infrared-renormalized amplitude $\Hhard^{(-)}_{\rm LL}$, 
which, as in case of ${\bf Z}^{(+)}_{\rm LL}$, is determined 
by single-Reggeon exchange, compare again with 
\eqns{Mreal}{alphag1}:
\be\label{Hll}
\Hhard^{(-)}_{\rm LL} = e^{ \frac{\bn{0}(\eps)-1}{2\eps} \, x \, C_A} 
\MM^{\rm (tree)}\,,
\ee
where we have substituted $\Tt \to C_A$, given that
in $\Hhard^{(-)}_{\rm LL}$ the operator $\Tt$ acts 
on the tree-level amplitude. 

By this point we collected all ingredients needed to explicitly 
write down the first term in \eqn{getH}. The only missing term 
on the r.h.s. of this equation is thus the even amplitude itself, 
$\MM^{(+)}_{\rm NLL}$. As explained above, in order to 
determine $\MM^{(+)}_{\rm NLL}$ by means of BFKL evolution, 
we wish to express it in terms of  the reduced amplitude 
${\cal \hat M}^{(+)}_{\rm NLL}$ of \eqn{Mreduced}. Substituting
eqs.~(\ref{Mreduced}),~(\ref{Zll}) and~(\ref{Hll}) into 
\eqn{getH} we get 
\begin{equation} \label{getH2}
	\Hhard^{(+)}_{\rm NLL} 
	= - e^{-\frac{x}{2\eps} \T_t^2} {\bf Z}^{(-)}_{\rm NLL} 
	\, e^{\frac{\bn{0}(\eps)-1}{2\eps} \, x \, C_A} \MM^{\rm (tree)}
	+e^{\frac{B_0(\eps)-1}{2\eps} \, x \, \T_t^2} {\cal \hat M}^{(+)}_{\rm NLL}\,\,,
\end{equation}
where the factor $e^{-\frac{x}{2\eps} \T_t^2} {\bf Z}^{(-)}_{\rm NLL}$ 
of~(\ref{Znll}) can be readily substituted as well (this will be done in 
section~\ref{hardFdef}). Eq.~(\ref{getH2}) is an important step 
because (given that $B_0(\eps)-1 = {\cal O}(\eps^2)$, eq.~(\ref{B0}))
it clearly shows that the hard function $\Hhard^{(+)}_{\rm NLL} $ at 
$\epsilon\to 0$ is completely determined once the BFKL-motivated 
reduced amplitude ${\cal \hat M}^{(+)}_{\rm NLL}$ is known to 
$\ord(\eps^0)$, which is the result anticipated at the beginning 
of this section. With this in mind, we proceed to compute 
${\cal \hat M}^{(+)}_{\rm NLL}$ to $\ord(\eps^0)$.

\subsection{Soft and hard wavefunction and amplitude}
\label{sec:soft_hard_split}

Our strategy to compute the finite part of the reduced 
amplitude ${\cal \hat M}^{(+)}_{\rm NLL}$ at higher orders 
is to separate soft and hard components of the wavefunction 
and truncate the latter to two transverse dimensions ($\eps=0$), 
where BFKL evolution is much more tractable (see section~\ref{2d-bfkl}).

As demonstrated in ref.~\cite{Caron-Huot:2017zfo}, 
the soft limit of the wavefunction, where one of the two 
Reggeons has a small momentum, e.g., $k^2\ll (p-k)^2\simeq p^2$, 
fully determines all the singular parts in $\epsilon$.
This was used to obtain the all-order result for the 
renormalisation factor ${\bf Z}^{(-)}_{\rm NLL} $ in 
\eqn{Znll}. In addition, the soft limit generates some 
${\cal O}(\eps^0)$ finite contributions, which must be 
added to those generated by the complementary hard 
region, where both $k^2$ and $(p-k)^2$ are of order $p^2$.

To control ${\cal O}(\eps^0)$ terms a clear separation 
between the two regions is necessary. We choose to do 
this at the level of the wavefunction $\Wf(p,q)$.  Recall 
that $\Wf(p,q)$ is a finite function\footnote{This is a direct 
consequence of the fact that we have removed  the factor 
of the gluon Regge trajectory in defining the reduced 
amplitude in~\eqn{Mreduced}.} of $\eps$~\cite{Caron-Huot:2017zfo}, 
i.e. any singularities in the reduced amplitude are 
generated through the final integration over the 
wavefunction in (\ref{ReducedAmpNLL}).
To proceed we split the wavefunction into 
two terms:
\beq \label{OmegaSplitDef}
\Wf(p,k) = \Ws(p,k) + \Wh(p,k)\,,
\eeq
such that the second term, the hard component, 
vanishes in soft limits:
\beq
\label{h_definition}
\lim_{k\to 0}\Wh(p,k) = \lim_{k\to p}\Wh(p,k) =0.
\eeq
It then follows from (\ref{ReducedAmpNLL}) that no 
singularities can be generated upon integrating $\Wh(p,k)$ 
(i.e. all the singularities in $\Mhnllp$ are generated upon 
integrating  $\Ws(p,k)$) and hence only the $\eps\to 0$ 
limit of $\Wh$ contributes to the finite part of the 
reduced amplitude. Denoting the wavefunction 
in this limit as 
\beq \label{WhardTwod}
\Wh^{({\rm 2d})}(p,k)\,\equiv \, \lim_{\eps \to 0} \Wh 
= \Wtd(p,k) - \Ws^{({\rm 2d})}(p,k),
\eeq
the reduced amplitude (\ref{ReducedAmpNLL}), 
through order ${\cal O}(\eps^0)$, is then given as 
a sum of soft and hard components:
\bea \label{eq:redampSplit} 
\Mhnllp \left(\frac{s}{-t}\right) = 
\hat {\cal M}^{(+)}_{\rm NLL,s}\left(\frac{s}{-t}\right) 
+ \hat {\cal M}^{(+)}_{\rm NLL,h}\left(\frac{s}{-t}\right) 
\eea
with
\begin{subequations} 
\label{Msh}
\begin{align}
	\hat {\cal M}^{(+)}_{\rm NLL,s}\left(\frac{s}{-t}\right) 
	&=-i\pi  \int \Dk \,\frac{p^2}{k^2(p-k)^2} \Ws(p,k)\,\,  \Tsu \Mtree_{ij\to ij}\,, \\
\label{Mhard}
	\hat {\cal M}^{(+)}_{\rm NLL,h}\left(\frac{s}{-t}\right) 
	&= -i\pi\lim_{\eps\to 0}\int \Dk \,\frac{p^2}{k^2(p-k)^2}\,
	\Wh^{({\rm 2d})}(p,k)\,\, \Tsu \Mtree_{ij\to ij}\, .
\end{align}
\end{subequations}
Equations (\ref{eq:redampSplit}) and (\ref{Msh}) are 
central to our approach and will guide our computations 
in what follows. They show that, to compute the finite part 
of the reduced amplitude, we must treat the soft wavefunction 
exactly as a function of $\eps$, but we are allowed to truncate 
the hard wavefunction to ${\cal O}(\eps^0)$. Note that in 
(\ref{Mhard}) we have already substituted the two-dimensional 
limit of the hard wavefunction, so taking the $\eps\to 0$ limit 
simply amounts to taking the integration momentum $k$ to 
be two-dimensional. These finite integrals will be done in 
section~\ref{amplitude}.

Let us briefly summarise our plan for the reminder of this section.
After reviewing the main arguments of~\cite{Caron-Huot:2017zfo}, 
our aim in section~\ref{sec:softwave} is to present a symmetrized 
version of the soft wavefunction in dimensional regularization, 
eq.~(\ref{Well-1-ansatz-sym}), which simultaneously captures 
the two regions where either of the two Reggeons is soft. We 
then extract the ${\cal O}(\epsilon^0)$ terms in the wavefunction 
and resum them; these will be used in section~\ref{amplitude} to 
determine the two-dimensional hard wavefunction 
$\Wh^{({\rm 2d})}(p,k)$ from the full one according to 
eq.~(\ref{WhardTwod}). Subsequently in section~\ref{sec:softAmpl} 
we use the soft wavefunction, computed to all-orders in $\epsilon$,
to determine the corresponding contributions to the reduced
$2\to 2$ amplitude. We also present an analytic formula 
resumming these corrections in eq.~(\ref{ReducedAmpNLLresum2}).
Finally, in section~\ref{hardFdef} we determine the soft wavefunction 
contribution to the infrared-renormalized 
amplitude~$\Hhard^{(+)}_{\rm NLL} $ 
using eq.~(\ref{getH2}).

\subsection{The soft wavefunction}\label{sec:softwave}

The central property of the wavefunction $\Omega(p,k)$ 
highlighted in \cite{Caron-Huot:2017zfo} and already 
mentioned above, is the fact that it is finite for $\eps \to 0$, 
to all orders in perturbation theory. This has far reaching 
consequences, because it means that all singularities in 
the amplitude  must arise from the last integration in 
(\ref{ReducedAmpNLL}), and originate from the soft limits 
$k \to 0$ and $k \to p$ of $\Omega(p,k)$.
One  finds that it is particularly easy to calculate the 
wavefunction in these limits: as it turns out, the soft 
approximation is closed under BFKL evolution, i.e., 
starting with $\Wt{j}(p,k)$, with $k$ soft, implies that 
the momentum $k'$ in $\Wt{j-1}(p,k')$, which has one 
rung fewer, can also be taken soft, $k' \to 0$, without 
affecting the result for $\Wt{j}(p,k)$. In other words, 
starting with $\Wt{j}(p,k)$ where $k$ is soft is equivalent 
to considering the entire side rail of the ladder consisting 
of  soft momenta, $k'$, $k''$, $\ldots\to 0$. Similarly, 
starting with $k \to p$ implies that all momenta $(p-k)$, 
$(p-k')$, $(p-k'')$, $\ldots$, are soft. The symmetry of 
\eqn{left-right-symmetry}, then, implies that 
$\Omega(p,k)$ in the two limits $k \to 0$ and 
$k \to p$ must be the same. 

In the soft limit the BFKL hamiltonian becomes
\cite{Caron-Huot:2017zfo}
\bea \nn
\Omega^{(\ell-1)}_s(p,k)&=&\hat H_s \, \Omega^{(\ell-2)}_s(p,k)\,, \\ \nn
\hat H_s \Psi(p,k) &=& (2C_A-\Tt)\int [Dk']
\frac{2(k\cdot k')}{k'^2(k - k')^2} \, \Big[ \Psi(p,k')-\Psi(p,k)\Big] \\&& 
+\, (C_A-\Tt)\, J_s(p,k) \, \Psi(p,k)\,, \label{softH}
\eea
where 
\be\label{JpSoft}
J_s(p,k) = \frac{1}{2\eps} \left[1-\mratio{p}{k}^{\eps}\,\right], 
\ee
is the soft approximation of \eqn{Jp-def2}. One finds that 
the wavefunction becomes a polynomial in $ \xi \equiv (p^2/k^2)^{\eps}$, 
i.e., the soft limit turns BFKL evolution into a one-scale problem.
The integrals involved in \eqn{softH} are simple bubble integrals 
of the type 
\be\label{bubbleGeneral1}
\int [{\rm D}k'] \, \frac{2(k\cdot k')}{k'^2(k - k')^2} \, 
\mratio{p}{k'}^{n \eps} =
-\frac{1}{2\eps}\frac{\bn{n}(\eps)}{\bn{0}(\eps)}\mratio{p}{k}^{(n+1)\eps}\,,
\ee
where the integration measure is given in eq.~(\ref{measure}),
and the class of bubble functions $\bn{n}(\eps)$ is
\be \label{bubbleGeneral2}
\bn{n}(\eps) = e^{\eps\gamma_{\rm E}}  \frac{\Gamma(1-\eps)}{\Gamma(1+n\eps)}
\frac{\Gamma(1+\eps + n\eps) \Gamma(1-\eps - n\eps)}{\Gamma(1-2 \eps - n\eps)}\,.
\ee
Note that $B_0$ of~(\ref{B0}) appearing in the gluon 
Regge trajectory and in the measure (\ref{measure}) 
corresponds to the special case of (\ref{bubbleGeneral2}) 
with $n=0$.

Using eq.~(\ref{bubbleGeneral1}) one can write the action 
of the soft Hamiltonian (\ref{softH}) on any monomial ($m\geq 0$):
\bea
\label{softHpower}
\hat H_s \,\xi^m &=& \frac{\xi^m}{2\eps} \left((1-\xi)(C_A-\Tt) 
+ \xi \hbn{m}(\eps)(2C_A-\Tt)\right)\,  \\
&=& \frac{(C_A-\Tt)}{2\eps} \left(\xi^m - \xi^{m+1}
\left[ 1- \hbn{m}(\eps)\frac{2C_A-\Tt}{C_A-\Tt}\right]\right)\,,\nn
\eea
where we have introduced the notation
\be\label{bubblehat} 
\hat B_n(\eps) \equiv 1- \frac{B_n(\eps)}{\bzero(\eps)}
\,=\,
2 n (2 + n) \zeta_3 \epsilon^3 
+ 3 n (2 + n) \zeta_4 \epsilon^4 
+\ldots\,.
\ee 
By making repeated use of 
\eqn{softHpower} one finds that the wavefunction at 
order $(\ell-1)$ can be expressed in a closed-form, 
as follows~\cite{Caron-Huot:2017zfo}: 
\be \label{Well-1-ansatz}
\Wts{\ell-1}(p,k) =
\frac{\Ctwo^{\ell-1}}{(2\eps)^{\ell-1}} \sum_{n=0}^{\ell-1} 
(-1)^n \binom{\ell-1}{n} \mratio{p}{k}^{n\eps} 
\prod_{m=0}^{n-1} 
\left\{1 - \hbn{m}(\eps) \frac{\Cone}{\Ctwo}\right\}\,.
\ee
As discussed in \cite{Caron-Huot:2017zfo}, this expression 
can be easily integrated, obtaining an expression for the 
amplitude which correctly describes its singular part 
to all orders in perturbation theory.

While \eqn{Well-1-ansatz} is perfectly valid in the soft 
limit, it breaks explicitly the symmetry of 
\eqn{left-right-symmetry} between the two soft limits. 
As we will see below, it is 
advantageous to work with expressions where 
this symmetry is manifest. In this paper, we thus 
introduce a different soft wavefunction, obtained 
by symmetrising \eqn{Well-1-ansatz} under 
$k \leftrightarrow (p-k)$:
\begin{align} \label{Well-1-ansatz-sym} \nn
	\Wts{\ell-1}(p,k) &=
	\frac{\Ctwo^{\ell-1}}{(2\eps)^{\ell-1}} \sum_{n=0}^{\ell-1} 
	(-1)^n \binom{\ell-1}{n} \mratio{p}{k}^{n\eps} 
	\bigg(\frac{p^2}{(p-k)^2}\bigg)^{n\eps} 
	\\ &\hspace{3.0cm} \times
	\, \prod_{m=0}^{n-1} 
	\left\{1 - \hbn{m}(\eps) \frac{\Cone}{\Ctwo}\right\}\,.
\end{align}
This formula simultaneously captures the correct 
behaviour of $\Omega(p,k)$ in both soft limits $k \to 0$ 
and $k \to p$. It will be used in section~\ref{sec:softAmpl} 
below to compute the soft contributions to the reduced 
$2\to 2$ amplitude. Before doing that let us have a 
closer look at the $\epsilon$ expansion of the soft 
wavefunction we obtained.

We recall~\cite{Caron-Huot:2017zfo} that all the negative 
powers of $\epsilon$ in (\ref{Well-1-ansatz-sym}) cancel 
upon performing the sum over $n$, leading to a finite 
wavefunction at any loop order. 
While positive powers of $\epsilon$ in (\ref{Well-1-ansatz-sym}) 
do play a role in the computation of the amplitude, the leading 
$\ord(\eps^0)$ have a special role: according to 
eq.~(\ref{WhardTwod}) it is precisely what must be 
subtracted from the full two-dimensional wavefunction 
to obtain the hard wavefunction $\Wh^{({\rm 2d})}$.
With this in mind, let us write down explicitly the leading 
terms in $\epsilon$ in the first few orders of the soft 
wavefunction in (\ref{Well-1-ansatz-sym}):
\begin{subequations}
\label{Wsoft2dim}
\begin{align}
	\Wsl{0}(p,k) \big|_{\ord(\eps^0)} &= 0, \\
	\Wsl{1}(p,k) \big|_{\ord(\eps^0)} &= 
	\frac{\Ctwo}{2} \log \pfrac{k^2(p-k)^2}{(p^2)^2}, \\
	\Wsl{2}(p,k) \big|_{\ord(\eps^0)} &= 
	\frac{\Ctwo^2}{4} \log^2 \pfrac{k^2(p-k)^2}{(p^2)^2}, \\
	\Wsl{3}(p,k) \big|_{\ord(\eps^0)} &= 
	\frac{\Ctwo^3}{8} \log^3 \pfrac{k^2(p-k)^2}{(p^2)^2} 
	+ \frac{\Cone \Ctwo^2}{2} \zeta_3, \\ \nn
	\Wsl{4}(p,k) \big|_{\ord(\eps^0)} &= 
	\frac{\Ctwo^4}{16} \log^4 \pfrac{k^2(p-k)^2}{(p^2)^2} \\ 
	& +\, \Cone \Ctwo^3 \log \pfrac{k^2(p-k)^2}{(p^2)^2} \zeta_3, \\ \nn
	\Wsl{5}(p,k) \big|_{\ord(\eps^0)} &= 
	\frac{\Ctwo^5}{32} \log^5 \pfrac{k^2(p-k)^2}{(p^2)^2} \\ 
	&+\, 
	\frac{\Cone \Ctwo^4}{4} \left[ 5 \log^2 \pfrac{k^2(p-k)^2}{(p^2)^2} \zeta_3 
	+ 6 \zeta_5 \right]\,.
\end{align}
\end{subequations}
In fact, these terms exponentiate and can be resummed 
into the following all-order expression using 
(\ref{OmegaEven}) for $\epsilon=0$, yielding
\beq \label{eq:wffullsoftresummed} 
\Ws(p,k) \big|_{\ord(\eps^0)} = \frac{\as}{\pi} 
\left[ e^{-x \gE \Ctwo} \frac{\Gamma \left(1 - \frac{x}{2} 
	\Ctwo \right)}{\Gamma \left(1 + \frac{x}{2} \Ctwo \right)}
\right]^\frac{2\Ca - \Tt}{\Ca - \Tt} 
{\pfrac{k^2(p-k)^2}{(p^2)^2 }}^{\frac{x}{2} \Ctwo },
\eeq
with $x = L\, \as/\pi$.

\subsection{Soft contributions to the $2\to2$ amplitude\label{sec:softAmpl}}

Next, let us consider the soft contribution to the reduced
$2\to 2$ scattering amplitude $\hat{\cal M}$. It is straighforward to 
insert \eqn{Well-1-ansatz-sym} into \eqn{ReducedAmpNLL2}, 
perform the last integration and derive the $\ell$-th 
order contribution to the amplitude. In particular, given 
the symmetrised form of \eqn{Well-1-ansatz-sym}, the 
last integration can be done with the integration measure 
$[{\rm D}k]$ in \eqn{measure}, i.e. avoiding the need to introduce 
a cut-off as in ref.~\cite{Caron-Huot:2017zfo}. 
After some arrangement we get 
\bea \label{MellReggeSoft-S} \nn
\Mreduced_{{\rm NLL,s}}^{(+,\ell)} &=& i \pi \, 
\frac{1}{(2\eps)^{\ell}} \, 
\frac{\left(\bzero(\eps)\right)^{\ell}}{\ell!} \,
\, \Ctwo^{\ell-1} \sum_{n=1}^{\ell-1} 
(-1)^{n+1} \, \binom{\ell}{n}  \frac{\tilde B_{n}(\eps) }{\bzero(\eps)} \\ 
&&\hspace{3.0cm} \times \,
\prod_{m=0}^{n-2}\bigg[ 1 - \hbn{m}(\eps) \frac{\Cone}{\Ctwo} \bigg]
\, \Tsu \, \Mtree,
\eea
where the functions $B_n(\eps)$ and $\hbn{n}(\eps)$
have been defined respectively in \eqns{bubbleGeneral2}{bubblehat}, 
and we have introduced 
\be
\tilde B_{n}(\eps) = e^{\eps \gamma_E} 
\frac{\Gamma^2 \big(1-n \eps\big) \Gamma\big(1-\eps+2 n \eps \big)}{
\Gamma\big(1-2 n \eps \big) \Gamma^2\big(1-\eps + n \eps \big) }.
\ee

The coefficients $\Mreduced_{{\rm NLL,s}}^{(+,\ell)}$ in 
(\ref{MellReggeSoft-S}) are of course polynomial in the colour 
factors. For illustration, we expand \eqn{MellReggeSoft-S} 
to the first few orders in perturbation theory, obtaining
\begin{subequations}
\renewcommand{\Cone}{C_1}
\renewcommand{\Ctwo}{C_2}
\begin{align}
	\label{eq:MsoftExpanded_1} 
	\Mreduced_{\rm NLL,s}^{(1)} &= 
	i\pi \Bn{0}  \bigg\{\frac{1}{2\eps} \bigg\}  \Tsu \Mtree, \\ 
	\Mreduced_{\rm NLL,s}^{(2)} &= 
	i\pi \frac{\Bn{0}^2}{2} \bigg\{ \frac{\Ctwo}{4 \eps^2} \bigg\} \Tsu \Mtree, \\  
	\Mreduced_{\rm NLL,s}^{(3)} &= 
	i\pi \frac{\Bn{0}^3}{3!} 
	\bigg\{ \Ctwo^2 \left( \frac{1}{8 \eps^3} 
	- \frac{11\zeta_3}{4}\right) 
	- \Cone \Ctwo\frac{3\zeta_3}{4}\bigg\} \Tsu \Mtree, \\  
	\Mreduced_{\rm NLL,s}^{(4)} &= 
	i\pi \frac{\Bn{0}^4}{4!}
	\bigg\{ \Ctwo^3 \left( \frac{1}{16 \eps^4} 
	+ \frac{\zeta_3}{8\eps} + \frac{3\zeta_4}{16}\right) 
	+ \Cone \Ctwo^2 \left( -\frac{\zeta_3}{8 \eps} 
	- \frac{3\zeta_4}{16}\right) \bigg\} \Tsu \Mtree, \\  \nn
	\Mreduced_{\rm NLL,s}^{(5)} &= 
	i\pi \frac{\Bn{0}^5}{5!}
	\bigg\{ \Ctwo^4 \left( \frac{1}{32 \eps^5} 
	+ \frac{\zeta_3}{16\eps^2} + \frac{3\zeta_4}{32 \eps} 
	- \frac{717\zeta_5}{16} \right) \\  
	&\hspace{1.0cm} +\, \Cone \Ctwo^3 \left( -\frac{\zeta_3}{16 \eps^2} 
	- \frac{3\zeta_4}{32 \eps} 
	- \frac{27\zeta_5}{16} \right) \bigg\} \Tsu \Mtree, \\ \nn  
	\Mreduced_{\rm NLL,s}^{(6)} &= 
	i\pi \frac{\Bn{0}^6}{6!}
	\bigg\{ \Cone^2 \Ctwo^3  \bigg( -\frac{39 \zeta_3^2}{16} \bigg) 
	+ \Cone \Ctwo^4 \bigg(-\frac{\zeta_3}{32 \eps^3} 
	- \frac{3\zeta_4}{64 \eps^2}  
	- \frac{3\zeta_5}{32 \eps}
	- \frac{963\zeta_3^2}{32} 
	+ \frac{5 \zeta_6}{32} \bigg) \\ 
	&\hspace{1.0cm}
	+ \Ctwo^5 \bigg(\frac{1}{64 \eps^6} 
	+ \frac{\zeta_3}{32 \eps^3} 
	+ \frac{3 \zeta_4}{64 \eps^2}  
	+ \frac{3 \zeta_5}{32 \eps} 
	- \frac{2879 \zeta_3^2}{32} 
	+ \frac{5 \zeta_6}{32} \bigg)
	\bigg\} \Tsu \Mtree, \\  \nn
	\Mreduced_{\rm NLL,s}^{(7)} &= 
	i\pi \frac{\Bn{0}^7}{7!}
	\bigg\{ \Cone^2 \Ctwo^4 \bigg( \frac{\zeta_3^2}{32 \eps} 
	+ \frac{3\zeta_3 \zeta_4}{32} \bigg) 
	+ \Cone \Ctwo^5 \bigg(
	- \frac{\zeta_3}{64 \eps^4} 
	- \frac{3\zeta_4}{128 \eps^3} 
	- \frac{3\zeta_5}{64 \eps^2}\\ \nn
	&\hspace{1.0cm}
	-\, \frac{3\zeta_3^2}{64 \eps} 
	- \frac{5 \zeta_6}{64 \eps} 
	- \frac{9 \zeta_3 \zeta_4}{64}
	- \frac{729 \zeta_7}{64} \bigg)  
	+ \Ctwo^6 \bigg( \frac{1}{128 \eps^7}
	+ \frac{\zeta_3}{64 \eps^4} 
	+ \frac{3 \zeta_4}{128 \eps^3} 
	+ \frac{3 \zeta_5}{64 \eps^2} \\ 
	&\hspace{1.0cm}
	+\, \frac{\zeta_3^2}{64 \eps} 
	+ \frac{5 \zeta_6}{64 \eps}  
	+ \frac{3 \zeta_3 \zeta_4}{64} 
	- \frac{90711 \zeta_7}{64} \bigg)
	\bigg\} \Tsu \Mtree, \\   \nn
	\label{eq:MsoftExpanded_8}
	\Mreduced_{\rm NLL,s}^{(8)} &= 
	i\pi \frac{\Bn{0}^8}{8!}
	\bigg\{ \Cone^2 \Ctwo^5 \bigg( 
	\frac{\zeta_3^2}{64 \eps^2} 
	+ \frac{3\zeta_3 \zeta_4}{64 \eps}
	- \frac{1341 \zeta_3 \zeta_5}{32} 
	+ \frac{21\zeta_8}{512} \bigg)
	+ \Cone \Ctwo^6 \bigg(
	-\frac{\zeta_3}{128 \eps^5} \\ \nn
	&\hspace{1.0cm}
	-\, \frac{3\zeta_4}{256 \eps^4}    
	- \frac{3\zeta_5}{128 \eps^3} 
	- \frac{3 \zeta_3^2}{128 \eps^2} 
	- \frac{5 \zeta_6}{128 \eps^2} 
	- \frac{9 \zeta_3 \zeta_4}{128 \eps} 
	- \frac{9\zeta_7}{128 \eps} 
	- \frac{96777 \zeta_3 \zeta_5}{64} \\ \nn
	&\hspace{1.0cm} 
	-\, \frac{189 \zeta_8}{1024} \bigg)
	+ \Ctwo^7 \bigg( \frac{1}{256 \eps^8} 
	+ \frac{\zeta_3}{128 \eps^5} 
	+ \frac{3\zeta_4}{256 \eps^4}
	+ \frac{3\zeta_5}{128 \eps^3} 
	+ \frac{\zeta_3^2}{128 \eps^2}
	+ \frac{5 \zeta_6}{128 \eps^2}  \\ 
	&\hspace{1.0cm} 
	+ \frac{3 \zeta_3 \zeta_4}{128 \eps} 
	+ \frac{9 \zeta_7}{128 \eps} 
	- \frac{483837 \zeta_3 \zeta_5}{64} 
	+ \frac{147 \zeta_8}{1024} \bigg)
	\bigg\} \Tsu \Mtree\, ,
	\renewcommand{\Cone}{(2\Ca-\Tt)}
	\renewcommand{\Ctwo}{(\Ca-\Tt)}
\end{align}
\end{subequations}
where we used the shorthand notation for the colour 
factors, $C_1=\Cone$ and $C_2=\Ctwo$. 
We note that the expansion coefficients display uniform 
transcendental weight (where, as usual $1/\epsilon$ has 
weight 1) and involve exclusively single zeta values
(sometimes referred to as ordinary zeta values, namely 
the values of the Riemann zeta function at integer arguments).
We further notice that $\zeta_2$ (or $\zeta_2$ times other 
zeta values, e.g. $\zeta_2\zeta_3$ at weight 5, etc.) factors 
do not appear in \eqnss{eq:MsoftExpanded_1}{eq:MsoftExpanded_8} 
($\zeta_2$ terms would be present if we were to expand the factor
factor $B_0^{\ell}(\eps)$). Higher even zeta numbers do appear, 
but we will see below that they have a distinct origin as compared 
to the odd ones.

Given that the expansion coefficients $\Mreduced_{{\rm NLL,s}}^{(+,\ell)}$ 
involve just single zeta values, and are moreover of uniform 
weight, it is interesting to explore the possibility to sum up the 
series to all orders. Indeed, such summation was achieved for 
the singular terms in Ref.~\cite{Caron-Huot:2017zfo}, so let us 
compare \eqn{MellReggeSoft-S} above with the result obtained 
in~\cite{Caron-Huot:2017zfo}. There we proved that the singular 
terms of the reduced amplitude admit a simplified form  
\be \label{MellReggeSoft-0}
\Mreduced_{{\rm NLL,s_{\rm simpl.}}}^{(+,\ell)} \!\!=\!  \frac{i \pi}{(2\eps)^{\ell}} \, 
\frac{\left(\bzero(\eps)\right)^{\ell}}{\ell!} \,\Ctwo^{\ell-1} \, \frac{B_{-1}(\eps)}{B_0(\eps)} 
\left( 1- \hbn{-1}(\eps) \frac{\Cone}{\Ctwo} \right)^{-1} \!
\! \Tsu \, \Mtree.
\ee
The latter, however, differs from the original soft amplitude 
obtained from \eqn{Well-1-ansatz-sym} starting at 
${\cal O}(\eps^{0})$ (compare eqs.~(3.13) and~(3.15)
of~\cite{Caron-Huot:2017zfo}). A nice feature of 
\eqn{MellReggeSoft-0} is that the loop functions in the 
amplitude at order $\ell$ do not depend on the index $\ell$, 
apart from the factor $\left(\bzero(\eps)\right)^{\ell}/\ell!$, and this allows 
one to easily resum \eqn{MellReggeSoft-0} to all orders in 
perturbation theory, obtaining an expression for the 
integrated soft amplitude $\Mreduced_{{\rm NLL,s_{\rm simpl.}}}$:
\begin{align} \label{MellReggeSoft-res}
\begin{split}
	\Mreduced_{{\rm NLL,s_{\rm simpl.}}} =
	\frac{i\pi}{L \Ctwo} \Bigg\{ &
	\Big( e^{\frac{B_0}{2\eps} \Ctwo x}-1 \Big) \frac{B_{-1}(\eps)}{B_0(\eps)} 	\\
	&\quad\times \left( 1- \hbn{-1}(\eps) \frac{\Cone}{\Ctwo} \right)^{-1}
	\bigg\} \, \Tsu \, \Mtree + \ord(\eps^0),
\end{split}
\end{align}
with $x =  L \,\as/\pi $ (see also eq.~(3.18) 
of~\cite{Caron-Huot:2017zfo}). 
This formula however does not correctly capture the non-singular 
terms obtained with a cut-off, for which no similar simplification 
was found. We nevertheless show that an all-order resummation 
formula can be found for the ${\cal O}(\epsilon^0)$ corrections 
to the amplitude defined in our current symmetric scheme,
\eqn{MellReggeSoft-S}. To this end we consider the 
coefficients defined as the finite part of the 
\emph{difference} between those in soft amplitude, 
eq.~(\ref{MellReggeSoft-S}), and in its simplified 
version, eq.~(\ref{MellReggeSoft-0}):
\begin{align}\label{deltadef}
\begin{split}
	\Mreduced_{{\rm NLL,s}}^{(+,\ell)} - 
	\Mreduced_{{\rm NLL,s_{\rm simpl.}}}^{(+,\ell)}
	&\equiv \, i \pi \,  
	\hat\Delta^{(+,\ell)}_{\rm NLL} \, \Tsu \, \Mtree +{\cal O}(\epsilon^1)\\ 
	&\equiv i \pi \, 
	\delta^{(\ell)} \, \Ctwo^{\ell-1} \,\Tsu \, \Mtree +{\cal O}(\epsilon^1)\,.
\end{split}
\end{align}
After some arrangement the coefficients 
$\hat\Delta^{(+,\ell)}_{\rm NLL}$ can be put 
into the form 
\bea \label{Delta_s3} 
\hat\Delta^{(+,\ell)}_{\rm NLL} &=& 
\frac{1}{(2\eps)^{\ell}} \, 
\frac{1}{\ell!} \,
\,  \Ctwo^{\ell-1} \bigg\{ \sum_{n=0}^{\ell-1} 
(-1)^{n+1} \, \binom{\ell}{n}  
{\tilde B_{n}(\eps)} \\ \nn
&&\hspace{-1.0cm}
\times \, \left( 1- \hbn{n-1}(\eps) \frac{\Cone}{\Ctwo} \right)^{-1}
\prod_{m=0}^{n-2}\bigg[ 1 - \hbn{m}(\eps) \frac{\Cone}{\Ctwo} 
\bigg] \bigg\},
\eea
where we discarded powers of $B_{0}(\epsilon)$, which do 
not affect the finite terms. From eq.~(\ref{Delta_s3}) the 
coefficients $\delta^{(\ell)}$ of (\ref{deltadef}) can be 
determined explicitly in terms of odd $\zeta$ numbers and 
the ratio of colour factors $r=\frac{\Cone}{\Ctwo}$. 
They are found to exponentiate in terms 
of the following rescaled odd $\zeta$ numbers: 
\be
\label{rescaled_zeta}
\tilde\zeta_{1+2 n} = \frac{2-2^{1-2n}}{1+2n} \zeta_{1+2n} 
\left(1+ \frac{r}{2^{1+2n}-2}  \right)\,,
\ee
such that the sum:
\begin{align}
\label{summing_soft_finite}
\begin{split}
	\sum_{\ell=1}^{\infty} \frac{X^{\ell}}{\ell !}\delta^{(\ell)} 
	=\,&1-\exp\left(\sum_{n=1}^{\infty} X^{2n+1} \tilde\zeta_{2n+1}\right)
	\\=\,& 1-e^{-\gamma_E r X}
	\frac{\Gamma\Big(1 - X\Big)}{\Gamma\Big(1 + X\Big)} 
	\frac{\Big[\Gamma\Big(1+\frac{X}{2}\Big)\Big]^{2-r}}{
	\Big[\Gamma\Big(1 -  \frac{X}{2}\Big)\big]^{2-r}} 
\end{split}
\end{align}
with $X\equiv \Ctwo x$ and $x = L \,\as/\pi $, where we used
\be
2 \sum _{n=1}^{\infty}\frac{x^{2 n+1} \zeta_{2 n+1}}{2 n+1}
=-2 x \gamma_E  +\log (\Gamma (1-x))-\log (\Gamma (x+1)) \,.
\ee
We conclude that the series 
$\hat\Delta^{(+,\ell)}_{\rm NLL}$ exponentiate to 
\begin{multline}\label{DeltaResum2} 
	\hat\Delta^{(+)}_{\rm NLL} =
	\frac{1}{L \Ctwo} \Bigg[ 1- e^{-\gamma_E \Cone \, x } 
	\frac{\Gamma\Big(1 - \Ctwo x\Big)}{\Gamma\Big(1 + \Ctwo x\Big)} \\ 
	\times \,
	\left(\frac{\Gamma\Big(1+\Ctwo \frac{x}{2}\Big)}{\Gamma
	\Big(1 - \Ctwo \frac{x}{2}\Big)}\right)^{-\frac{{\mathbf T}_t^2}{C_A-{\mathbf T}_t^2}}
	\Bigg].
\end{multline}
Using now the fact that that the simplified amplitude
$\Mreduced_{{\rm NLL,s_{\rm simpl.}}}^{(+,\ell)} $ in 
\eqn{MellReggeSoft-0} exponentiates independently,
see \eqn{MellReggeSoft-res}, we obtain
\begin{multline}\label{ReducedAmpNLLresum2} 
	\Mreduced_{\rm NLL,s} = \\
	i\pi \,\Bigg\{ \frac{e^{\frac{B_0}{2\eps} \Ctwo x}-1}{L \Ctwo}
	\frac{B_{-1}(\eps)}{B_0(\eps)} 
	\left( 1- \hbn{-1}(\eps) \frac{\Cone}{\Ctwo} \right)^{-1}\,+ 
	\hat\Delta^{(+)}_{\rm NLL} \Bigg\} \Tsu \, \Mtree  \\ 
	=
	i\pi \,\Bigg\{ \frac{e^{\frac{B_0}{2\eps} \Ctwo x}-1}{L \Ctwo}
	\left( 1- \frac{C_A}{\Ctwo} R(\eps)\right)^{-1}\,+ 
	\hat\Delta^{(+)}_{\rm NLL} \Bigg\} \Tsu \, \Mtree\, ,
\end{multline}
where in the second line we expressed the amplitude in 
terms of the function $R(\eps) ={\bzero(\eps)}/{\bn{-1}(\eps)} -1 $ 
of  \eqn{Rdef}. Writing the reduced amplitude as in the second 
line of \eqn{ReducedAmpNLLresum2} makes it easier to extract 
the infrared-renormalized amplitude from the reduced amplitude, 
as we will see in section \ref{hardFdef}. Writing explicitly the 
factor $\hat\Delta^{(+)}_{\rm NLL} $, 
the reduced amplitude reads 
\begin{multline}\label{ReducedAmpNLLresum2B} 
	\Mreduced_{\rm NLL,s} 
	= \frac{i\pi}{L \Ctwo} \Bigg\{\left( e^{\frac{B_0}{2\eps} \Ctwo x}-1 \right)  
	\left( 1- \frac{C_A}{\Ctwo} R(\eps)\right)^{-1}\,+ 1 \\
	- e^{-\gamma_E \Cone \, x} \,\,
	\frac{\Gamma\Big(1 - \Ctwo x\Big)}{\Gamma\Big(1 + \Ctwo x\Big)}
	\left(\frac{\Gamma\Big(1+\Ctwo \frac{x}{2}\Big)}{\Gamma
	\Big(1 - \Ctwo \frac{x}{2}\Big)}\right)^{-\frac{{\mathbf T}_t^2}{C_A-{\mathbf T}_t^2}}
	\Bigg\} \Tsu \, \Mtree .
\end{multline}
Of course, upon expansion (\ref{ReducedAmpNLLresum2B}) 
yields back the coefficients of (\ref{MellReggeSoft-S}) we listed 
in (\ref{eq:MsoftExpanded_1}) through (\ref{eq:MsoftExpanded_8}).
Having at hand a resummed expression we can gain further insight 
on number-theoretical features of the expansion coefficients in 
\eqnss{eq:MsoftExpanded_1}{eq:MsoftExpanded_8}. We already 
know based on the derivation above that the $\hat\Delta^{(+)}_{\rm NLL}$ 
component in (\ref{ReducedAmpNLLresum2}) gives rise to odd zeta 
values only. It then transpires that the sole origin of even ones is the 
function $R(\epsilon)$ in the first term. Further number-theoretical 
features will be discussed in section~\ref{amplitude}, once we have 
computed the hard contribution to the reduced amplitude.
The possibility to resum the series for the amplitude to all orders 
including finite ${\cal O}(\epsilon^0)$ terms is highly nontrivial, and 
it is an additional advantage of $k \leftrightarrow (p-k)$ symmetric 
scheme we adopted here for the soft approximation. It will be used 
below in deriving a resummed expression for the contribution of the 
soft region to the infrared-renormalized amplitude.

\subsection{From the reduced amplitude to the infrared-renormalized amplitude}\label{hardFdef}

Now that we have determined the soft wavefunction
and the corresponding reduced amplitude, we are in a position to 
consider again the infrared-renormalized amplitude, as defined in 
\eqn{getH2}. Following \eqns{OmegaSplitDef}{eq:redampSplit}
we split the infrared-renormalized amplitude into a soft and a 
hard component:
\be\label{Hsplit}
\mathcal{H}^{(+)}_{\rm NLL} 
= \mathcal{H}^{(+)}_{\rm NLL,s} 
+\mathcal{H}^{(+)}_{\rm NLL,h}\,.
\ee
Then, from \eqn{getH2} it follows that
\begin{subequations}
\label{getH2both}
\begin{align} \label{getH2-s} 
	\Hhard^{(+)}_{\rm NLL,s} &=
	- e^{-\frac{x}{2\eps} \T_t^2} {\bf Z}^{(-)}_{\rm NLL} 
	\, e^{\frac{\bn{0}(\eps)-1}{2\eps} \, x \, C_A} \MM^{\rm (tree)}
	+e^{\frac{B_0(\eps)-1}{2\eps} \, x \, \T_t^2} {\cal \hat M}^{(+)}_{\rm NLL,s}, \\
	\label{getH2-h}
	\Hhard^{(+)}_{\rm NLL,h} &= {\cal \hat M}^{(+)}_{\rm NLL,h}\,,
\end{align}
\end{subequations}
where in (\ref{getH2-h}) we neglected positive powers of $\epsilon$ 
originating in the expansion of $(1-B_0(\epsilon))$, using the fact that 
${\cal \hat M}^{(+)}_{\rm NLL,h}$ is itself finite. Of course such a 
simplification cannot be applied to (\ref{getH2-s}) where there is 
an interplay between positive powers of $\epsilon$ and negative 
ones. In section \ref{sec:softAmpl} we have determined 
the reduced soft amplitude, thus we are in a position to explicitly 
write down the soft part of the infrared-renormalized amplitude at 
$\epsilon\to 0$, according to \eqn{getH2-s}. Inserting 
eqs.~(\ref{Znll}) and (\ref{ReducedAmpNLLresum2}) 
into \eqn{getH2-s} we get 
\begin{multline}\label{getH3}
	\Hhard^{(+)}_{\rm NLL,s} = i\pi \Bigg\{ 
	- e^{\frac{\bn{0}(\eps)-1}{2\eps} \, x \, C_A} 
	\bigg[\frac{ e^{\frac{x}{2\eps} (\CA-\Tt)}}{L(\CA-\Tt)}
	\left( 1 - \frac{C_A}{C_A -\Tt} R(\eps) \right)^{-1} \bigg]_{\rm poles} \\ 
	+e^{\frac{B_0(\eps)-1}{2\eps} \, x \, \T_t^2} 
	\Bigg[ \frac{e^{\frac{B_0}{2\eps} x \Ctwo}-1}{L \Ctwo}
	\left( 1- \frac{C_A}{\Ctwo} R(\eps) \right)^{-1}\,+ 
	\hat\Delta^{(+)}_{\rm NLL} \bigg]\Bigg\} \Tsu \, \Mtree\,,
\end{multline}
where we recall that $x = L \, \as/\pi$, and in the first line, 
corresponding to ${\bf Z}^{(-)}_{\rm NLL}$, we have dropped 
the $-1$ term in the numerator inside the square brackets, 
which does not generate any poles (see discussion following 
\eqn{Znll}). This expression can be rearranged as follows: 
first of all, by collecting a factor $e^{\frac{\bn{0}(\eps)-1}{2\eps} 
\, x \, C_A}$ we get
\begin{multline}\label{getH4first}
	\Hhard^{(+)}_{\rm NLL,s} = i\pi 
	e^{\frac{\bn{0}(\eps)-1}{2\eps} \, x \, C_A}
	\Bigg\{-\bigg[\frac{ e^{\frac{x}{2\eps} (\CA-\Tt)}}{L(\CA-\Tt)}
	\left( 1 - \frac{C_A}{C_A -\Tt} R(\eps) \right)^{-1} \bigg]_{\rm poles} \\ 
	+ \Bigg[ \frac{e^{\frac{x}{2\eps} \Ctwo}
	-e^{\frac{1-B_0(\eps)}{2\eps} \, x \, \Ctwo}}{L \Ctwo}
	\left( 1- \frac{C_A}{\Ctwo} R(\eps) \right)^{-1} \\ 
	+ e^{\frac{1-B_0(\eps)}{2\eps} \, x \, \Ctwo}
	\hat\Delta^{(+)}_{\rm NLL} \bigg]\Bigg\} \Tsu \, \Mtree\,.
\end{multline}
We see at this point that the second line nicely cancel the 
poles from the first line. Furthermore, given that 
$1-B_0(\eps) ={\cal O}(\eps^2)$, see  eq.~(\ref{B0}), and both 
$\big[ 1 - C_A/(C_A -\Tt) \, R(\eps) \big]^{-1}
= 1 + \ord(\eps^3)$ and $\hat\Delta^{(+)}_{\rm NLL} = \ord(\eps^0)$, 
it is safe to set to one all exponentials containing the factor 
$1-B_0(\eps)$. We thus obtain
\begin{multline}\label{getH4}
	\Hhard^{(+)}_{\rm NLL,s} = i\pi 
	\Bigg\{ \bigg[\frac{ e^{\frac{x}{2\eps} (\CA-\Tt)}-1}{L(\CA-\Tt)}
	\left( 1 - \frac{C_A}{C_A -\Tt} R(\eps) \right)^{-1} \bigg]_{\eps^0} 
	+ \hat\Delta^{(+)}_{\rm NLL} \Bigg\} \Tsu \, \Mtree\,,
\end{multline}
with $\hat\Delta^{(+)}_{\rm NLL}$ given in \eqn{DeltaResum2}.
For later reference we also expand \eqn{getH4} to 
the first few orders in perturbation theory, obtaining 
(recall $C_2=C_A-\Tt$):
\begin{subequations}
\renewcommand{\Cone}{C_1}
\renewcommand{\Ctwo}{C_2}
\begin{align}
	\label{eq:HsoftExpanded_1} 
	{\cal H}_{\rm NLL,s}^{(1)} &= 0, \\ 
	{\cal H}_{\rm NLL,s}^{(2)} &= 0, \\  
	{\cal H}_{\rm NLL,s}^{(3)} &= \frac{i\pi}{3!}  
	\bigg\{ - C_A \Ctwo \frac{3\zeta_3}{4}
	-\Ctwo^2 \frac{7\zeta_3}{2} \bigg\} \Tsu \Mtree, \\  
	{\cal H}_{\rm NLL,s}^{(4)} &= \frac{i\pi}{4!}
	\bigg\{ - C_A \Ctwo^2 \frac{3\zeta_4}{16} \bigg\} \Tsu \Mtree, \\ 
	{\cal H}_{\rm NLL,s}^{(5)} &= \frac{i\pi}{5!}
	\bigg\{- C_A \Ctwo^3 \frac{27\zeta_5}{16}
	-\Ctwo^4 \frac{93\zeta_5}{2} \bigg\} \Tsu \Mtree, \\ 
	{\cal H}_{\rm NLL,s}^{(6)} &= \frac{i\pi}{6!}
	\bigg\{ - C_A^2 \Ctwo^3  \frac{39 \zeta_3^2}{16} 
	-C_A \Ctwo^4 \bigg(
	\frac{1119\zeta_3^2}{32} 
	+ \frac{5\zeta_6}{32} \bigg) 
	- \Ctwo^5 \frac{245 \zeta_3^2}{2} \bigg\} \Tsu \Mtree, \\  
	{\cal H}_{\rm NLL,s}^{(7)} &= \frac{i\pi}{7!}
	\bigg\{ C_A^2 \Ctwo^4 \frac{3\zeta_3 \zeta_4}{32} 
	+ C_A \Ctwo^5 \bigg( 
	\frac{3\zeta_3 \zeta_4}{64}
	- \frac{729 \zeta_7}{64} \bigg)  
	- \Ctwo^6 \frac{5715 \zeta_7}{4} \bigg\} \Tsu \Mtree, \\   \nn
	\label{eq:HsoftExpanded_8}
	{\cal H}_{\rm NLL,s}^{(8)} &= \frac{i\pi}{8!}
	\bigg\{ C_A^2 \Ctwo^5 \bigg(
	- \frac{1341 \zeta_3 \zeta_5}{32} 
	+ \frac{21\zeta_8}{512} \bigg)
	+ C_A \Ctwo^6 \bigg(
	- \frac{102141 \zeta_3 \zeta_5}{64}  
	- \frac{105\zeta_8}{1024} \bigg) \\
	&\hspace{5.0cm} \,- \Ctwo^7 \, 9114 \zeta_3 \zeta_5\bigg\} \Tsu \Mtree\, .
\renewcommand{\Cone}{(2\Ca-\Tt)}
\renewcommand{\Ctwo}{(\Ca-\Tt)}
\end{align}
\end{subequations}
It is interesting note that $\zeta_n$ values with even~$n$ 
originate solely from the expansion of the factor 
$R(\eps)$ in \eqn{getH4}, while the expansion of the factor 
$\hat\Delta^{(+)}_{\rm NLL}$ generates only $\zeta_n$ values 
with odd~$n$. The latter property of $\hat\Delta^{(+)}_{\rm NLL}$ 
makes this function compatible with the class of zeta values we 
will encounter considering the two-dimensional amplitude in 
section~\ref{amplitude}.

In summary, according to (\ref{Hsplit}) the infrared-renormalized 
amplitude is given as a sum of two terms: $\mathcal{H}_{\rm s}$, 
computed in this section using the soft approximation, plus 
$\mathcal{H}_{\rm h}$, which is identical to the hard part 
of the reduced amplitude (see \eqn{getH2-h}). The latter is 
infrared finite and originates in the hard wavefunction, 
which can be computed directly in two transverse dimensions. 
Let us turn now to evaluate it.


\section{BFKL evolution in two transverse dimensions} \label{2d-bfkl}

As discussed in the introduction and in section~\ref{chap:bfkl}, 
much of the complication of solving the \ac{bfkl} evolution 
stems from the $d$-dimensionality of the Hamiltonian. 
Recalling that the two-reggeon wavefunction is finite at 
any loop order and that singularities are exclusively 
created by integration near the soft limit, it should be 
clear that no regularisation is required if we (a) only 
care about finite terms, and (b) remove any soft 
kinematics from the last integration. The latter condition is fulfilled 
by construction, having defined the split between the hard and 
soft wavefunctions (\ref{OmegaSplitDef}) subject to the condition 
(\ref{h_definition}): the vanishing of $\Wh^{({\rm 2d})}(p,k)$ in the 
soft limits guaranties that the corresponding amplitude 
$\hat {\cal M}^{(+)}_{\rm NLL,h}\left(\frac{s}{-t}\right)$ 
of~eq.~(\ref{Mhard}) is finite.

Our task in this section is therefore to compute 
$\Wh^{({\rm 2d})}(p,k)$. We do so by iteratively applying the 
Hamiltonian of eqs.~(\ref{Hdef1}) and (\ref{Hamil}), according to 
eq.~(\ref{Hdef0}). We keep the kinematics general, but in contrast to 
section~\ref{chap:bfkl}, we work strictly in two transverse dimensions.
To exploit the advantage of two-dimensional kinematics let us view 
the Euclidean momentum vectors $k$, $k'$ and~$p$ as complex 
numbers
\beq
k = k_x + i k_y, \quad k' = k_x' + i k_y' \quad \text{and} 
\quad p = p_x + i p_y,
\eeq
where the real and imaginary parts are the components 
of the corresponding momenta and introduce new variables 
$z,w \in \mathbb{C}$ according to
\beq
\label{eq:zwdef} \frac{k_x + i k_y}{p_x + i p_y} = 
\frac{z}{z-1} \qquad \text{and} \qquad \frac{k_x' + i k_y'}{p_x + i p_y} 
= \frac{w}{w-1}\,.
\eeq
Since the wavefunction is a function of Lorentz scalars 
(i.e.\ squares of momenta) it will be symmetric under the 
exchange $z \leftrightarrow \zb$ with $\zb$ the complex 
conjugate of $z$.  In particular, $\Wh^{({\rm 2d})}(p,k)$ 
depends on the two ratios
\begin{equation}
	\label{k2zzb}
	\frac{k^2}{p^2}= \frac{z\zb}{(1-z)(1-\zb)}\,,\quad\qquad 
	\frac{(p-k)^2}{p^2}= \frac{1}{(1-z)(1-\zb)}\,.
\end{equation}
These relations also clarify that the symmetry under interchanging 
the two Reggeons, eq.~(\ref{left-right-symmetry}), corresponds to 
$z\to1/z$, and specifically, the two soft limits where one or the other 
Reggeon is soft correspond respectively to $z\to 0$ and $z\to \infty$.
The limit $z\to 1$ instead represents maximally hard kinematics, 
where both $k^2$ and $(p-k)^2$ are much larger than $p^2$. 

In the new variables the \ac{bfkl} kernel \eqref{bfkl-kernel} 
reads
\beq
\label{eq:f2d} p^2 f(p,k,k') \longrightarrow (1-w)^2 (1-\wb)^2 K(w,\wb,z,\zb),
\eeq
where
\beq
\label{eq:K2d} K(w,\wb,z,\zb) = \frac{z\wb+w\zb}{w\wb(z-w)(\zb-\wb)} 
= \frac{1}{\wb(z-w)} + \frac{2}{(z-w)(\zb-\wb)} + \frac{1}{w(\zb-\wb)}\,. 
\eeq
Furthermore, in the limit $\eps \to 0$, $J(p,k)$ of eq.~\eqref{Jp-def2} 
becomes
\beq
\label{eq:j2d} J(p,k) \longrightarrow j(z,\zb) \equiv 
\frac12 \log \left[ \frac{z}{(1-z)^2} \frac{\zb}{(1-\zb)^2} \right],
\eeq
and the measure reads
\beq
\label{eq:dw2d} \frac{\dd^2 k'}{p^2} \longrightarrow 
\frac{\dd^2 w}{(1-w)^2 (1-\wb)^2}.
\eeq
Here, $\dd^2 w \equiv \dd \RE(w)\,\dd \IM(w)$ where the real and 
imaginary part of $w$ are to be integrated from $-\infty$ to $+\infty$, 
in accordance with eq.~\eqref{eq:zwdef}.

In applying BFKL evolution we employ the same notation as in the 
$d$-dimensional case but add the subscript ``2d'' to avoid confusion. 
In particular, from here on we express the two-dimensional 
hard wavefunction as $\Wh^{({\rm 2d})}(p,k)= \Wtd(z,\zb)$. 
We expand it as in eq.~(\ref{OmegaEven}), where we take 
$B_0(0)= 1$, i.e.
\beqa
\label{OmegaEven_hard}
\Wtd(z,\zb) &=& \sum_{\ell=0}^{\infty} 
\left( \frac{\alpha_s}{\pi} \right)^{\ell+1} 
\frac{L^{\ell} }{\ell !} \, \Wtdl{\ell}(z,\zb)\,, 
\eeqa
where the coefficients of increasing orders are related by 
the action of the Hamiltonian according to eq.~(\ref{Hdef0}), 
which now reads: 
\beq \label{eq:Htdaction}
\Wtdl{\ell}(z,\zb) = \Htd \Wtdl{\ell-1}(z,\zb),
\eeq
where
\beq
\Htd \Psi(z,\zb) = 
\Cone \Hitd \Psi(z,\zb) 
+ \Ctwo \Hmtd \Psi(z,\zb).
\eeq
Plugging in the above expressions we find the 
two parts of the Hamiltonian to be
\begin{subequations}
\begin{align}
	\label{eq:hi2d} \Hitd \Psi(z,\zb) &= \frac{1}{4\pi} 
	\int \dd^2 w K(w,\wb,z,\zb) \left[ \Psi(w,\wb) - \Psi(z,\zb) \right], \\
	\label{eq:hm2d} \Hmtd \Psi(z,\zb) &= j(z,\zb) \Psi(z,\zb),
\end{align}
\end{subequations}
where $\Wtdl{0}(z,\zb) = \Wfl{0}(p,k) = 1$.

In the next section we proceed to solve for the wavefunction 
$\Wtd$ by iterating the two-dimensional Hamiltonian 
\eqref{eq:Htdaction}.

\subsection{The two-dimensional wavefunction}
\label{sec:wf2d}
It is useful to settle on a language before diving into the 
iteration of the two-dimensional wavefunction. To this end 
we introduce the class of iterated integrals dubbed 
\emph{single-valued harmonic polylogarithms} (\ac{svhpls}), 
which were first described by Brown in ref.~\cite{Brown:2004ugm}. 
Since then, several applications of \ac{svhpls} in computing 
scattering amplitudes have been found, in particular in the 
context of the high-energy limit, e.g.~\cite{Pennington:2012zj,Dixon:2012yy,DelDuca:2013lma,Dixon:2014voa,DelDuca:2016lad,DelDuca:2018hrv}, 
and in the context of infared singularities in general 
kinematics~\cite{Almelid:2017qju,Dixon:2019lnw}. 
Here we will show that these functions also form a 
suitable basis for expressing the two-dimensional 
wavefunction $\Wtdl{\ell}(z,\zb)$ defined above.

As the name suggests, single-valued harmonic polylogarithms 
are single-valued functions which can be written as linear 
combinations of products of harmonic polylogarithms (\ac{hpls}) 
of $z$ with \ac{hpls} of $\zb$. We shall denote \ac{svhpls} by 
$\El_\sigma(z,\zb)$ where $\sigma$ is a sequence of 
\emph{letters}, typically zeros and ones.\footnote{For the most 
	part of this section we will use the standard letters, 0 and 1. 
	Only in section~\ref{sec:asalphabet} we introduce a new 
	alphabet to simplify the two-dimensional evolution.} 
The \emph{letters} are said to form an \emph{alphabet}, 
$\{0,1\}$, and $\sigma$ is, by analogy, referred to as a 
\emph{word}. The length of a word is often called the 
(transcendental) \emph{weight} of the \ac{svhpl}. 

\ac{svhpls} are the natural choice for the two-dimensional 
\ac{bfkl} evolution, since $j(z,\zb)$ of eq.~\eqref{eq:j2d} 
belongs to this class,
\beq
j(z,\zb) = \frac12 \El_{0}(z,\zb) + \El_{1}(z,\zb),
\eeq
and the action of the Hamiltonian $\Hitd$ preserves 
single-valuedness when acting on a single-valued function. 
This can be expected on general grounds: any complex pair 
$z,\zb$ identifies a point in the Euclidean transverse momentum 
plane. Physically there cannot be branch cuts in the Euclidean 
region; this, by definition, guarantees single-valued results. 
Indeed, single-valuedness may be confirmed at every step of the 
iteration. Determining the wavefunction is greatly simplified by 
working directly with SVHPLs; we briefly summarise their main 
properties, which will be used below, in Appendix~\ref{app:svhpls}.

As noted upon introducing the variables $z$ and $\bar{z}$ in 
(\ref{eq:zwdef}), the two-dimensional wavefunction is symmetric 
under $z \leftrightarrow \zb$. In addition, as mentioned following 
(\ref{k2zzb}), owing to the symmetry upon interchanging the two 
Reggeons in \eqn{left-right-symmetry}, the wavefunction is 
invariant under simultaneously swapping $z \leftrightarrow 1/z$ 
and $\zb \leftrightarrow 1/\zb$. Both these symmetries are easily 
verified by looking at eqs.~\eqref{eq:j2d} and \eqref{eq:hi2d}, 
where, for the latter symmetry, one changes the integration 
variables $w \to 1/w$, $\wb \to 1/\wb$. We will use these 
properties to simplify the iteration of the wavefunction as 
well as its results in section~\ref{sec:asalphabet}.

The evolution of the wavefunction in strictly two transverse 
dimensions according to~(\ref{eq:Htdaction}) has the following 
basic characteristics. Firstly, iterating $\Hmtd$ amounts to 
multiplying by $j(z,\zb)$ and therefore evaluating shuffle 
products of \ac{svhpls}. Secondly, each application of 
$\Hitd$ adds one layer of integration such that $\Wtdl{\ell-1}$ 
can be written as a linear combination of \ac{svhpls} of weight 
$\ell-1$.  A method to calculate the convolution in eq.~(\ref{eq:hi2d}) 
in terms of residues was described in chapter 6 of 
Ref.~\cite{DelDuca:2018hrv}.  Here we develop an 
alternative method: we translate the action of the 
Hamiltonian into a set of differential equations, 
which we then solve in terms of \ac{svhpls}. 

Suppose we wish to compute the action of a linear operator 
$\hat O$, which may involve integration, on a function $\Psi(z,\zb)$. 
Assume now that we find a differential operator $\Delta$, 
which is linear in logarithmic derivatives with respect to 
$z$ and $\zb$, with the following properties:
\begin{subequations}
\label{eq:deltacrit}
\begin{align}
	\label{eq:deltacriti} i.\ &\Delta \text{ commutes with } \hat{O} \\
	\label{eq:deltacritii} ii.\ &\Delta \Psi \text{ is a pure function with a 
	weight that is lower than } \Psi~\text{by one unit}.
\end{align}
\end{subequations}
Then,
\beq
\label{eq:deltadiffeq} \Delta \left[ \hat{O} \Psi(z,\zb) \right] 
= \hat{O} \left[ \Delta \Psi(z,\zb) \right],
\eeq
and we can compute $\hat{O} \left[ \Psi(z,\zb) \right]$ by integrating 
the differential equation \eqref{eq:deltadiffeq}, assuming that the 
r.h.s.\ is known explicitly. If it is not the procedure can be applied 
recursively, i.e.
\beq
\Delta \left[ \hat{O} \left[ \Delta \Psi(z,\zb) \right] \right] 
= \hat{O} \left[ \Delta^2 \Psi(z,\zb) \right],
\eeq
until the r.h.s.\ is simple enough to be calculated. 
After each integration a constant has to be fixed, 
e.g.\ by matching to known boundary conditions.

Importantly, because $\Delta$ is assumed to be linear in 
derivatives with respect to $z$ and~$\zb$, solving the 
differential equation amounts to computing a one-dimensional 
integral. This may be contrasted with the original integral in 
(\ref{eq:hi2d}) which is two-dimensional. Given 
\eqref{eq:deltacritii}, solving this differential equation 
is straightforward, and the result remains in the class 
of \ac{hpls}, (see eq.~\eqref{eq:hpldef}). The same 
applies for the class of \ac{svhpls}: to solve the 
differential equation within this class, we first 
integrate its holomorphic part according to 
eq.~\eqref{eq:holopartdef}, and subsequently 
recover the full result, depending on both $z$ 
and $\zb$, by applying the single-valued map 
$\mathbf{s}$ defined in eq.~\eqref{eq:sdef}. 
Having outlined the general approach let us see 
how it is implemented in practice to solve for the 
wavefunction in (\ref{eq:hi2d}).

Let us start by considering the $\hat{O}$ in 
eq.~(\ref{eq:deltadiffeq})  to coincide with the 
two-dimensional Hamiltonian $\Hitd$ (we will 
see below that the final procedure involves 
considering parts of the Hamiltonian $\Hitd$ 
in turn). The most natural candidate for the 
operator $\Delta$ in eq.~(\ref{eq:deltadiffeq}) 
is $\Delta_1=z \ddz$, since condition 
(\ref{eq:deltacriti}) is satisfied, as we 
now show.

For generic values of $w$ and $z$ one finds 
using eq.~\eqref{eq:K2d}
\beq
\label{eq:zddzKsym} z\ddz K(w,\wb,z,\zb) 
= -\ddw w K(w,\wb,z,\zb) \qquad \text{(for generic $w,z$)}.
\eeq
This implies that $z \ddz$ commutes with the Hamiltonian,
\begin{align}
	z\ddz \left[ \Hitd \Psi(z,\zb) \right] &= \frac{1}{4\pi} 
	\int \dd^2 w \left\{ \left( -\ddw w K(w,\wb,z,\zb) \right) \left[ \Psi(w,\wb) - \Psi(z,\zb) \right] \right. \nn \\
	&\hspace{45mm} \left. - K(w,\wb,z,\zb) \left( z\ddz \Psi(z,\zb) \right) \right\} \nn \\
	&= \frac{1}{4\pi} \int \dd^2 w K(w,\wb,z,\zb) \left[ w\ddw \Psi(w,\wb) - z\ddz \Psi(z,\zb) \right] \nn \\
	\label{eq:zddzhisym} &= \Hitd \left[ z\ddz \Psi(z,\zb) \right] \qquad \text{(for generic $w,z$)}.
\end{align}
fulfilling condition (\ref{eq:deltacriti}). However, some extra 
caution is needed here: the complex-conjugate pairs 
$w,\wb$ and $z,\zb$ cannot be treated as independent 
variables everywhere. Derivatives w.r.t.\ those variables 
receive additional contributions from the non-holomorphic 
or singular points of the function they act on. These 
``anomalies'' are captured by the two-dimensional 
Poisson equation
\beq
\label{eq:2dpoisson} \partial_w \partial_{\wb} \log (w \wb) = \pi \delta^2(w)
\eeq
namely, by contributions of the form
\beq
\label{eq:cterms} \ddw \frac{1}{\wb-c} = \pi \delta^2(w-c) 
\eeq
with $c$ a complex number. The two-dimensional $\delta$ 
function in the above equations fixes both the real and the 
imaginary part of its argument such that
\beq
\int \dd^2 w\, \delta^2(w-c)\, f(w,\wb) = f(c,\bar{c})
\eeq
for some function $f$, \emph{cf.}\ the remark below 
eq.~\eqref{eq:dw2d}.

For easy bookkeeping let us split a derivative into its 
\emph{regular part} (``reg''), which is correct in the 
holomorphic regime, and its \emph{contact terms} (``con''), 
governed by eq.~\eqref{eq:cterms}. Eq.~\eqref{eq:zddzKsym} 
therefore correctly reads
\begin{align}
	z\ddz K(w,\wb,z,\zb) &\!= \!\left[ z\ddz K(w,\wb,z,\zb) \right]_\reg \!\!
	+ \left[ z\ddz K(w,\wb,z,\zb) \right]_\con \nn \\
	&\!=\! -\left[ \ddw w K(w,\wb,z,\zb) \right]_\reg \!\!
	+ \left[ z\ddz K(w,\wb,z,\zb) \right]_\con \nn \\
	&\!=\! -\ddw \left[ w K(w,\wb,z,\zb) \right] 
	+ \left[ \ddw w K(w,\wb,z,\zb) \right]_\con \!\!
	+ \left[ z\ddz K(w,\wb,z,\zb) \right]_\con
\end{align}
which modifies eq.~\eqref{eq:zddzhisym} to give
\begin{multline}
	\label{eq:zddzhisym2} z\ddz \left[ \Hitd \Psi(z,\zb) \right] 
	= \Hitd \left[ z\ddz \Psi(z,\zb) \right] \\
	+ \frac{1}{4\pi} \int \dd^2 w \left\{ \left[ \ddw w K(w,\wb,z,\zb) \right]_\con 
	+ \left[ z\ddz K(w,\wb,z,\zb) \right]_\con \right\} \\
	\times \left[ \Psi(w,\wb) - \Psi(z,\zb) \right].
\end{multline}
We shall continue to refer to the behaviour in 
eq.~\eqref{eq:zddzhisym2} as the commutativity 
of $z\ddz$ and $\Hitd$ even though we implicitly 
mean commutativity \emph{modulo contact terms}. 
Note, that the presence of the contact terms does 
not conflict with the strategy outlined above; each 
contact term contains a (two-dimensional) 
$\delta$-function which makes the integral on 
the r.h.s.\ of eq.~\eqref{eq:zddzhisym2} easy 
to evaluate.

We will derive the explicit form of the contact 
terms towards the end of this section, at which 
point eq.~(\ref{eq:zddzhisym2}) will become 
directly usable for determining the action of 
$\Hitd$ on the wavefunction $\Psi$.  Before 
doing that, however, we turn our attention to 
condition (\ref{eq:deltacritii}).  Concretely in 
eq.~(\ref{eq:zddzhisym2}) the requirement is 
that $z\frac{d}{dz} \Psi$  should be a pure 
function of weight ones less than $\Psi$ itself. 
We find that the operator $z\frac{d}{dz}$, upon 
acting on any \ac{svhpl} of the form 
$\El_{0,\sigma}(z,\zb)$, does indeed 
yield such a pure function, so 
eq.~(\ref{eq:zddzhisym2}) 
becomes: 
\beq
z\ddz \left[ \Hitd \El_{0,\sigma}(z,\zb) \right] 
= \Hitd \left[ \El_\sigma(z,\zb) \right] 
+ \ct \label{eq:zddzL0},
\eeq
where we have used eq.~\eqref{eq:ddzL}. 
On the other hand, $z\ddz$ does not have the 
same effect when acting on an 
\ac{svhpl} $\El_{1,\sigma}(z,\zb)$, 
where one obtains instead
\beq
z\ddz \left[ \Hitd \El_{1,\sigma}(z,\zb) \right] 
= \Hitd \left[ \frac{z}{1-z} \El_\sigma(z,\zb) \right] + \ct,
\eeq
which does not fulfil the condition~\eqref{eq:deltacritii}. 
One may be tempted to use $(1-z)\ddz$ instead but, 
unfortunately, this operator does not commute with 
$\Hitd$. 

The solution is to first split the Hamiltonian 
$\Hitd = \Hitdone + \Hitdtwo$ with
\beq
\Hitdn \Psi(z,\zb) = \frac{1}{4\pi} \int \dd^2 w 
K_n(w,\wb,z,\zb) \left[ \Psi(w,\wb) - \Psi(z,\zb) \right]
\eeq
and
\begin{subequations}
\begin{align}
	K_1(w,\wb,z,\zb) &= 
	\left(\frac{1}{w-z}-\frac{1}{w}\right) \frac{1}{\wb-\zb} \\
	K_2(w,\wb,z,\zb) &= 
	\frac{1}{w-z} \left(\frac{1}{\wb-\zb}-\frac{1}{\wb}\right)
\end{align}
\end{subequations}
where $K_1(w,\wb,z,\zb) + K_2(w,\wb,z,\zb) 
= K(w,\wb,z,\zb)$, \emph{cf.}\ eq.~\eqref{eq:K2d}.
This split is useful because it opens the possibility of identifying 
different differential operators $\Delta_i$ that commute with the 
separate components of the Hamiltonian $\Hitdone$ and $\Hitdtwo$, 
\emph{and} yield a pure function when acting directly on 
$\El_{0,\sigma}(z,\zb)$ or on $\El_{1,\sigma}(z,\zb)$, thus 
simultaneously fulfilling both conditions in~\eqref{eq:deltacrit}.

Regarding the commutation relations, condition \eqref{eq:deltacriti}, 
it is straightforward to verify that the following four relations hold, 
up to contact terms:
\begin{subequations}
\label{eq:Hicommut}
\begin{align}
	\left[ z\ddz, \Hitdone \right] &= \ct, & 
	\left[ z(1-z)\ddz, \Hitdone \right] &= \ct \label{eq:Hi1ops}, \\
	\left[ z\ddz, \Hitdtwo \right] &= \ct, & 
	\left[ (1-z)\ddz, \Hitdtwo \right] &= \ct. \label{eq:Hi2ops} 
\end{align}
\end{subequations}
Let us therefore define the following three differential operators:
\beq
\label{Deltai}
\Delta_i = f_i(z) \ddz \qquad \text{with} \qquad f_i(z) 
= \left\{ \begin{array}{ll} z& \,\text{for}\,\,i=1 
\\ 1-z& \,\text{for}\,\,i=2  \\ z(1-z) &\, \text{for}\,\,i=3  \end{array} \right.\,,
\eeq
and show that we can arrange the wavefunction, which 
is a linear combination of $\El_{0,\sigma}(z,\zb)$ and 
$\El_{1,\sigma}(z,\zb)$, such that condition 
(\ref{eq:deltacritii}) would also be fulfilled. 

To this end, let us first note that upon acting on 
$\El_{0,\sigma}(z,\zb)$ with either of the two parts 
of the Hamiltonian we have (using \eqref{eq:Hicommut}):
\begin{equation}
	\label{eq:hi1diffeq2} z\ddz \left[ \Hitdn \El_{0,\sigma}(z,\zb)  \right] 
	= \Hitdn \left[ \El_\sigma(z,\zb) \right] + \ct\,,
\end{equation}
just as in (\ref{eq:zddzL0}). Thus, the remaining 
challenge is to handle terms containing $\El_{1,\sigma}(z,\zb)$; 
this is where the additional flexibility of splitting the Hamiltonian 
pays off. Let us consider first the simplest case of $\Hitdtwo$ 
where we obtain
\beq
\label{eq:hi2diffeq}
(1-z)\ddz \left[ \Hitdtwo \El_{1,\sigma}(z,\zb) \right] 
= \Hitdtwo \left[ \El_\sigma(z,\zb) \right] + \ct.
\eeq
Now $\Hitdtwo \Psi$ can be readily integrated for any 
$\Psi$ using (\ref{eq:hi1diffeq2}) and (\ref{eq:hi2diffeq}). 
Turning to consider $\Hitdone$, let us write
\beq
\El_{1,\sigma}(z,\zb) = (\El_{1,\sigma}(z,\zb) +
\El_{0,\sigma}(z,\zb)) - \El_{0,\sigma}(z,\zb) \label{eq:L1split}
\eeq
and use the linearity of the Hamiltonian to act with it on 
$(\El_{1,\sigma} + \El_{0,\sigma})$ and $(-\El_{0,\sigma})$ 
separately. We may now apply respectively the differential 
operators $\Delta_3$ and $\Delta_1$ of \eqref{Deltai} to 
these terms. With eq.~\eqref{eq:Hi1ops} and \eqref{eq:ddzL} 
one can easily verify that they produce the desired pure 
functions of lower weight in accordance with
\eqref{eq:deltacritii}:
\begin{align}
	\label{eq:hi1diffeq1} z(1-z)\ddz \left[ \Hitdone \left( \El_{0,\sigma}(z,\zb) 
	+ \El_{1,\sigma}(z,\zb) \right) \right] 
	&= \Hitdone \left[ \El_\sigma(z,\zb) \right] + \ct 
\end{align}
Using (\ref{eq:hi1diffeq1}) along with (\ref{eq:hi1diffeq2}) 
we see that also $\Hitdone \Psi$ can be integrated for any 
$\Psi$. Thus, by splitting the Hamiltonian and the wavefunction 
in a convenient  way, we were able to identify linear differential 
operators that admit both requirements in (\ref{eq:deltacrit}).

In order to complete the process of setting up the differential 
equations let us now return to derive the explicit form of the 
contact terms. First, let us write eq.~\eqref{eq:zddzhisym2} 
for general $\Delta_i = f_i(z) \ddz$ and the two parts of the 
split Hamiltonian,
\begin{multline}
	\label{eq:fddzhinsym} f_i(z) \ddz \left[ \Hitdn \Psi(z,\zb) \right] 
	= \Hitdn \left[ f_i(z) \ddz \Psi(z,\zb) \right] \\
	+ \frac{1}{4\pi} \int \dd^2 w \left\{ \left[ \ddw 
	f_i(w) K_n(w,\wb,z,\zb) \right]_\con 
	+ \left[ f_i(z) \ddz K_n(w,\wb,z,\zb) \right]_\con \right\} \\
	\times \left[ \Psi(w,\wb) - \Psi(z,\zb) \right]
\end{multline}
where, according to eqs.~(\ref{eq:hi1diffeq2}), (\ref{eq:hi2diffeq}) 
and~(\ref{eq:hi1diffeq1}), the relevant combinations of $i$ and 
$n$ are 
\begin{subequations}
\begin{align}
	n = 1 \quad &\longrightarrow \quad i = 1 \text{ or } 3 \\
	n = 2 \quad &\longrightarrow \quad i = 1 \text{ or } 2\,.
\end{align}
\end{subequations}
In computing the contact terms in (\ref{eq:fddzhinsym}) we note 
that the $f_i(z)$~\eqref{Deltai} are functions of $z$ only whilst 
being independent of the complex conjugate $\zb$. According 
to eq.~\eqref{eq:cterms} this implies that
\beq
\label{eq:pulloutf} \left[ \ddw f_i(w) K_n(w,\wb,z,\zb) \right]_\con 
= f_i(w) \left[ \ddw K_n(w,\wb,z,\zb) \right]_\con.
\eeq
for $n=1,2$, and thus (\ref{eq:fddzhinsym}) becomes:
\begin{multline}
	\label{eq:fddzhinsym_simpl} f_i(z) \ddz \left[ \Hitdn \Psi(z,\zb) \right] 
	= \Hitdn \left[ f_i(z) \ddz \Psi(z,\zb) \right] \\
	+ \frac{1}{4\pi} \int \dd^2 w \left\{ \left[f_i(w) \ddw 
	K_n(w,\wb,z,\zb) \right]_\con 
	+ \left[ f_i(z) \ddz K_n(w,\wb,z,\zb) \right]_\con \right\} \\
	\times \left[ \Psi(w,\wb) - \Psi(z,\zb) \right]\,.
\end{multline}
Consequently, we only have to consider the following four derivatives,
\begin{subequations}
\begin{align}
	\label{eq:ddwk1ct} \left[ \ddw K_1(w,\wb,z,\zb) \right]_\con 
	&= \pi \left[ \delta^2(w-z) - \delta^2(w-\infty) \right] \frac{z}{w(w-z)} \\
	\label{eq:ddzk1ct} \left[ \ddz K_1(w,\wb,z,\zb) \right]_\con 
	&= -\pi \delta^2(z-w) \frac{z}{w(w-z)} \\[3mm]
	\label{eq:ddwk2ct} \left[ \ddw K_2(w,\wb,z,\zb) \right]_\con 
	&= \pi \left[ \delta^2(w-z) - \delta^2(w) \right] \frac{1}{w-z} \\  
	\label{eq:ddzk2ct} \left[ \ddz K_2(w,\wb,z,\zb) \right]_\con 
	&= -\pi \delta^2(z-w) \frac{1}{w-z},
\end{align}
\end{subequations}
where in eqs.~\eqref{eq:ddzk1ct} and \eqref{eq:ddzk2ct} we 
have dropped terms proportional 
to $\delta^2(z)$, restricting our calculation to $z \neq 0$ (we 
emphasise that $z$ is an external variable so this can be 
consistently done). Due to the sum of contact terms inside 
the curly brackets in eq.~\eqref{eq:fddzhinsym_simpl}  the 
terms proportional to $\delta^2(w-z) = \delta^2(z-w)$ in 
eqs.~\eqref{eq:ddwk1ct}--\eqref{eq:ddzk2ct} cancel identically, 
so the remaining contact-term contributions are only at $w=\infty$ 
for $K_1$ and at $w=0$ for $K_2$.  Using the corresponding 
$\delta$ functions to turn the integrals over $w$ in 
(\ref{eq:fddzhinsym_simpl} ) into evaluation of limits at 
infinity and  at zero respectively we finally obtain:
\begin{align}
	\label{eq:fddzhinsym_H1} f_i(z) \ddz 
	\left[ \Hitdone \Psi(z,\zb) \right] 
	&= \Hitdone \left[ f_i(z) \ddz \Psi(z,\zb) \right] 
	-\frac{1}{4} \lim_{w\to\infty}  \frac{z f_i(w)}{w(w-z)}  
	\left[ \Psi(w,\wb) - \Psi(z,\zb) \right]\,, \\
	\label{eq:fddzhinsym_H2} f_i(z) \ddz 
	\left[ \Hitdtwo \Psi(z,\zb) \right] 
	&= \Hitdtwo \left[ f_i(z) \ddz \Psi(z,\zb) \right] 
	- \frac{1}{4}  \lim_{w\to 0}  \frac{f_i(w) }{w-z}  
	\left[ \Psi(w,\wb) - \Psi(z,\zb) \right]\,.
\end{align}
This equations will be used in the next section to 
determine the wavefunction.

\subsection{Differential equations and an iterative solution for the wavefunction}
\label{sec:diffeq} Finding the differential equations is now 
simply a matter of compiling together the results of the 
previous section. Starting with the easiest case, 
$\Delta_1 \Hitdn \El_{0,\sigma}$, we notice that with 
$f_1(w) = w$ both the $w\to \infty$ limit in eq.~(\ref{eq:fddzhinsym_H1}) 
and the $w\to 0$ limit in eq.~(\ref{eq:fddzhinsym_H2}) 
vanish, and thus there are no contributions from contact 
terms in either of these cases. Dividing by $f_1(z) = z$ 
to arrive at
\beq
\ddz \Hitdn \El_{0,\sigma}(z,\zb) = 
\frac{\Hitdn \El_{\sigma}(z,\zb)}{z}\,. \label{eq:diffeq0}
\eeq
Next consider the case $\Delta_2 \Hitdtwo \El_{1,\sigma}$, 
corresponding to eq.~\eqref{eq:hi2diffeq}. Here $f_2(w) = 1-w$ 
and eq.~(\ref{eq:fddzhinsym_H2}) yields
\beq \label{eq:diffeq1b}
\ddz \Hitdtwo \El_{1,\sigma}(z,\zb) 
= \frac{\Hitdtwo \El_{\sigma}(z,\zb)}{1-z} 
- \frac14 \frac{\El_{1,\sigma}(z,\zb) 
- [\El_{1,\sigma}(w,\wb)]_{w,\wb \rightarrow 0}}{z(1-z)}. 
\eeq
where we have divided by $f_2(z) = 1-z$ and used the 
shorthand $[\ldots]_{w,\wb \rightarrow 0}$ to denote the 
$w,\wb \rightarrow 0$ limit of the functions inside the 
square brackets. This term can, in fact, be 
dropped as it always contains a single \ac{svhpl} 
whose indices feature (at least) one ``1'' and, thus, 
is equal to zero in the limit.

The last case, $\Delta_i \Hitdone \El_{1,\sigma}$, 
is governed by eqs.~\eqref{eq:hi1diffeq1} and 
\eqref{eq:hi1diffeq2}, using the wavefunction 
split of eq.~(\ref{eq:L1split}). 
Considering in turn the action of eq.~(\ref{eq:fddzhinsym_H1})  
on $(\El_{1,\sigma}(z,\zb) +\El_{0,\sigma}(z,\zb))$ with $f_3(w) 
= w(1-w)$ and on $(- \El_{0,\sigma}(z,\zb))$ with $f_1(w) = w$, 
we derive two separate equations, which we then combine 
using the linearity of operators $\Hitdone$ and~$\ddz$ to obtain
\begin{multline} \label{eq:diffeq1a}
	\ddz \Hitdone \El_{1,\sigma}(z,\zb) 
	= \frac{\Hitdone \El_{\sigma}(z,\zb)}{1-z} \\
	- \frac14 \frac{\El_{0,\sigma}(z,\zb) 
	+ \El_{1,\sigma}(z,\zb) - [\El_{0,\sigma}(w,\wb) 
	+ \El_{1,\sigma}(w,\wb)]_{w,\wb \rightarrow \infty}}{1-z} 
\end{multline}
with $[\ldots]_{w,\wb \rightarrow \infty}$ the 
$w,\wb \rightarrow \infty$ limit of the functions inside 
the square brackets. Taking this limit requires some 
careful analytic continuation of the relevant \ac{hpls} 
to ensure that $w$ and $\wb$ stay complex-conjugate 
as they approach infinity. 

Because the Hamiltonian $\Hitd$ and its components 
$\Hitdn$ are linear operators one can sum up the above 
equations \eqref{eq:diffeq0}--\eqref{eq:diffeq1a} 
and recombine $\Hitdone + \Hitdtwo \to \Hitd$ 
obtaining more compact expressions:
\begin{subequations}
\label{diffeq}
\begin{align}
	\ddz \Hitd \El_{0,\sigma}(z,\zb) 
	&= \frac{\Hitd \El_{\sigma}(z,\zb)}{z}\,, 
	\label{eq:diffeq0new} \\
	\ddz \Hitd \El_{1,\sigma}(z,\zb) 
	&= \frac{\Hitd \El_{\sigma}(z,\zb)}{1-z} 
	- \frac14 \frac{\El_{1,\sigma}(z,\zb)}{z} \nn \\
	&\hspace{-5mm} - \frac14 \frac{\El_{0,\sigma}(z,\zb) 
	+ 2\El_{1,\sigma}(z,\zb) - [\El_{0,\sigma}(w,\wb) 
	+ \El_{1,\sigma}(w,\wb)]_{w,\wb \rightarrow \infty}}{1-z}\,. 
	\label{eq:diffeq1new}
\end{align}
\end{subequations}
These differential equations compactly represent the action 
of the Hamiltonian $\hat{H}_{{\rm 2d},{\rm i}}$ according to 
eq.~(\ref{eq:hi2d}). By solving them we are able to effectively 
bypass the computation of the two-dimensional integrals in 
the latter equation.

Since the differential equations only fix the $z$ 
dependence of the (wave)function --- which is a 
function of both $z$ and $\zb$ --- a small detour 
is necessary to recover the action of $\Hitd$ on 
\ac{svhpls}: we take the holomorphic part of a 
given \ac{svhpl}, integrate it w.r.t.~$z$ according 
to the differential equations in (\ref{diffeq}), and then
reconstruct the functional dependence on~$\zb$ 
by requiring the result be single-valued. 
This ultimately amounts to simply replacing
\beq
\int_0^z \dd t \frac{\El_\sigma(t,\bar{t})}{t} 
\longrightarrow \El_{0,\sigma}(z,\zb) \qquad 
\text{and} \qquad \int_0^z \dd t \frac{\El_\sigma(t,\bar{t})}{1-t} 
\longrightarrow \El_{1,\sigma}(z,\zb) \,.
\eeq
For more details on this procedure see 
appendix~\ref{app:holomorphicpart}.

After each integration we need to fix an 
integration constant. We find that this is 
conveniently done by matching with the soft limit. 
Specifically, it is convenient to consider the soft 
limit where $k^2/p^2=z\zb$ tend to zero.
For small $z,\zb$, only \ac{svhpls} 
with all-zero indices can give non-zero 
contributions; these correspond to 
powers of logarithms:
\begin{equation}
	\El_{\vec{0}_n}(z,\zb) = \frac{\log^n(z\zb)}{n!} 
	\qquad \text{with} \qquad \vec{0}_n = 
	\underbrace{0,\dots,0}_{n\ \text{zeros}} \,.
\end{equation}
In eq.~\eqref{softHpower} we calculated the 
action of the small-$k$ (or soft) Hamiltonian 
$\Has$ on powers of $\xi = (k^2/p^2)^{-\eps}$. 
The action of $\Hi$ in the soft limit can be 
isolated by looking at the coefficient of 
$2\Ca - \Tt$ and thus is
\begin{equation}
	\Hi|_{\text{soft}} \left( \frac{k^2}{p^2} \right)^{-m\eps} 
	= \frac{\hat B_m(\eps)}{2\eps} 
	\left( \frac{k^2}{p^2} \right)^{-(m+1)\eps} \label{eq:Hik2p2}
\end{equation}
where $\hat B_m(\eps)$ is given in eq.~(\ref{bubblehat}).
Expanding both sides in $\eps$ and matching powers of 
$\delta=m\epsilon$ in the limit $\eps \rightarrow 0$ lets us 
extract the action of $\Hi$ in the soft limit on any given 
power of $\log (k^2/p^2) = \log (z\zb)$. For reference, 
we find
\begin{subequations}
\begin{align}
	\Hi|_{\text{soft}} \El_0(z,\zb) 
	&= \mathcal{O}(\eps) \label{HiL0} \\
	\Hi|_{\text{soft}} \El_{0,0}(z,\zb) 
	&= \zeta_3 + \mathcal{O}(\eps) \\
	\Hi|_{\text{soft}} \El_{0,0,0}(z,\zb) 
	&= \zeta_3 \El_0(z,\zb) + \mathcal{O}(\eps) \\
	\Hi|_{\text{soft}} \El_{0,0,0,0}(z,\zb) 
	&= \zeta_3 \El_{0,0}(z,\zb) + \zeta_5 + \mathcal{O}(\eps) \\
	\Hi|_{\text{soft}} \El_{0,0,0,0,0}(z,\zb) 
	&= \zeta_3 \El_{0,0,0}(z,\zb) + \zeta_5 \El_0(z,\zb) 
	+ \mathcal{O}(\eps) \label{HiL00000} 
\end{align}
\end{subequations}
etc., from which we observe that the integration 
constants exhibit a very simple pattern. Specifically, 
they only contribute single (ordinary) zeta numbers 
because they are generated upon expanding 
$\Bh{m}(\eps)$ which is a product of
gamma functions. 

We can now calculate the action of $\Hitd$ on 
any \ac{svhpl} by iteratively solving the differential 
equations \eqref{eq:diffeq0new} and \eqref{eq:diffeq1new}, 
starting from the lowest-weight functions, $\El_0$ and $\El_1$. 
Effectively, we have set up an algorithm for calculating the 
two-dimensional wavefunction to any loop order. Due to 
the finiteness of the wavefunction it is straightforward to 
verify the results numerically: We integrate 
eq.~\eqref{eq:hi2d} numerically and compare 
to the analytical result for a number of randomly 
generated pairs $z,\zb$. Specifically, with 
$w = w_1 + i w_2$ and $z = z_1 + i z_2$ the 
action of $\Hitd$ \eqref{eq:hi2d} can be written
\begin{multline}
	\Hitd \Psi(z,\zb) = \frac{1}{2\pi} \int_{-\infty}^\infty 
	\dd w_1 \int_{-\infty}^\infty \dd w_2 \frac{w_1 z_1 
	+ w_2 z_2}{(w_1^2 + w_2^2)((w_1 - z_1)^2 + (w_2 - z_2)^2)} \\
	\times \left[ \Psi(w_1 + i w_2,w_1 - i w_2) 
	- \Psi(z_1 + i z_2,z_1 - i z_2) \right].
\end{multline}
where $\Psi(z,\zb)$ is a (linear combination of) 
\ac{svhpl}(s). This type of integral is readily 
evaluated numerically in e.g.\ \texttt{Mathematica}.

For the wavefunction up to weight four we find
\begin{subequations}
\begin{align}
	\Wtdl{1} &= \frac{1}{2} C_2 \left(\El_0+2 \El_1\right) \label{eq:wtd1} \\
	\Wtdl{2} &= \frac{1}{2} C_2^2 \left(\El_{0,0}+2 \El_{0,1}
	+2 \El_{1,0}+4 \El_{1,1}\right)+\frac{1}{4} C_1 C_2 \left(-\El_{0,1}
	-\El_{1,0}-2 \El_{1,1}\right) \\
	\Wtdl{3} &=
	\frac{3}{4} C_2^3 \left(\El_{0,0,0}+2 \El_{0,0,1}
	+2 \El_{0,1,0}+4 \El_{0,1,1}+2 \El_{1,0,0}
	+4 \El_{1,0,1}+4 \El_{1,1,0}
	+8 \El_{1,1,1}\right)
	\nn \\	&\hspace{4mm}+
	\frac{1}{4} C_1 C_2^2 \left(2 \zeta_3 -2 \El_{0,0,1}
	-3 \El_{0,1,0}-7 \El_{0,1,1}-2 \El_{1,0,0}
	-7 \El_{1,0,1}-7 \El_{1,1,0}
	-14 \El_{1,1,1}\right)
	\nn \\&\hspace{4mm}
	+\frac{1}{16} C_1^2 C_2 \left(\El_{0,0,1}
	+2 \El_{0,1,0}+4 \El_{0,1,1} 
	+\El_{1,0,0}+4 \El_{1,0,1}
	+4 \El_{1,1,0}+8 \El_{1,1,1}\right)\\
	\Wtdl{4} &= 
	\frac{3}{2} C_2^4 \left(\El_{0,0,0,0}
	+2 \El_{0,0,0,1}+2 \El_{0,0,1,0}+4 \El_{0,0,1,1}
	+2 \El_{0,1,0,0}+4 \El_{0,1,0,1}\right. \nn \\
	&\hspace{4mm} \left. +4 \El_{0,1,1,0}+8 \El_{0,1,1,1}
	+2 \El_{1,0,0,0}+4 \El_{1,0,0,1}+4 \El_{1,0,1,0}
	+8 \El_{1,0,1,1}\right. \nn \\
	&\hspace{4mm} \left. +4 \El_{1,1,0,0}+8 \El_{1,1,0,1}
	+8 \El_{1,1,1,0}+16 \El_{1,1,1,1}\right)
	\nn \\
	&\hspace{4mm}
	+\frac{1}{8} C_1 C_2^3 \left(-9 \El_{0,0,0,1}
	-14 \El_{0,0,1,0}-34 \El_{0,0,1,1}-14 \El_{0,1,0,0}
	-42 \El_{0,1,0,1}\right. \nn \\
	&\hspace{4mm} \left. -44 \El_{0,1,1,0}-92 \El_{0,1,1,1}
	-9 \El_{1,0,0,0}-34 \El_{1,0,0,1}-42 \El_{1,0,1,0}
	-92 \El_{1,0,1,1}\right. \nn \\
	&\hspace{4mm} \left. -34 \El_{1,1,0,0}-92 \El_{1,1,0,1}
	-92 \El_{1,1,1,0}-184 \El_{1,1,1,1}+8 \El_0 \zeta_3
	+28 \El_1 \zeta_3\right) \nn \\
	&\hspace{4mm} +\frac{1}{32} C_1^2 C_2^2 \left(7 \El_{0,0,0,1}
	+15 \El_{0,0,1,0}+34 \El_{0,0,1,1}+15 \El_{0,1,0,0}
	+56 \El_{0,1,0,1}\right. \nn \\
	&\hspace{4mm} \left. +56 \El_{0,1,1,0}+116 \El_{0,1,1,1}
	+7 \El_{1,0,0,0}+40 \El_{1,0,0,1}+56 \El_{1,0,1,0}\right. \nn \\
	&\hspace{4mm} \left. +116 \El_{1,0,1,1}+34 \El_{1,1,0,0}
	+116 \El_{1,1,0,1}+116 \El_{1,1,1,0}+232 \El_{1,1,1,1}
	-44 \El_1 \zeta_3\right) \nn \\
	&\hspace{4mm}
	+\frac{1}{64} C_1^3 C_2 \left(-\El_{0,0,0,1}
	-3 \El_{0,0,1,0}-6 \El_{0,0,1,1}-3 \El_{0,1,0,0}\right. \nn \\
	&\hspace{4mm} \left. -12 \El_{0,1,0,1}-12 \El_{0,1,1,0}
	-24 \El_{0,1,1,1}-\El_{1,0,0,0}-8 \El_{1,0,0,1}
	-12 \El_{1,0,1,0}\right. \nn \\
	&\hspace{4mm} \left. -24 \El_{1,0,1,1}-6 \El_{1,1,0,0}
	-24 \El_{1,1,0,1}-24 \El_{1,1,1,0}-48 \El_{1,1,1,1}
	+12 \El_1 \zeta_3\right) 
	\label{eq:wtd4}
\end{align}
\end{subequations}
where we introduced the notation $C_1 = 2\Ca - \Tt$, 
$C_2 = \Ca - \Tt$ and wrote $\Wtdl{\ell} \equiv \Wtdl{\ell}(z,\zb)$ 
and $\El_\sigma \equiv \El_\sigma(z,\zb)$ for brevity. 
Further results up to weight 14 can be found in the 
ancillary file \texttt{2Reggeon-wavefunction-L01-Basis.txt}.

Interestingly, a new type of transcendental number 
appears for the first time in the twelve-loop 
wavefunction --- a so-called multiple zeta value 
(\ac{mzv}). While it is no surprise that \ac{mzvs} 
do not appear at lower loop orders as we explain 
in the following two paragraphs, the fact that they 
\emph{do} appear starting at twelve loops is a 
non-trivial statement with number-theoretical 
implications. 

\ac{mzvs} are the values of \ac{hpls} evaluated 
at special points, typically their branch points 
$z=1$ or $z \to \infty$, for example\footnote{\ac{mzvs} 
use the collapsed notation, \emph{cf.}\ eq.~\eqref{eq:collapsedn} 
in appendix~\ref{app:hpls}.} $H_{0,0,0,0,1,0,0,1}(1) = 
H_{5,3}(1) = \zeta_{5,3}$. It turns out that \ac{svhpls} only 
cover a subset of all \ac{mzvs} when evaluated at 
$z = \zb = 1$ or $z,\zb \to \infty$ and we refer to this 
subset as single-valued multiple zeta values. They 
are discussed in detail in refs.~\cite{Schnetz:2013hqa,Brown:2013gia} 
where the authors show that, up to weight ten, 
the algebra of single-valued \ac{mzvs} is 
generated by ordinary (odd) zeta numbers 
$\zeta_n$. At weight eleven, however, a new 
type of number appears, alongside the expected 
$\zeta_{11}$. We shall call it\footnote{Brown~\cite{Brown:2013gia} 
refers to it as $\zeta_{\rm sv}(3,5,3)$ while 
Schnetz~\cite{Schnetz:2013hqa} calls it $g_{335}$.} 
$g_{5,3,3}$ and it is defined by
\beq
g_{5,3,3} = -\frac47 \zeta_2^3 \zeta_5 +\frac65 \zeta_2^2 \zeta_7 
+ 45 \zeta_2 \zeta_9 + \zeta_{5,3,3}\,, \label{eq:g533}
\eeq
where $\zeta_{5,3,3} = H_{5,3,3}(1)$. 

There are two sources that contribute (multiple) zeta 
values to the wavefunction: the integration constants 
fixed by the soft limit and the $w,\wb \to \infty$ limit 
in eq.~\eqref{eq:diffeq1new}. The former are generated 
by expanding gamma functions, \emph{cf.}\ eq.~\eqref{eq:Hik2p2} 
with eq.~\eqref{bubblehat}, and can therefore contribute 
only \emph{single} (ordinary) zeta numbers. The value of 
the large-$w,\wb$ limit instead does generally involve (single-valued)  
\emph{multiple} zeta values. We note that it is guaranteed 
to multiply the weight-one \ac{svhpl} $\El_1(z,\zb)$ 
which is generated by the denominator, $1-z$, upon 
integrating the differential equation \eqref{eq:diffeq1new}. 
Being the sole source of (single-valued) \ac{mzvs}, we 
conclude that such zeta values of weight $w$ can only 
occur starting at the next loop order, i.e. $\ell=w+1$. 
Specifically, this explains why $g_{5,3,3}$, which is 
weight 11, cannot appear at loop orders $\ell < 12$. 
Indeed, we find that $g_{5,3,3}$ is accompanied by 
$\El_1$ in the twelve-loop wavefunction:
\begin{multline}
	\Wtdl{12}(z,\zb) \supset \frac{1}{80} \left( 
	\frac{88653 C_2^2 C_1^{10}}{2048}
	-\frac{1021171 C_2^3 C_1^9}{4096}
	-\frac{3517129 C_2^4 C_1^8}{1024} \right. \\
	+\frac{43378313 C_2^5 C_1^7}{1024}
	-\frac{5951395 C_2^6 C_1^6}{32}
	+\frac{1583033 C_2^7 C_1^5}{4} \\
	\left. -\frac{6320709 C_2^8 C_1^4}{16}
	+ 135513 C_2^9 C_1^3 \right) \times g_{5,3,3} \, \El_1(z,\zb)\,.
\end{multline}
According to ref.~\cite{Brown:2013gia} (\emph{cf.}\ eq.~(7.4) there) 
two more such numbers have to be introduced at weight 13 
and, using the same logic, we anticipate that they make an 
appearance in the 14-loop wavefunction. Indeed, defining
\beq
\label{eq:g553} g_{5,5,3} =  10 \zeta_2^2 \zeta_9 
+ \frac{275}{2} \zeta_2 \zeta_{11} 
+ 5 \zeta_5 \zeta_{5,3} + \zeta_{5,5,3}
\eeq
and
\beq
\label{eq:g733} g_{7,3,3} = -\frac{32}{35} \zeta_2^3 \zeta_7 
+ \frac{56}{5} \zeta_2^2 \zeta_9 + \frac{407}{2} \zeta_2 \zeta_{11} 
+ 6 \zeta_5 \zeta_{5,3} + \zeta_{7,3,3}
\eeq
we observe that the 14-loop wavefunction contains the term
\begin{multline}
	\Wtdl{14}(z,\zb) \supset \frac{1}{2240}\left( 
	- \frac{132291047 C_2^2 C_1^{12}}{20480} 
	+ \frac{7701138629 C_2^3 C_1^{11}}{183500800} \right. \\
	- \frac{21177619993 C_2^4 C_1^{10}}{81920} 
	- \frac{141869475599 C_2^5 C_1^9}{40960} 
	+ \frac{144180124197 C_2^6 C_1^8}{4096} \\
	- \frac{1550199662073 C_2^7 C_1^7}{10240} 
	+ \frac{941115705999 C_2^8 C_1^6}{2560} 
	- \frac{41630406511 C_2^9 C_1^5}{80} \\ \left. 
	+ \frac{15828500247 C_2^{10} C_1^4}{40} 
	- 120229353 C_2^{11} C_1^3 \right) \times g_{5,5,3} \, \El_1(z,\zb)
\end{multline}
as well as
\begin{multline}
	\Wtdl{14}(z,\zb) \supset \frac{1}{896} \left( 
	\frac{557319 C_2^2 C_1^{12}}{256} 
	- \frac{296956417 C_2^3 C_1^{11}}{16384} \right. \\
	- \frac{3811324785 C_2^4 C_1^{10}}{16384} 
	+ \frac{36358896425 C_2^5 C_1^9}{8192} 
	- \frac{125984665967 C_2^6 C_1^8}{4096} \\
	+ \frac{241764230539 C_2^7 C_1^7}{2048} 
	- \frac{139303244409 C_2^8 C_1^6}{512} 
	+ \frac{11897473261 C_2^9 C_1^5}{32} \\ \left. 
	- \frac{2180551359 C_2^{10} C_1^4}{8} 
	+ 79134813 C_2^{11} C_1^3 \right) \times g_{7,3,3} \, \El_1(z,\zb).
\end{multline}
The observed term $g_{5,3,3}\, \El_1(z,\zb)$ at twelve loops 
immediately rules out the possibility to find a closed-form 
expresson for the two-dimensional wavefunction in terms 
of gamma functions as was done in the soft limit 
\eqref{Well-1-ansatz}. The non-zero coefficients of 
$g_{5,5,3}\, \El_1(z,\zb)$ and $g_{7,3,3}\, \El_1(z,\zb)$ 
at 14 loops may be seen as hint that indeed \emph{all} 
single-valued \ac{mzvs} appear in the two-dimensional 
wavefunction --- when and as soon as the weight, i.e.\ 
loop order, allows for it.

We will, in fact, encounter a 
contribution proportional to $g_{5,3,3}$ in the amplitude 
at eleven loops. We will thus return to discuss 
single-valued \ac{mzvs} when interpreting our results for 
the amplitude in section~\ref{sec:finiteamp}.

Before we press ahead and compute the amplitude it is 
worthwhile exploring the aforementioned 
symmetries of the wavefunction in some more detail and we 
do so in the next subsection. This will ultimately lead to a better 
understanding of the iteration in two dimensions and enable 
us to calculate it to even higher loop orders.

\subsection{Alphabets and symmetries}
\label{sec:asalphabet}
Throughout this paper we have tried 
to exploit the symmetries of the \ac{bfkl} evolution to aide 
calculations and simplify expressions. In this section we 
explore to what extent symmetries can guide us in the 
two-dimensional limit. As mentioned in section~\ref{sec:wf2d}, 
in two dimensions, the wavefunction is invariant under two 
transformations: complex conjugation and inversion of the 
arguments. The latter, i.e.\ the fact that $\Wtd(z,\zb) = \Wtd(1/z,1/\zb)$, 
corresponds to eq.~(\ref{left-right-symmetry}), i.e. to swapping the 
two reggeons, and was used, for example, to identify the two 
soft limits in section~\ref{soft}. In the present context, it 
inspired us to introduce a new alphabet for \ac{svhpls}, 
as we now explain. Instead of 0 and 1, corresponding 
to integration over $\dd \log z$ and $\dd \log (1-z)$, 
respectively, we shall use $a$ and $s$. They are 
associated with integration over $\dd \log z$ and 
$\dd \log z/(1-z)^2$ and thus behave antisymmetrically 
and symmetrically, respectively, under $z \to 1/z$. 
In particular 
\begin{equation}
	\label{ElDef}
	\El_s(z,\zb)=\log \frac{z\zb}{(1-z)^2(1-\zb)^2}
	\qquad\Longrightarrow\qquad 
	\El_s(1/z,1/\zb)=\El_s(z,\zb)\,.
\end{equation}
The leading-order wavefunction  simplifies to 
$\Wtdl{1} = \frac{1}{2} C_2 \El_s(z,\zb)$, and 
at higher orders, the $z \to 1/z$ symmetry 
implies that the antisymmetric letter $a$ would 
only ever appear an even number of times.

Let us now consider the evolution directly in 
terms of this alphabet.
Using the letters $a$ and $s$ simplifies 
$j(z,\zb) = \El_s(z,\zb)/2$ of eq.~\eqref{eq:j2d} and 
hence the action of $\Hmtd$ in eq.~\eqref{eq:hm2d}, 
which now amounts to shuffling an $s$ into 
the indices of the function it acts on (and multiplying 
by a $\frac12$), for example
\beq
\Hmtd \El_{a,s,a,s}(z,\zb) = \frac12 \El_{s,a,s,a,s}(z,\zb) 
+ \El_{a,s,s,a,s}(z,\zb) + \El_{a,s,a,s,s}(z,\zb).
\eeq
The action of $\Hitd$ has a much 
richer and more complicated structure. However, we 
notice that at symbol level, i.e.\ keeping only the 
highest-weight \ac{svhpls}, it simply amounts to 
replacing $s \to ss - aa$ and multiplying by $-\frac14$, for
example,
\begin{multline} 
	\label{eq:hisymbollevel} \Hitd \El_{a,s,a,s}(z,\zb) 
	= -\frac14 \Big( \El_{a,s,s,a,s}(z,\zb) - \El_{a,a,a,a,s}(z,\zb) 
	+ \El_{a,s,a,s,s}(z,\zb) - \El_{a,s,a,a,a}(z,\zb) \Big) \\
	+ \Sigma_{\text{sub}}\,,
\end{multline}
where $\Sigma_{\text{sub}}$ contains products of 
subleading-weight \ac{svhpls} and zeta numbers, 
i.e.\ terms like $\El_\sigma \zeta_{n_1} \cdots \zeta_{n_m}$ 
with $|\sigma| + n_1 + \dots + n_m = 5$ and $|\sigma| < 5$ 
in the above example. This replacement rule can be derived 
from the differential equations \eqref{eq:diffeq0new} and 
\eqref{eq:diffeq1new}, as we now explain.

To this end, let us consider the two cases $\Hitd \El_{a,\sigma}$ 
and $\Hitd \El_{s,\sigma}$ in turn. Considering the former, due 
to the equivalence of the letters 0 and $a$, 
eq.~\eqref{eq:diffeq0new} immediately 
gives the action on $\El_{a,\sigma}$
\beq
\label{eq:diffeqaleading} \ddz \Hitd \El_{a,\sigma}(z,\zb) 
= \frac{\Hitd \El_\sigma(z,\zb)}{z}.
\eeq
The simple recursive nature of this equation implies 
that $\Hitd$ does not affect the $a$ indices of a 
\ac{svhpl} and can, at most, generate subleading 
terms $\Sigma_{\text{sub}}$ through integration 
constants, \emph{cf.}\ eqs.~\eqref{HiL0}--\eqref{HiL00000}.

The action on $\El_{s,\sigma}$ can be written as
\begin{align}
	\ddz \Hitd \El_{s,\sigma}(z,\zb) 
	&= \ddz \Hitd \left[ \El_{0,\sigma}(z,\zb) 
	+ 2\El_{1,\sigma}(z,\zb) \right] \nn\\
	&= \frac{1+z}{z(1-z)} \Hitd \El_\sigma(z,\zb) \nn \\
	&\hspace{10mm} - \frac12 \left( \frac{\El_{0,\sigma}(z,\zb) 
		+ 2\El_{1,\sigma}(z,\zb)}{1-z} 
	+ \frac{\El_{1,\sigma}(z,\zb)}{z} \right) 
	+ \Sigma_{\text{sub}} \nn \\
	&= \frac{1+z}{z(1-z)} \Hitd \El_\sigma(z,\zb) \nn \\
	&\hspace{10mm} - \frac14 \frac{1+z}{z(1-z)} 
	\left( \El_{0,\sigma}(z,\zb) + 2\El_{1,\sigma}(z,\zb) \right) 
	+ \frac14 \frac{\El_{0,\sigma}(z,\zb)}{z} + \Sigma_{\text{sub}} \nn \\
	&= \frac{1+z}{z(1-z)} \Hitd \El_\sigma(z,\zb) \nn \\
	&\hspace{10mm} - \frac14 \left( \frac{1+z}{z(1-z)} \El_{s,\sigma}(z,\zb) 
	- \frac{\El_{a,\sigma}(z,\zb)}{z} \right) + \Sigma_{\text{sub}}, \label{eq:diffeqsleading}
\end{align}
where at each step we have used $\Sigma_{\text{sub}}$ 
to collect subleading terms into.  The first term in the 
final expression is again an inert term, like the one 
encountered in eq.~\eqref{eq:diffeqaleading}. The 
following term however, creates two leading-weight 
terms which, upon integration, yield $-\frac14 (\El_{s,s,\sigma} 
- \El_{a,a,\sigma})$ and hence confirm the pattern 
described above eq.~\eqref{eq:hisymbollevel}. 
Note that by the recursive nature of the differential 
equation this applies (separately) to \emph{every} 
letter $s$ in the word $(s,\sigma)$, not just the first 
one (see e.g. eq.~(\ref{eq:hisymbollevel})).

In the following we show that it is possible to 
unravel the recursive definition of $\Hitd$ beyond 
symbol level. The $\Sigma_{\text{sub}}$ terms in the 
above equations are generated by two 
independent and additive sources: the 
$w,\wb \to \infty$ limit in eq.~\eqref{eq:diffeq1new} 
and the constants of integration as shown in 
eqs.~\eqref{HiL0}--\eqref{HiL00000}. Let us 
denote them $\Sigma_{\text{sub}(\infty)}$ and 
$\Sigma_{\text{sub}(0)}$, respectively, with 
their sum equalling $\Sigma_{\text{sub}}$. 
Empirically we observe that $\Sigma_{\text{sub}(0)}$ 
follows a simple pattern when using the $\{a,s\}$ 
alphabet:
\beq
\Hi \El_{w_1,\dots,w_{\ell-1}}(z,\zb) 
= \Sigma_{\text{lead}} + \Sigma_{\text{sub}(\infty)}
+ \sum_{j \geq 3 \text{, odd}}^\ell \zeta_j \El_{w_1,\dots,w_{\ell-j}}(z,\zb).  
\label{eq:subtermszero}
\eeq
with $\Sigma_{\text{lead}}$ now the leading-weight 
\ac{svhpls} governed by eq.~\eqref{eq:diffeqsleading}. 
$\Sigma_{\text{sub}(\infty)}$ in turn can be summarised by
\begin{multline}
	\Hi \El_{w_1,\dots,w_{\ell-1}}(z,\zb) 
	= \Sigma_{\text{lead}} + \Sigma_{\text{sub}(0)} 
	+ \frac18 \sum_{j=1}^{\ell-1}  \left( \El_{w_1,\dots,w_j}(z,\zb) 
	-\El_{w_1,\dots,w_{j-1},a}(z,\zb) \right) \\
	\times \left[ \El_{a,w_{j+1},\dots,w_{\ell-1}}(z,\zb) 
	+ \El_{s,w_{j+1},\dots,w_{\ell-1}}(z,\zb) \right]_{z,\zb \to \infty}. 
	\label{eq:subtermsinf}
\end{multline}
In both these equations the final term in the sum needs to 
be interpreted with care: in eq.~(\ref{eq:subtermszero}), 
for $j=\ell$ one obtains $\El_{w_1,\ldots,w_0}\equiv1$  
and in eq.~(\ref{eq:subtermsinf}) for $j=\ell-1$ one obtains 
in the second factor $\El_{a,w_{\ell},\dots,w_{\ell-1}}(z,\zb) 
+ \El_{s,w_{\ell},\dots,w_{\ell-1}}(z,\zb) \equiv
\El_{a}(z,\zb) 	+ \El_{s}(z,\zb)$.
Observe that \hbox{$w_j \overset{!}{=} s$} in 
eq.~(\ref{eq:subtermsinf}) is a necessary yet not 
sufficient requirement for a non-zero contribution. 
Being based on observations, the patterns described 
in eqs.~\eqref{eq:subtermszero} and \eqref{eq:subtermsinf} 
need to be verified against the wavefunctions computed in 
the previous section. We find perfect agreement with the 
wavefunction up to and including 13 loops, and are thus 
confident that the above description is correct.

By introducing the $\{a,s\}$ alphabet we have 
accounted for the symmetry of the wavefunction 
under inversion, $z \to 1/z$, at symbol level, i.e.\ 
as far as leading-weight terms are concerned. 
Our basis of \ac{svhpls} respects neither this 
nor the invariance under complex conjugation 
at function level: in general $\El_\sigma(z,\zb) 
\neq \El_\sigma(1/z,1/\zb)$ and $\El_\sigma(z,\zb) 
\neq \El_\sigma(\zb,z)$. Expecting further 
simplifications we will therefore construct 
a set of symmetrised functions in the 
remainder of this section. 

In the following we heavily use relations 
between \ac{svhpls} under a standard set 
of variable transformations. We summarise 
the most important aspects of these relations 
in appendix~\ref{app:svhplvariables}. Quintessentially, 
these relations determine the coefficients $c_w$ in 
$\El_\sigma(g(z),g(\zb)) = \sum_w c_w \El_w(z,\zb)$ 
where the sum runs over all words up to weight 
$|\sigma|$ and, in the present case, $g(x) = 1/x$ 
or $g(x) = \bar x$.

Let us define
\beq
\label{eq:Fdef} \Ef_\sigma(z,\zb) 
\equiv \frac14 \left( \El_\sigma(z,\zb) 
+ \El_\sigma(\zb,z) + \El_\sigma(1/z,1/\zb) 
+ \El_\sigma(1/\zb,1/z) \right)
\eeq
with $\sigma$ a word belonging to an alphabet 
of one's choosing. In the following we stick with 
the $\{a,s\}$ alphabet. We stress that the set of $\Ef$s does 
not span the space of \ac{svhpls} but 
it does cover the entire space of wavefunctions.

Due to the symmetries of the wavefunction
\beq
\label{eq:wfsyms1} \Wtd(z,\zb) = \Wtd(\zb,z) 
= \Wtd(1/z,1/\zb) = \Wtd(1/\zb,1/z)
\eeq
and thus
\beq
\label{eq:wfsyms2} \Wtd(z,\zb) 
= \frac14 \left( \Wtd(z,\zb) + \Wtd(\zb,z) 
+ \Wtd(1/z,1/\zb) + \Wtd(1/\zb,1/z) \right)
\eeq
one can simply replace $\El_\sigma(z,\zb) 
\to \Ef_\sigma(z,\zb)$ to go from the $\El$ 
to the $\Ef$ basis. It may therefore not be 
immediately obvious how eq.~\eqref{eq:Fdef} 
simplifies the results. Indeed, it requires a few 
more steps to showcase the advantages of a 
symmetrised basis.

Firstly, the wavefunction in the $\El$ basis 
contains functions whose indices feature an 
odd number of the letter $a$. Their leading-weight 
components are antisymmetric under $z \to 1/z$ 
because 
\beq
\dd \log z = - \dd \log 1/z
\eeq
Converted to $\Ef$ functions they are hence 
zero at symbol level or, in other words, equal 
to products of lower-weight \ac{svhpls} and 
zeta numbers. This can be turned into a 
recursive algorithm that successively 
removes all odd-$a$ functions. 
Schematically,
\begin{enumerate}
	\item Consider the wavefunction at a given 
	order and replace 
	$\El_\sigma(z,\zb) \to \Ef_\sigma(z,\zb)$
	\item Choose an $\Ef_\sigma(z,\zb)$ 
	where $\sigma$ contains an odd number 
	of $a$ letters. Plug in definition 
	\eqref{eq:Fdef} and rewrite \ac{svhpls} 
	as functions of $z,\zb$ using the rules 
	in appendix~\ref{app:svhplvariables}. 
	The resulting \ac{svhpls} will be of 
	lower weight than the original $\Ef_\sigma$, 
	multiplied by zeta numbers.
	\item Replace again $\El_\sigma(z,\zb) 
	\to \Ef_\sigma(z,\zb)$
	\item Repeat steps 2 \& 3 until a fixed point is 
	reached and only functions with an even 
	number of $a$ letters remain. 
\end{enumerate}
Note that step 3 is valid for the same reason 
it was legitimate to replace $\El_\sigma(z,\zb) 
\to \Ef_\sigma(z,\zb)$ in the wavefunction, 
\emph{cf.}\ eqs.~\eqref{eq:wfsyms1} and 
\eqref{eq:wfsyms2}. To give a few 
examples for odd-$a$ functions,
\begin{subequations}
	\begin{align}
		\Ef_{a}(z,\zb) &= 0 \\
		\Ef_{a,s}(z,\zb) = \Ef_{s,a}(z,\zb) &= 0 \\
		\Ef_{a,s,s}(z,\zb) = \Ef_{s,s,a}(z,\zb) &= 4 \zeta_3 \\
		\Ef_{s,a,s}(z,\zb) &= -8 \zeta_3 \\
		\Ef_{s,s,s,a}(z,\zb) = \Ef_{s,a,s,s}(z,\zb) &= 4 \zeta_3 \Ef_s(z,\zb).
	\end{align}
\end{subequations}

Secondly, we may combine $\Ef_\sigma(z,\zb)$ 
and $\Ef_{\tilde \sigma}(z,\zb)$ with $\tilde \sigma$ 
the word $\sigma$ reversed, at the cost of 
generating subleading terms. This is due to 
the following identity of \ac{svhpls}:
\beq
\El_\sigma(z,\zb) = \El_{\tilde \sigma}(\zb,z) 
+ \Sigma_{\text{sub}}
\eeq
For a function $\Ef_\sigma$ this entails
\beq
\Ef_\sigma(z,\zb) = \Ef_{\tilde \sigma}(z,\zb) 
+ \Sigma_{\text{sub}} \label{eq:Fwordrev}
\eeq
due to the invariance under complex conjugation. 
Besides removing nearly half of the $\Ef$ functions 
we find the generated subleading terms to sometimes 
reduce but never increase the complexity of a given 
expression. For the procedure to be algorithmic 
one chooses which letter to cumulate in the left 
(or right) half of a word.

For the wavefunction up to four loops and with 
the same abbreviations as in 
eqs.~\eqref{eq:wtd1}--\eqref{eq:wtd4} we find
\begin{subequations}
\begin{align}
	\Wtdl{1} &= \frac12 C_2 \Ef_s \label{eq:wtd1f} \\
	\Wtdl{2} &= \frac{1}{8} C_1 C_2 \left(\Ef_{a,a}
	-\Ef_{s,s}\right)+\frac{1}{2} C_2^2 \Ef_{s,s} \\
	\Wtdl{3} &= \frac{1}{16} C_1 C_2^2 \left(\Ef_{a,s,a}
	+6 \Ef_{s,a,a}-7 \Ef_{s,s,s}+8 \zeta_3\right) 
	+\frac{1}{16} C_1^2 C_2 
	\left(\Ef_{s,s,s}-\Ef_{s,a,a}\right)
	+\frac{3}{4} C_2^3 \Ef_{s,s,s} \\
	\begin{split}
	\Wtdl{4} &= \frac{1}{16} C_1 C_2^3 \left(\Ef_{a,s,s,a}
	+6 \Ef_{s,a,a,s}+4 \Ef_{s,a,s,a}+12 \Ef_{s,s,a,a}
	-23 \Ef_{s,s,s,s}\right.  \\
	&\hspace{4mm} \left.+20 \zeta_3 \Ef_s\right)
	+\frac{1}{64} C_1^2 C_2^2 \left(-\Ef_{a,s,s,a}
	-9 \Ef_{s,a,a,s}-2 \Ef_{s,a,s,a}-24 \Ef_{s,s,a,a}\right.  \\
	&\hspace{4mm} \left.+7 \Ef_{a,a,a,a}+29 \Ef_{s,s,s,s}
	-4 \zeta_3 \Ef_s\right)+\frac{1}{64} C_1^3 C_2 \left(\Ef_{s,a,a,s}
	+3 \Ef_{s,s,a,a}\right.  \\
	&\hspace{4mm} \left.-\Ef_{a,a,a,a}-3 \Ef_{s,s,s,s}\right)
	+\frac{3}{2} C_2^4 \Ef_{s,s,s,s} \label{eq:wtd4f}
\end{split}
\end{align}
\end{subequations}
where we used eq.~\eqref{eq:Fwordrev} in favour of 
words that start rather then end with the letter $s$.
Further results up to weight 13 can be found in the 
ancillary file \texttt{2Reggeon-wavefunction-\\Fsa-Basis.txt}.

Indeed, comparing the results in eqs.~\eqref{eq:wtd1f}--\eqref{eq:wtd4f} 
to the wavefunction in terms of standard \ac{svhpls} (and the standard 
$\{0,1\}$ alphabet) in eqs.~\eqref{eq:wtd1}--\eqref{eq:wtd4} shows the 
benefits of the new basis. In terms of $\Ef$ functions the wavefunction 
takes not only a very compact form and is expressed in terms of fewer 
functions, it also removes subleading terms in some cases, like the 
$-\frac{3}{16} \El_1 \zeta_3$ in the coefficient of $C_1^3 C_2$ 
at four loops \eqref{eq:wtd4}.


\section{Finite corrections to the amplitude from two-dimensional evolution}
\label{amplitude}

We now have  an algorithm for the calculation of the 
wavefunction $\Omega_{\rm 2d}$ to any loop order, and 
we shall use it for the computation of the amplitude. 
Let us recall from section \ref{soft} that the soft 
part has been fully determined, and our goal here
is the calculation of the hard part of the amplitude,
as defined in \eqn{Mhard}. This, in turn, requires
the hard part of the two-dimensional wavefunction, 
which according to \eqn{WhardTwod} is obtained by 
subtracting the 
$d=2$ limit of the soft wavefunction from the full 
(two-dimensional) wavefunction $\Omega_{\rm 2d}$ of 
the previous section. To this end we first define 
\begin{equation}
	\label{Wsoft2d}
	\Ws^{({\rm 2d})}(z,\zb)\,\equiv \, \left.\lim_{\eps \to 0} 
	\Ws(p,k)\right|_{\log \pfrac{k^2(p-k)^2}{(p^2)^2}\,\to\, \El_s(z,\zb)}\,,
\end{equation}
where taking the limit simply corresponds to selecting the 
leading $\ord(\eps^0)$ terms in $\Ws(p,k)$. Notice that within 
the $d=2$ limit we switch to the two-dimensional variables $z$ 
and $\zb$ of eq.~\eqref{eq:zwdef}, and the single-valued 
logarithm $\El_s(z,\zb)=\log\frac{z\zb}{(1-z)^2(1-\zb)^2}$ 
defined in eq.~(\ref{ElDef}). 
Having used the symmetrised soft wavefunction we land 
directly in the class of SVHPLs used in the two-dimensional 
computation of section~\ref{sec:diffeq}, and in this way the 
computations of $\Omega_{\rm 2d}$ and $\Ws^{({\rm 2d})}$ are 
entirely compatible. We note that with the replacement 
in (\ref{Wsoft2d}) the two-dimensional soft wavefunction 
in~\eqref{Wsoft2dim} becomes a polynomial in $\El_s(z,\zb)$ 
at any given order. According to eq.~(\ref{eq:wffullsoftresummed}), 
these terms exponentiate and can be resummed to 
all-orders. Upon applying the changed of variable of (\ref{Wsoft2d}) 
this resummed expression is
\beq
\label{eq:wffullsoftresummed_SVHPL} 
\Ws^{({\rm 2d})}(z,\zb)= \frac{\as}{\pi} \left[ e^{-x \gE \Ctwo} 
\frac{\Gamma \left(1 - \frac{x}{2} \Ctwo \right)}{\Gamma \left(1 + \frac{x}{2} 
\Ctwo \right)} \right]^\frac{2\Ca - \Tt}{\Ca - \Tt} e^{\frac{x}{2} \Ctwo \El_s(z,\zb)}\, ,
\eeq
with $x = L \, \as/\pi$.

Using the results in section \ref{2d-bfkl} for $\Omega_{\rm 2d}$
and the expansion of \eqn{eq:wffullsoftresummed_SVHPL} for $\Ws^{({\rm 2d})}$
we determine $\Wh^{({\rm 2d})} = \Omega_{\rm 2d} 
-\Ws^{({\rm 2d})}$, and we can proceed to determine the 
hard part of the amplitude order by order, according to \eqn{Mhard}.
To this end, recall that the hard wavefunction $\Wh^{({\rm 2d})}$ 
is guaranteed to integrate to finite terms only, hence 
it can be integrated in strictly two dimensions. Applying the  
limit $\eps \to 0$ in \eqn{Mhard} to the integrand and the 
integration measure using the variables $z$ and $\bar{z}$ 
(\emph{cf.} eqs. (\ref{k2zzb}) and (\ref{eq:dw2d})) we obtain:
\beq
\hat {\cal M}^{(+)}_{\rm NLL,h}\left(\frac{s}{-t}\right) = -i\pi 
\left[ \frac{1}{4\pi} \int \frac{\dd^2 z}{z\zb} \Wh^{({\rm 2d})}(z,\zb) \right] 
\Tsu \Mtree_{ij\to ij}, \label{eq:hardAmp2d}
\eeq
where, in practice, we loop-expand both the wavefunction 
and amplitude, as was done in eq.~\eqref{ReducedAmpNLL2}. 
The next two subsections are dedicated to the computation 
of the integral in eq.~(\ref{eq:hardAmp2d}), thus 
determining the hard component of the reduced amplitude order 
by order. In section~\ref{sec:finiteamp} we combine the soft and 
hard components of the reduced amplitude according to 
eq.~(\ref{eq:redampSplit}), and finally in section~\ref{sec:hardAmpl} 
we similarly combine the soft and hard components of the 
infrared-renormalized amplitude using eqs.~(\ref{Hsplit}) 
and (\ref{getH2both}). 

To set up the computation of eq.~(\ref{eq:hardAmp2d}) let us define
\beq
I \equiv \frac{1}{4\pi} \int \frac{\dd^2 z}{z\zb} \Wh^{({\rm 2d})}(z,\zb) 
\label{eq:Idef}
\eeq
and introduce in turn two independent methods for computing 
these integrals. For the sake of simplicity of notation, given that 
the entire computation is done in two dimensions, we shall now 
drop the $({\rm 2d})$ superscript, and refer to the integrand in 
(\ref{eq:Idef}) as $\Wh(z,\zb)$. Similarly, while (\ref{eq:Idef}) is 
applied order-by-order, in describing the methods we refrain 
from using an index for the loop order on either side of 
(\ref{eq:Idef}).

The first method, described in section~\ref{sec:method1} below, 
is based on using the known analytic structure of the wavefunction, 
in order to convert the two-dimensional integral into an integral over 
the discontinuity of the wavefunction. It was inspired by the 
calculations described in section 7.1 of ref.~\cite{Schwartz:2014wha}. 
The second method, presented in section~\ref{sec:method2} below, 
relies on the symmetry of the wavefunction under inversion, 
$z \rightarrow 1/z$, $\zb \rightarrow 1/\zb$, and the action 
of $\Hitd$ at fixed external points.

\subsection{Method I: final integration using 
	the discontinuity of the wavefunction} \label{sec:method1}
Let us define a regularised version of the 
integral $I$ in eq.~\eqref{eq:Idef}:
\begin{equation}
	I_\reg = \frac{1}{4\pi} \int_{\delta^2 < z \zb < 1/\delta^2} 
	\frac{\dd^2 z}{z \zb} \Wh(z,\zb) \,,
	\label{eq:Iregdef}
\end{equation}
where the cutoff $\delta$ is assumed to be small, 
$\delta \ll 1$. The introduction of $\delta$ may seem 
superfluous at this point as $\lim_{z,\zb \to 0} \Wtd = 
\lim_{z,\zb \to \infty} \Wtd = \lim_{\eps \to 0} \Ws$ and 
thus, using eq.~(\ref{eq:wffullsoftresummed_SVHPL}), 
$\lim_{z,\zb \to 0} \Wh = \lim_{z,\zb \to \infty} \Wh= 0$; 
more precisely, $\Wh$ vanishes linearly in $z\bar{z}$
in the soft limit, up to logarithms, rendering the integral 
in (\ref{eq:Idef}) convergent, and the difference 
$I - I_\reg = \ord(\delta^2)$ (up to logarithms). The 
necessity of this cutoff despite this good convergence 
will become clear shortly.  

The exclusion of the points $\{0,\infty\}$ in (\ref{eq:Iregdef}) 
enables us to introduce polar coordinates such that 
$z\zb=r^2$ and $\frac{z}{\zb}=e^{2i\theta}$,
as now all points in the integration region have 
a non-vanishing Jacobian:
\begin{equation} 
	\label{eq:lastintpol0} I_\reg = \frac{1}{4\pi} 
	\int_\delta^{1/\delta} \frac{\dd r}{r} \int_0^{2\pi} d\theta\,  
	\Wh \left( re^{i\theta},re^{-i\theta} \right)\,. 
\end{equation}
To proceed we express the angular integral in the latter 
as an integration in the complex~$y$ plane where 
$y\equiv e^{i\theta}$, getting
\begin{equation} 
	\label{eq:lastintpol} I_\reg 
	= \frac{1}{4\pi i} \int_\delta^{1/\delta} \frac{\dd r}{r} 
	\oint_{|y|=1} \frac{\dd y}{y} \, \Wh(ry,r/y)\,,
\end{equation}
where the contour runs along the unit circle. The method 
outlined in the following is based on deforming the contour 
in the complex $y$ plane. Essential to this is the fact that 
the integrand, at any order, is expressed in terms of 
\ac{svhpls}, whose analytic structure is well understood. 
These functions are single-valued as long as their 
arguments are complex conjugates of one another, 
namely as long as the contour in eq.~\eqref{eq:lastintpol} 
runs along the unit circle. Outside of this region, i.e.\ upon 
deformation the contour, the \ac{hpls} in $\Wh(z,\zb)$ 
exhibit branch cuts
where $z \in [1,\infty]$ and $\zb \in [1,\infty]$. In the $r,y$ 
coordinates of eq.~\eqref{eq:lastintpol} they correspond 
to cuts along the real axis in the complex $y$ plane where 
$y \in [1/r,\infty]$ and $y \in [0,r]$, respectively. 
\begin{figure}[ht]
	\centering
	\includegraphics{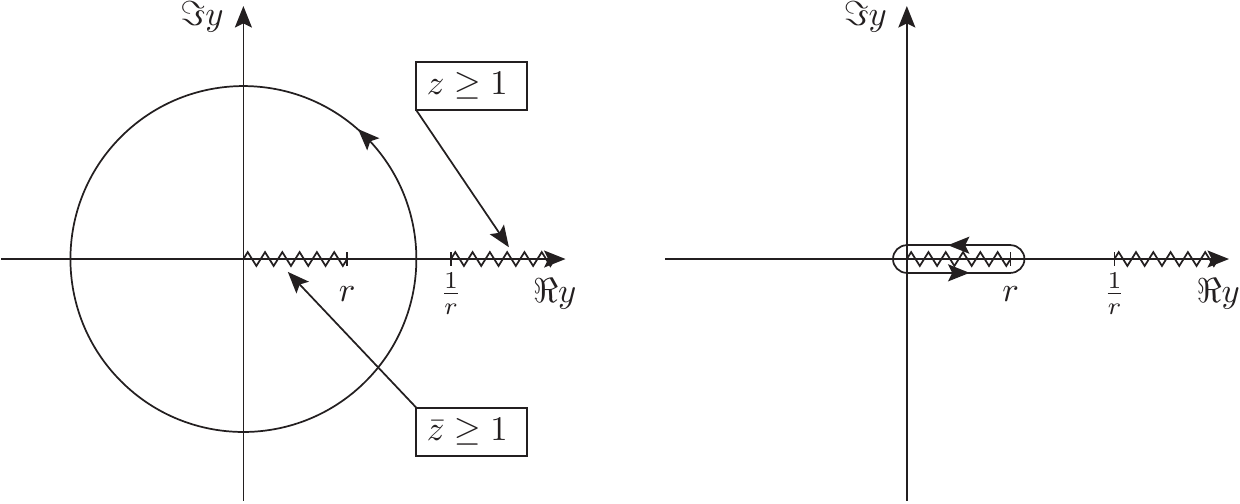}
	\caption{Position of the branch cuts in $z$ and $\zb$ in 
		the complex $y$-plane for $r<1$ (left). The contour 
		along the unit circle in eq.~\eqref{eq:lastintpol} can be 
		deformed and, consequently, identified with the integral 
		of the $\zb$-discontinuity (right), as written in eq.~\eqref{eq:intdisc}.}
	\label{fig:ycontour}
\end{figure}

For $r<1$ there is a branch cut-free interval $(r,1/r)$ 
through which the contour along the unit circle passes, 
\emph{cf.}\ the l.h.s.\ of figure~\ref{fig:ycontour}. The contour can 
consequently be shrunk until it corresponds to integrating 
the $\zb$-discontinuity of the wavefunction over $y$ from 
$0$ to $r$, \emph{cf.}\ the r.h.s.\ of figure~\ref{fig:ycontour}. We 
can now understand why it is necessary to work with the 
regularised integral $I_\reg$ of eq.~\eqref{eq:Iregdef} instead 
of the original $I$ of eq.~\eqref{eq:Idef}: while the hard 
wavefunction $\Wh(z,\zb)$ vanishes at 0 and $\infty$, its 
discontinuity, in general, does not. In other words, the 
contour deformation introduces spurious divergent terms 
and the cutoff introduced in eq.~\eqref{eq:Iregdef} 
regularises them. .

For $r>1$ the branch cuts of $z$ and $\zb$ overlap. 
However, the discontinuity 
cancels identically in the interval 
$(1/r,r)$. Repeating the procedure, we again identify the 
contour integration with integrating the $\zb$-discontinuity 
of $\Wh(z,\zb)$ over $y$, this time, from $0$ to $1/r$. 

In total, having modified the contour in (\ref{eq:lastintpol}) we find
\begin{align}
	I_\reg &= \frac{1}{4\pi i} \left( \int_\delta^1 \frac{\dd r}{r} \int_0^r \frac{\dd y}{y} \disc{\zb} [\Wh(ry,r/y)] 
	+ \int_1^{1/\delta} \frac{\dd r}{r} \int_0^{1/r} \frac{\dd y}{y} \disc{\zb} [\Wh(ry,r/y)] \right) \nonumber \\
	&= \frac{1}{4\pi i} \left( \int_\delta^1 \frac{\dd r}{r} \int_1^\infty \frac{\dd \zb}{\zb} \disc{\zb} [\Wh(r^2/\zb,\zb)] 
	+ \int_1^{1/\delta} \frac{\dd r}{r} \int_0^1 \frac{\dd z}{z} \disc{\zb} [\Wh(z,r^2/z)] \right) \nonumber \\
	&= \frac{1}{8\pi i} \left( \int_1^\infty \frac{\dd \zb}{\zb} \int_{\delta^2/\zb}^{1/\zb} \frac{\dd z}{z} \disc{\zb} [\Wh(z,\zb)] 
	+ \int_0^1 \frac{\dd z}{z} \int_{1/z}^{1/(\delta^2 z)} \frac{\dd \zb}{\zb} \disc{\zb} [\Wh(z,\zb)] \right) \nonumber \\
	&= \frac{1}{8\pi i} \left( \int_0^1 \frac{\dd x}{x} \int_{\delta^2 x}^x \frac{\dd z}{z} \disc{\zb} [\Wh(z,1/x)] 
	+ \int_0^1 \frac{\dd z}{z} \int_{\delta^2 z}^z \frac{\dd x}{x} \disc{\zb} [\Wh(z,1/x)] \right) \label{eq:intdisc}
\end{align}
where the two terms correspond respectively to $r<1$ and $r>1$. 
It is clear from the outset that they are equal: this corresponds to 
splitting (\ref{eq:Iregdef}) at $z\bar{z}=1$, which admits 
$(z,\zb)\leftrightarrow (1/z,1/\zb)$ symmetry.
In the second line of (\ref{eq:intdisc}) we reverted to the variable 
$\zb=r/y$ in the first integral and $z=ry$ in the second; in the third 
we changed the order of integration before reverting to $z=r^2/\zb$ 
in the first integral and $\zb=r^2/z$ in the second; finally in the last 
line we defined $x=1/\zb$ in the both integrals. 
\begin{figure}[ht]
	\centering
	\includegraphics{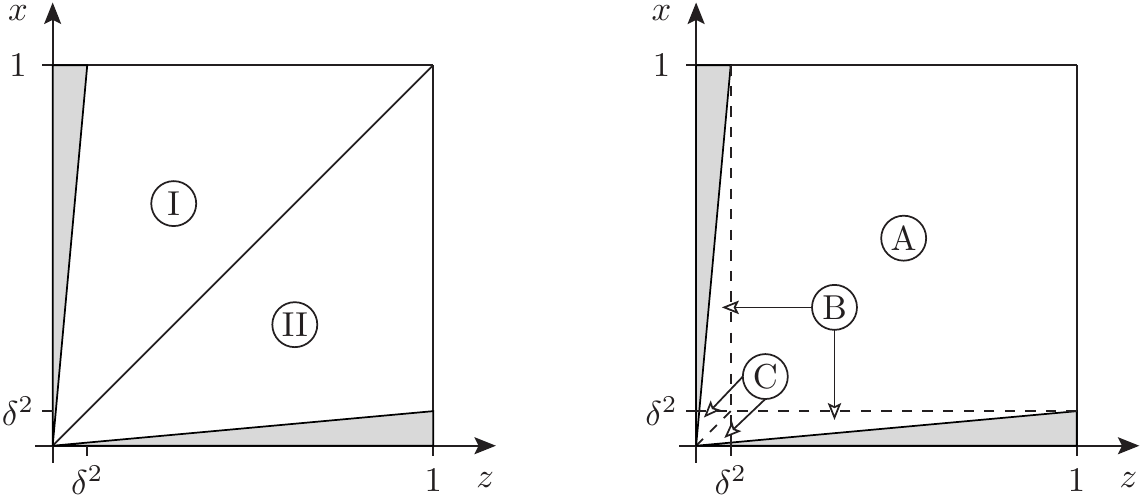}
	\caption{Illustration of the integrations in the $r<1$ (I) 
		and $r>1$ (II) contribution to $I_\reg$ of eq.~\eqref{eq:lastintpol} 
		(white triangles, l.h.s.). They can be viewed as the integral 
		over a square (A) plus two wedges (B) minus two small 
		triangles (C) (delimited by dashed lines, r.h.s.).}
	\label{fig:method1}
\end{figure}

Let us now discuss the evaluation of the final expression 
in eq.~(\ref{eq:intdisc}), where the integration region of 
two terms is depicted as the white area in figure~\ref{fig:method1}. 
In order to perform the integration it is useful to view the 
integrals (\emph{cf.}\ the r.h.s.\ of figure~\ref{fig:method1}) 
as the integral over a square
\begin{equation}
	\label{eq:square} I_A(\delta) = \frac{1}{8\pi i} \int_{\delta^2}^1 
	\frac{\dd z}{z} \int_{\delta^2}^1 \frac{\dd x}{x} \disc{\zb} [\Wh(z,1/x)]\,,
\end{equation}
plus (the integrals over) two wedges
\beq
\label{eq:wedges} I_B(\delta)  = 
\frac{1}{8\pi i} \left( \int_0^1 \frac{\dd x}{x} \int_{\delta^2 x}^{\delta^2} \frac{\dd z}{z} \disc{\zb} [\Wh(z,1/x)] 
+ \int_0^1 \frac{\dd z}{z} \int_{\delta^2 z}^{\delta^2} \frac{\dd x}{x} \disc{\zb} [\Wh(z,1/x)] \right)\,,
\eeq
minus two small triangles
\beq
\label{eq:kite} I_C(\delta)  = 
\frac{1}{8\pi i} \left( \int_0^{\delta^2} \frac{\dd x}{x} \int_{\delta^2 x}^x \frac{\dd z}{z} \disc{\zb} [\Wh(z,1/x)] 
+ \int_0^{\delta^2} \frac{\dd z}{z} \int_{\delta^2 z}^z \frac{\dd x}{x} \disc{\zb} [\Wh(z,1/x)] \right)\,,
\eeq
where both $z$ and $x$ are small. 
Next we would like to evaluate each of these contributions, distinguishing 
between finite, $\delta$-independent terms, and logarithmically divergent 
cut-off dependent terms.

The discontinuity w.r.t.\ $\zb$ of $\Wh(z,1/x)$ evaluates to 
\ac{hpls} of $z$ and $x$. $I_A(\delta)$ of eq.~\eqref{eq:square} 
thus immediately evaluates to \ac{hpls} at 1, giving rise to \ac{mzvs}, 
and at $\delta^2$; the latter contain logarithmically divergent terms 
in $\delta$. The first (second) integral in the expression of $I_B(\delta)$ 
in~\eqref{eq:wedges} is calculated close to $z=0$ ($x=0$), \emph{cf.}\ 
figure~\ref{fig:method1}. One can therefore expand the discontinuity 
function in the integrand and discard terms suppressed by powers of 
$z$ ($x$) keeping only powers of $\log z$ ($\log x$). The inner 
integrals then yield powers of $\log \delta^2$, $\log \delta^2 x = 
\log x + \log \delta^2$ and $\log \delta^2 z = \log z + \log \delta^2$, 
respectively. The outer integrals thereupon generate \ac{mzvs} from 
their upper limits; in addition it contains logarithmically divergent 
terms in $\delta$. Contributions from the lower integration limits 
are dropped according to the (standard) regularisation of \ac{hpls}:
\beq
\lim_{z \to 0} \log z = 0. \label{eq:hplreg}
\eeq
A similar analysis of $I_C(\delta)$ in eq.~\eqref{eq:kite} 
reveals that only powers of $\log \delta^2$ are generated 
by the integrations over the two small triangles in 
figure~\ref{fig:method1}.

Since the original integral $I$~of eq.~(\ref{eq:Idef}) is 
finite and $I_\reg \to I$ for $\delta \to 0$ all terms 
proportional to $\log \delta^2$ have to cancel between 
the three contributions $I_A(\delta)$, $I_B(\delta)$ and 
$I_C(\delta)$. This enables us to derive a simplified 
integral in which the logarithmically divergent terms 
are absent altogether whilst giving the same finite 
terms:
\begin{multline}
	I_\reg = \frac{1}{8\pi i} \int_0^1 \frac{\dd z}{z} 
	\int_0^1 \frac{\dd x}{x} \disc{\zb} [\Wh(z,1/x)] \\
	+ \frac{1}{8\pi i} \left( - \int_0^1 \frac{\dd x}{x} 
	\int_0^x \frac{\dd z}{z} \disc{\zb=1} [\Wh(z,1/x)] \big|_{z \ll 1} \right. \\
	\left. - \int_0^1 \frac{\dd z}{z} \int_0^z \frac{\dd x}{x} 
	\disc{\zb=1} [\Wh(z,1/x)] \big|_{x \ll 1} \right)\,, \label{eq:Iregsimplified}
\end{multline}
where all integrals are regulated according to 
eq.~\eqref{eq:hplreg} and $\disc{\zb=1} [\Wh(z,1/x)] 
\big|_{z \ll 1}$ and $\disc{\zb=1} [\Wh(z,1/x)] \big|_{x \ll 1}$ 
refer to the aforementioned expansion of the integrand 
around small $z$ and $x$, respectively. The first integral 
in eq.~(\ref{eq:Iregsimplified}) reproduces all finite, cut-off 
independent terms in $I_A(\delta)$ of (\ref{eq:square}), 
while the second and third ones reproduce, respectively, 
the finite terms in the two integral in $I_B(\delta)$ in 
(\ref{eq:wedges}); finally, given that no cut-off independent 
terms are produce by $I_C(\delta)$, it has no trace in 
(\ref{eq:Iregsimplified}). 

The above calculation is biased towards the discontinuity 
with respect to~$\zb$ which is purely a matter of choice. 
A similar calculation can be performed to get an answer in terms 
of the discontinuity with respect to~$z$ or a mixed expression that 
features both discontinuities. 

This integration method was further checked as follows. 
Given a wavefunction (or \ac{svhpl}) we expand around 
$z = \zb = 0$ and change variables to the polar coordinates 
introduced above in eq.~\eqref{eq:lastintpol}. The result is 
a sum of terms of the form $r^a y^b \log^c(r^2)$ with  
rational constant coefficients and $a,c \geq 0$ and $b$ 
are integer powers. Integrating the azimuth over $[0,\, 2\pi]$ 
then removes all terms that explicitely depend on $y$, i.e.\ 
that have $b \neq 0$. Next, we determine the rational 
coefficients in terms of harmonic numbers\footnote{This step 
	requires some amount of creativity but is greatly helped by 
	The On-Line Encyclopedia of Integer Sequences (\ac{oeis}), 
	\texttt{https://oeis.org}.}. This enables us to perform the sum 
ad infinitum after we integrate term-by-term with respect to~$r$.

\subsection{Method II: final integration as an action of the Hamiltonian}
\label{sec:method2} The previous method, albeit 
straightforward on paper, is computationally demanding 
at high loop orders as it requires extensive use of analytic 
continuations of \ac{hpls} to calculate discontinuities. It turns 
out there is an easier way to perform the final integration, 
which lets us make use of our knowledge about the action 
of the Hamiltonian,  established upon computing the 
wavefunction in section~\ref{2d-bfkl}. 

Consider the action of $\Hitd$ \eqref{eq:hi2d} on the 
wavefunction $\Wh(1-z,1-\zb)$
\beq
\label{eq:method2int1} \Hitd \Wh(1-z,1-\zb) = 
\frac{1}{4\pi} \int \dd^2 w K(w,\wb,z,\zb) 
\Big[\Wh(1-w,1-\wb) - \Wh(1-z,1-\zb)\Big]
\eeq
and set $z = \zb = 1$ under the integral. 
Using $\Wh(0,0) = 0$ one gets on the right-hand side:
\beq
\lim_{z,\zb \to 1} K(w,\wb,z,\zb) \Big[\Wh(1-w,1-\wb) - \Wh(1-z,1-\zb)\Big] 
= K(w,\wb,1,1) \Wh(1-w,1-\wb) \label{eq:integrandat1}
\eeq
with the kernel
\beq
K(w,\wb,1,1) = \frac{1}{w\wb (1-w)(1-\wb)} + \frac{1}{(1-w)(1-\wb)} - \frac{1}{w\wb},
\eeq
\emph{cf.}\ eq.~\eqref{eq:K2d}. 
It thus follows that (\ref{eq:method2int1}), 
taken in the limit $z,\zb\to 1$, yields:
\begin{eqnarray}
	\label{eq:method2int1.5} 
	\begin{split}
		\hspace*{-10pt}&\Hitd \Wh(1-z,1-\zb) \big|_{z,\zb \to 1} =\\ 
		&\hspace*{20pt} =\, \frac{1}{4\pi} \int \dd^2 w \left[\frac{ \Wh(1-w,1-\wb) }{w\wb (1-w)(1-\wb)} 
		+ \frac{ \Wh(1-w,1-\wb) }{(1-w)(1-\wb)} - \frac{ \Wh(1-w,1-\wb)  }{w\wb} \right] 
		\\ &\hspace*{20pt} =\, \frac{1}{4\pi} \int \frac{\dd^2 w}{w \wb} 
		\left[ \Wh \left( \frac{1}{1-w},\frac{1}{1-w} \right) + \Wh(w,\wb) 
		- \Wh(1-w,1-\wb) \right]\,,
	\end{split}
\end{eqnarray}
where in the second line we changed the integration 
variables in the first two terms -- in the first using 
$w \rightarrow w/(w-1)$, and in the second using 
$w \rightarrow 1-w$, and then factored out a common 
denominator. Given that the wavefunction is symmetric 
under inversion, $\Wh(1/w,1/\wb) = \Wh(w,\wb)$, the 
first and third terms in the last equation cancel and 
we find 
\beq
\label{eq:method2int3} \Hitd \Wh(1-z,1-\zb) \big|_{z,\zb \to 1} 
= \frac{1}{4\pi} \int \frac{\dd^2 w}{w \wb} \Wh(w,\wb) = I\,, 
\eeq   
which can be readily identified with the integral in 
eq.~\eqref{eq:Idef} which we are interested to compute.

We thus conclude that the integral in 
eq.~\eqref{eq:hardAmp2d}, representing the hard 
wavefunction contribution to the reduced amplitude, integrated 
in exactly two dimensions, may be calculated with the 
methods we developed for the computation of the 
two-dimensional wavefunction itself, described in 
section~\ref{sec:wf2d}. In practice one rewrites the 
wavefunction $\Wh(1-z,1-\zb)$ in terms of \ac{svhpls} 
of $z$ and $\zb$, then applies the Hamiltonian by 
solving the corresponding differential equations, 
and finally evaluates the resulting expression at 
$z,\zb = 1$. The last step produces the anticipated 
\ac{mzvs}.

Method I, described in section~\ref{sec:method1}, 
and method II outlined here show perfect agreement when applied to 
the wavefunction. However, we emphasise that while former 
may be applied on individual \ac{svhpls}, the latter can only 
be applied to expressions which are symmetric under 
inversion of their arguments, \emph{cf.}\ eqs.~(\ref{eq:method2int1.5}) 
and (\ref{eq:method2int3}).

\subsection{Results for the reduced amplitude}
\label{sec:finiteamp}
With the methods described 
in the previous sections it is straightforward to integrate 
the two-dimensional wavefunction and thereby compute 
the hard contribution to the amplitude, namely the finite 
terms not captured by the soft limit. 

Before presenting our results let us recall the 
number-theoretic observations we made about 
the amplitude at the end of section~\ref{chap:bfkl}. 
There, we claimed that the $\ell$-loop amplitude (divided by 
$B_0^\ell$, \eqref{B0}) has two important number-theoretic 
properties: all of its terms have weight $\ell$ and there are 
no terms proportional to $\zeta_2$. We proved this statement 
for contributions from the soft limit in section~\ref{soft}, see 
below eq.~(\ref{eq:MsoftExpanded_8}). We now show that 
it holds also for the hard contributions.

We begin by noting that the integrand in (\ref{eq:Idef}) 
is expressed as a pure function of uniform weight, 
written as sums of products of HPLs. We note that 
both methods for the last integral, in sections~\ref{sec:method1} 
and \ref{sec:method2}, increase the weight of the functions 
they act on by one, before evaluating the result at 
$z = \zb = 1$. In method I the action of the discontinuity 
first lowers the weight of its argument by one; this is then 
compensated by two consecutive integrations of a $d\log$ 
form, each raising the weight by one. Method II, in turn, 
applies the Hamiltonian $\Hitd$ on the wavefunction after 
a variable transformation $z \to 1-z$. Changing the 
variables of an \ac{svhpl} does obviously not change 
its weight and the action of the Hamiltonian corresponds 
to integrating a first-order differential equation, which 
raises the weight of the operand by one. 

\ac{svhpls} at $z = \zb = 1$ evaluate to multiple zeta 
values (\ac{mzvs}) of the same weight, as discussed 
following eqs.~\eqref{eq:wtd1}--\eqref{eq:wtd4}. We remind 
the reader that the $(\ell-1)$-loop wavefunction consists 
of weight-$(\ell-1)$ \ac{svhpls} and weight-$(\ell-1)$ 
products of \ac{svhpls} and zeta numbers and conclude 
that the hard contributions to the $\ell$-loop amplitude 
therefore have uniform weight $\ell$. The absence of 
$\zeta_2$ in the hard component $\Mreduced_{{\rm NLL,h}}^{(+,\ell)}$  
is readily explained by the fact that \ac{svhpls} can, 
by construction, only ever evaluate to odd zeta 
numbers, for any argument.

We start the discussion of the results by presenting the 
contributions that originate in the hard region. They are 
the immediate result of the previous sections and, 
through eight loops, read 
\renewcommand{\Cone}{C_1}
\renewcommand{\Ctwo}{C_2}\,
\begin{subequations}
\label{reduced_hard_results}
\begin{align} 
	\Mreduced_{{\rm NLL,h}}^{(+,1)} &= 0,  \label{eq:m1hardfinite} \\
	\Mreduced_{{\rm NLL,h}}^{(+,2)} &= 0, \\
	\Mreduced_{{\rm NLL,h}}^{(+,3)}  &= \frac{i\pi}{3!} 
	\bigg\{ \frac{3\zeta_3}{4} \Cone \Ctwo \bigg\} \Tsu \Mtree,  \\
	\Mreduced_{{\rm NLL,h}}^{(+,4)} &= 0, \\
	\Mreduced_{{\rm NLL,h}}^{(+,5)} &= \frac{i \pi}{5!} 
	\bigg\{-\frac{5\zeta_5}{2} \Cone^2 \Ctwo^2   
	+\, \frac{45\zeta_5}{2} \Cone \Ctwo^3  \bigg\} \Tsu \Mtree,  \\
	\Mreduced_{{\rm NLL,h}}^{(+,6)} &= \frac{i\pi}{6!} 
	\bigg\{\frac{39\zeta_3^2}{16} \Cone^3 \Ctwo^2 
	- \frac{45\zeta_3^2}{2} \Cone^2 \Ctwo^3 
	+\,  \frac{225\zeta_3^2}{2} \Cone \Ctwo^4 \bigg\} \Tsu \Mtree,  \\
	\Mreduced_{{\rm NLL,h}}^{(+,7)} &= \frac{i\pi}{7!} 
	\bigg\{ -\frac{2135\zeta_7}{256} \Cone^4 \Ctwo^2 
	+ \frac{30135\zeta_7}{256} \Cone^3 \Ctwo^3  
	-\,  \frac{20111\zeta_7}{32} \Cone^2 \Ctwo^4 \nn  \\
	&\hspace{4.0cm}  
	+ \frac{6111\zeta_7}{4} \Cone \Ctwo^5\bigg\} \Tsu \Mtree, \\
	\Mreduced_{{\rm NLL,h}}^{(+,8)} &= \frac{i\pi}{8!} 
	\bigg\{ \frac{611\zeta_3 \zeta_5}{32} \Cone^5 \Ctwo^2 
	- \frac{643\zeta_3 \zeta_5}{2} \Cone^4 \Ctwo^3 
	+ \frac{8597\zeta_3 \zeta_5}{4} \Cone^3 \Ctwo^4 \nn \\ 
	&\hspace{2.0cm}  
	- 7086\zeta_3 \zeta_5 \, \Cone^2 \Ctwo^5    
	+\, 13230 \zeta_3 \zeta_5 \, \Cone \Ctwo^6\bigg\} \Tsu \Mtree,
	\label{eq:m8hardfinite} 
	\end{align}
\end{subequations}
where we again used the shorthand notation for the 
colour factors, $C_1=2\Ca-\Tt$ and $C_2=\Ca-\Tt$.
One may observe the aforementioned homogeneous 
weight property and absence of even zeta numbers. 
In fact, considering the first eight loop orders, 
one may get the false impression that each order 
contains just a single (product) of zeta numbers 
and that they are all single (ordinary) zeta numbers. 
Both these features are artefacts of looking at low 
weights, and a much richer structure will be revealed 
at higher loop orders, as we discuss shortly. 
\renewcommand{\Cone}{(2\Ca-\Tt)}
\renewcommand{\Ctwo}{(\Ca-\Tt)}

Given the identity in \eqn{getH2-h}, i.e. 
$\Hhard^{(+)}_{\rm NLL,h} = {\cal \hat M}^{(+)}_{\rm NLL,h}$,
the result of (\ref{reduced_hard_results}) is sufficient to compute 
the full infrared-renormalized amplitude $\Hhard^{(+)}_{\rm NLL}$ 
by combining it with the soft contribution $\Hhard^{(+)}_{\rm NLL,s}$ 
of eqs. (\ref{getH4}) and (\ref{eq:HsoftExpanded_1}). This will be 
done in the section \ref{sec:hardAmpl} below.
Before doing this let us combine the hard and soft components 
for the reduced amplitude itself, and comment further on some 
number-theoretic properties, as promised.

According to eqs.~(\ref{eq:redampSplit}) and (\ref{Msh}), the 
expressions for the full reduced amplitude through 
${\cal O}(\epsilon^0)$ can be easily obtained order-by-order 
summing the results for the soft amplitude provided in 
eqs.~(\ref{eq:MsoftExpanded_1})-(\ref{eq:MsoftExpanded_8})
and those for the hard amplitude in 
eqs.~(\ref{eq:m1hardfinite})--(\ref{eq:m8hardfinite}) 
above, where the former accounts for all infrared 
singularities plus some finite terms, and the latter 
for the remaining finite contributions. We obtain 
\renewcommand{\Cone}{C_1}
\renewcommand{\Ctwo}{C_2}
\begin{subequations}
\begin{align}
	\label{eq:m1finite} 
	\Mreduced_{\rm NLL}^{(1)} &= 
	i\pi \Bn{0} \bigg\{ \frac{1}{2\eps} \bigg\} \Tsu \Mtree, \\ 
	\Mreduced_{\rm NLL}^{(2)} &= 
	i\pi \frac{\Bn{0}^2}{2} \bigg\{ \frac{\Ctwo}{4 \eps^2} \bigg\} \Tsu \Mtree, \\  
	\Mreduced_{\rm NLL}^{(3)} &= 
	i\pi \frac{\Bn{0}^3}{3!} 
	\bigg\{ \Ctwo^2 \left( \frac{1}{8 \eps^3} 
	- \frac{11\zeta_3}{4}\right) \bigg\} \Tsu \Mtree, \\  
	\Mreduced_{\rm NLL}^{(4)} &= 
	i\pi \frac{\Bn{0}^4}{4!}
	\bigg\{ \Cone \Ctwo^2 \left(
	-\frac{\zeta_3}{8 \eps} 
	- \frac{3\zeta_4}{16}\right) 
	+ \Ctwo^3 \left( \frac{1}{16 \eps^4} 
	+ \frac{\zeta_3}{8\eps} 
	+ \frac{3\zeta_4}{16}\right) \bigg\} \Tsu \Mtree, \\  \nn
	\Mreduced_{\rm NLL}^{(5)} &= 
	i\pi \frac{\Bn{0}^5}{5!}
	\bigg\{ \Cone^2 \Ctwo^2 \left(
	- \frac{5\zeta_5}{2} \right) 
	+ \Cone \Ctwo^3 
	\left( -\frac{\zeta_3}{16 \eps^2} 
	- \frac{3\zeta_4}{32 \eps} 
	+ \frac{333\zeta_5}{16} \right) \\  
	&\hspace{1.0cm} 
	+\, \Ctwo^4 \left( \frac{1}{32 \eps^5} 
	+ \frac{\zeta_3}{16\eps^2} + \frac{3\zeta_4}{32 \eps} 
	- \frac{717\zeta_5}{16} \right) \bigg\} \Tsu \Mtree, \\ \nn  
	\Mreduced_{\rm NLL}^{(6)} &= 
	i\pi \frac{\Bn{0}^6}{6!}
	\bigg\{ \Cone^3 \Ctwo^2 \bigg( \frac{39 \zeta_3^2}{16} \bigg) 
	+ \Cone^2 \Ctwo^3  \bigg( -\frac{399 \zeta_3^2}{16} \bigg) \\ \nn 
	&\hspace{1.0cm} 
	+\, \Cone \Ctwo^4 \bigg(-\frac{\zeta_3}{32 \eps^3} 
	- \frac{3\zeta_4}{64 \eps^2}  
	- \frac{3\zeta_5}{32 \eps}
	+ \frac{2637\zeta_3^2}{32}  
	+ \frac{5 \zeta_6}{32} \bigg) \\ 
	&\hspace{1.0cm}
	+\, \Ctwo^5 \bigg(\frac{1}{64 \eps^6} 
	+ \frac{\zeta_3}{32 \eps^3} 
	+ \frac{3 \zeta_4}{64 \eps^2}  
	+ \frac{3 \zeta_5}{32 \eps} 
	- \frac{2879 \zeta_3^2}{32} 
	+ \frac{5 \zeta_6}{32} \bigg)
	\bigg\} \Tsu \Mtree, \\  \nn
	\Mreduced_{\rm NLL}^{(7)} &= 
	i\pi \frac{\Bn{0}^6}{6!}
	\bigg\{ \Cone^4 \Ctwo^2 \bigg( 
	\frac{2135\zeta_7}{256} \bigg)
	\Cone^3 \Ctwo^3 \bigg(  
	\frac{30135\zeta_7}{256} \bigg)
	+ \Cone^2 \Ctwo^4 \bigg( \frac{\zeta_3^2}{32 \eps} 
	+ \frac{3\zeta_3 \zeta_4}{32} 
	-\, \frac{20111\zeta_7}{32}\bigg)
	\\ \nn
	&\hspace{1.0cm}
	+ \Cone \Ctwo^5 \bigg(
	- \frac{\zeta_3}{64 \eps^4} 
	- \frac{3\zeta_4}{128 \eps^3} 
	- \frac{3\zeta_5}{64 \eps^2}
	- \frac{3\zeta_3^2}{64 \eps} 
	- \frac{5 \zeta_6}{64 \eps} 
	- \frac{9 \zeta_3 \zeta_4}{64} 
	+\, \frac{97047 \zeta_7}{64} \bigg)
	\\ \nn
	&\hspace{1.0cm}
	+ \Ctwo^6 \bigg( \frac{1}{128 \eps^7}
	+ \frac{\zeta_3}{64 \eps^4} 
	+ \frac{3 \zeta_4}{128 \eps^3} 
	+ \frac{3 \zeta_5}{64 \eps^2}
	+ \frac{\zeta_3^2}{64 \eps} 
	+ \frac{5 \zeta_6}{64 \eps}   
	\\ 	&\hspace{1.0cm}
	+\, \frac{3 \zeta_3 \zeta_4}{64} 
	- \frac{90711 \zeta_7}{64} \bigg)
	\bigg\} \Tsu \Mtree, \\   \nn
	\label{eq:m8finite}
	\Mreduced_{\rm NLL}^{(8)} &= 
	i\pi \frac{\Bn{0}^8}{8!}
	\bigg\{ \Cone^5 \Ctwo^2 \bigg(  
	\frac{611\zeta_3 \zeta_5}{32} \bigg)
	+\Cone^4 \Ctwo^3 \bigg(  
	- \frac{643\zeta_3 \zeta_5}{2} \bigg)
	+\Cone^3 \Ctwo^4 \bigg(  
	\frac{8597\zeta_3 \zeta_5}{4} \bigg) \\ \nn
	&\hspace{1.0cm}
	+\, \Cone^2 \Ctwo^5 \bigg( 
	\frac{\zeta_3^2}{64 \eps^2} 
	+ \frac{3\zeta_3 \zeta_4}{64 \eps}
	- \frac{228093 \zeta_3 \zeta_5}{32} 
	+ \frac{21\zeta_8}{512} \bigg)
	+ \Cone \Ctwo^6 \bigg(
	-\frac{\zeta_3}{128 \eps^5} \\ \nn
	&\hspace{1.0cm}
	-\, \frac{3\zeta_4}{256 \eps^4}    
	- \frac{3\zeta_5}{128 \eps^3} 
	- \frac{3 \zeta_3^2}{128 \eps^2} 
	- \frac{5 \zeta_6}{128 \eps^2} 
	- \frac{9 \zeta_3 \zeta_4}{128 \eps} 
	- \frac{9\zeta_7}{128 \eps} 
	+ \frac{749943 \zeta_3 \zeta_5}{64} \\ \nn
	&\hspace{1.0cm} 
	-\, \frac{189 \zeta_8}{1024} \bigg)
	+ \Ctwo^7 \bigg( \frac{1}{256 \eps^8} 
	+ \frac{\zeta_3}{128 \eps^5} 
	+ \frac{3\zeta_4}{256 \eps^4}
	+ \frac{3\zeta_5}{128 \eps^3} 
	+ \frac{\zeta_3^2}{128 \eps^2}
	+ \frac{5 \zeta_6}{128 \eps^2}  \\ 
	&\hspace{1.0cm} 
	+ \frac{3 \zeta_3 \zeta_4}{128 \eps} 
	+ \frac{9 \zeta_7}{128 \eps} 
	- \frac{483837 \zeta_3 \zeta_5}{64} 
	+ \frac{147 \zeta_8}{1024} \bigg)
	\bigg\} \Tsu \Mtree\, ,
\end{align}
\end{subequations}
\renewcommand{\Cone}{(2\Ca-\Tt)}
\renewcommand{\Ctwo}{(\Ca-\Tt)}

These result reproduce the one- to four-loop 
results of ref.~\cite{Caron-Huot:2013fea} and of our 
numerically-determined five-loop result in eq.~\eqref{eq:m5num}. 
In the ancillary file \texttt{NLL-reduced-amplitude.txt} 
we provide the result for the soft, the hard and the full 
reduced amplitude up to 13 loops. Furthermore, the amplitude 
can now be calculated to any number of loops with the methods 
presented in sections~\ref{soft}, \ref{2d-bfkl} and \ref{amplitude}.

Similarly to the wavefunction at twelve loops (and above), the 
hard contributions to the amplitude (and thus the full amplitude
itself) cannot be expressed in terms of ordinary zeta numbers 
beyond a certain loop order. In fact, most of what we discussed 
in the context of the wavefunction below 
eqs.~\eqref{eq:wtd1}--\eqref{eq:wtd4} applies to the 
amplitude as well: Either of the two methods presented in 
section~\ref{sec:method1} and \ref{sec:method2} requires us 
to evaluate \ac{svhpls} at $z = \zb = 1$ and we hence expect 
the presence of (single-valued) \ac{mzvs} starting 
from weight eleven. Indeed, the eleven-loop amplitude 
features a term proportional to $g_{5,3,3}$, defined in 
eq.~\eqref{eq:g533}:
\renewcommand{\Cone}{C_1}
\renewcommand{\Ctwo}{C_2}
\begin{multline}
	\Mhnllpl{11} \supset 
	\frac{1}{102400}\bigg(-\frac{149}{6720} \Cone^8 \Ctwo^2 
	+ \frac{26209}{60480} \Cone^7 \Ctwo^3 
	- \frac{14813}{4320} \Cone^6 \Ctwo^4 
	+ \frac{210383}{15120} \Cone^5 \Ctwo^5  \\
	- \frac{7549}{252} \Cone^4 \Ctwo^6
	+ \frac{39257}{1260} \Cone^3 \Ctwo^7 
	- 11 \Cone^2 \Ctwo^8 \bigg) \times g_{5,3,3}.
	\label{eq:m11g533}
\end{multline}
Of course this term is entirely due to the hard component of 
the amplitude, as the soft one consists exclusively of ordinary 
zeta values (non-single-valued ones), as discussed in 
section~\ref{soft}. At twelve loops the reduced amplitude is
again comprised of ordinary zeta numbers $\zeta_n$, as 
there are no weight 12 single-valued MZVs. Such numbers 
appear then again in the thirteen loop amplitude:
\begin{multline}
	\Mhnllpl{13} \supset 
	\frac{1}{2207744000}\bigg( \frac{5367943}{497664}  \Cone^{10} \Ctwo^2
	- \frac{32668315}{124416} \Cone^9 \Ctwo^3 
	+ \frac{6876365071}{2488320 }  \Cone^8 \Ctwo^4 \\
	- \frac{10213439791}{622080}  \Cone^7 \Ctwo^5 
	+ \frac{37444840199}{622080} \Cone^6 \Ctwo^6
	- \frac{10827306157}{77760}  \Cone^5 \Ctwo^7 \\
	+ \frac{3841520891}{19440}  \Cone^4 \Ctwo^8
	- \frac{503783639}{3240}  \Cone^3 \Ctwo^9
	+ 50459 \Cone^2 \Ctwo^{10} \bigg) \times g_{5,5,3}, 
	\label{eq:m13g553}
\end{multline}
and
\begin{multline}
	\Mhnllpl{13} \supset 
	\frac{1}{2649292800}\bigg( -\frac{1819475}{82944}  \Cone^{10} \Ctwo^2
	+ \frac{5621717}{10368}  \Cone^9 \Ctwo^3 
	- \frac{961202489}{165888}  \Cone^8 \Ctwo^4 \\
	+ \frac{482408111}{13824}  \Cone^7 \Ctwo^5 
	- \frac{5356152533}{41472}  \Cone^6 \Ctwo^6
	+ \frac{1551101681}{5184}  \Cone^5 \Ctwo^7 \\
	- \frac{543921901}{1296} \Cone^4 \Ctwo^8
	+ \frac{69045265}{216}  \Cone^3 \Ctwo^9
	- 96967 \Cone^2 \Ctwo^{10} \bigg) \times g_{7,3,3},
	\label{eq:m13g733}
\end{multline}
where the single-values zeta numbers 
$g_{5,5,3}$ and $g_{7,3,3}$ have been defined 
in eqs.~\eqref{eq:g553} and~\eqref{eq:g733}.
\renewcommand{\Cone}{(2\Ca-\Tt)}
\renewcommand{\Ctwo}{(\Ca-\Tt)}

The fact that the MZV terms in 
eqs.~\eqref{eq:m11g533},~\eqref{eq:m13g553}
and~\eqref{eq:m13g733} appear 
in the eleven- and thirteen-loop amplitude already 
excludes that there could be a simple all-order formula in 
terms of gamma functions for the reduced amplitude. This 
stands in sharp contrast to the contributions associated 
with the soft limit, both singular and finite, which can be 
resummed to all orders by means of gamma functions, 
as we have seen in section~\ref{soft}.

\subsection{The infrared-renormalized amplitude\label{sec:hardAmpl}}

We conclude this section by discussing 
the perhaps most physically relevant 
infrared-renormalized amplitude (or hard function),
which according to \eqn{Hsplit}, is obtained by 
summing the soft component, given to all orders 
by the closed expression in \eqn{getH4}, and the 
hard component, which according to \eqn{getH2-h}, 
coincides with the hard component of the 
reduced amplitude, ${\cal \hat M}^{(+)}_{\rm NLL,h}$.
The latter can be determined to any loop order by 
following the methods discussed in sections 
\ref{sec:method1} and \ref{sec:method2}, however
a closed-form expression cannot be obtained as 
in case of the soft part of the infrared-renormalized 
amplitude. Thus, in practice we limit ourselves to 
determine this amplitude to 13 loops, and the result is 
provided in the ancillary file \texttt{NLL-IR-renormalised-amplitude.txt}. 
Here we provide a sample of the result (with ${\cal H}$ 
defined in eq.~(\ref{IRfacteq}) and loop-expanded 
following eq.~(\ref{MhatEven})), up to eight loops:  
\begin{subequations}
\label{eq:HExpanded}
\renewcommand{\Cone}{C_1}
\renewcommand{\Ctwo}{C_2}
\begin{align}
	\label{eq:HExpanded_1} 
	{\cal H}_{\rm NLL}^{(1)} &= 0, \\ 
	{\cal H}_{\rm NLL}^{(2)} &= 0, \\  
	{\cal H}_{\rm NLL}^{(3)} &= \frac{i\pi}{3!}  
	\bigg\{-\Ctwo^2 \frac{11\zeta_3}{4} \bigg\} \Tsu \Mtree, \\  
	{\cal H}_{\rm NLL}^{(4)} &= \frac{i\pi}{4!}
	\bigg\{ - C_A \Ctwo^2 \frac{3\zeta_4}{16} \bigg\} \Tsu \Mtree, \\ 
	{\cal H}_{\rm NLL}^{(5)} &= \frac{i\pi}{5!}
	\bigg\{- C_A^2 \Ctwo^2 \frac{5\zeta_5}{2}
	+ C_A \Ctwo^3 \frac{253\zeta_5}{16}
	-\Ctwo^4 \frac{53\zeta_5}{2} \bigg\} \Tsu \Mtree, \\ \nn 
	{\cal H}_{\rm NLL}^{(6)} &= \frac{i\pi}{6!}
	\bigg\{ C_A^3 \Ctwo^2  \frac{39 \zeta_3^2}{16}
	- C_A^2 \Ctwo^3  \frac{141 \zeta_3^2}{8} 
	+ C_A \Ctwo^4 \bigg(\frac{1275\zeta_3^2}{32} 
	-\, \frac{5\zeta_6}{32} \bigg) 
	\\		& \hspace{2.0cm}
	- \Ctwo^5 \frac{481 \zeta_3^2}{16} \bigg\} \Tsu \Mtree, \\ \nn 
	{\cal H}_{\rm NLL}^{(7)} &= \frac{i\pi}{7!}
	\bigg\{ - C_A^4 \Ctwo^2 \frac{2135 \zeta_7}{256} 
	+ C_A^3 \Ctwo^3 \frac{21595\zeta_7}{256} 
	+ C_A^2 \Ctwo^4 \bigg( \frac{3\zeta_3 \zeta_4}{32} 
	-\frac{83293\zeta_7}{256} \bigg) 
	\\
	& \hspace{2.0cm}
	+\, C_A \Ctwo^5 \bigg( \frac{3\zeta_3 \zeta_4}{64}
	+ \frac{148277 \zeta_7}{256} \bigg)  
	- \Ctwo^6 \frac{13443 \zeta_7}{32} \bigg\} \Tsu \Mtree, \\   \nn
	\label{eq:HExpanded_8}
	{\cal H}_{\rm NLL}^{(8)} &= \frac{i\pi}{8!}
	\bigg\{ C_A^5 \Ctwo^2 \frac{611 \zeta_3 \zeta_5}{32} 
	-C_A^4 \Ctwo^3 \frac{7233 \zeta_3 \zeta_5}{32} 
	+C_A^3 \Ctwo^4 \frac{16867 \zeta_3 \zeta_5}{16} 
	+C_A^2 \Ctwo^5 \bigg(
	- \frac{77383 \zeta_3 \zeta_5}{32} \\
	&\hspace{1.0cm} 
	+\, \frac{21\zeta_8}{512} \bigg)  
	+ C_A \Ctwo^6 \bigg(
	\frac{174033 \zeta_3 \zeta_5}{64}  
	- \frac{105\zeta_8}{1024} \bigg) 
	- \Ctwo^7 \frac{35941\zeta_3 \zeta_5}{32}\bigg\} \Tsu \Mtree\, .
\end{align}
\end{subequations}
Given that $\Hhard^{(+)}_{\rm NLL,h} = 
{\cal \hat M}^{(+)}_{\rm NLL,h}$, the same 
number-theory properties discussed at the 
end of section \ref{sec:finiteamp} apply to 
the infrared-renormalized amplitude as well. 
In particular, resummation in terms of gamma 
functions is excluded.
\renewcommand{\Cone}{(2\Ca-\Tt)}
\renewcommand{\Ctwo}{(\Ca-\Tt)}


\section{Numerical analysis and convergence properties} \label{numerics}

The calculation developed in sections 
\ref{soft}, \ref{2d-bfkl} and \ref{amplitude}
has allowed us to determine symmetries 
and general features of the wavefunction 
and the amplitude, as well as their exact 
analytic structure to fourteen and thirteen 
orders in perturbation theory respectively. 
We are now interested to perform a numerical 
analysis, and focus on features which are not 
directly evident from the analytic expressions, 
such as the qualitative behaviour of the 
wavefunction, the relative size of the soft 
and hard contributions to the wavefunction, 
and the convergence properties of the 
infrared-renormalized amplitude as an 
expansion in $x \equiv L \,\as/\pi$.

\subsection{Wavefunction}

Let us begin by analysing the wavefunction. 
Given its finiteness, we consider here the 
leading term in the $\eps$ expansion, i.e. 
the two-dimensional soft, hard and full 
wavefunctions, defined respectively in 
eqs.~(\ref{Wsoft2d}), (\ref{WhardTwod}) 
and (\ref{eq:Htdaction}).

\begin{figure}[!ht]
\centering
\includegraphics[width=0.32\textwidth]{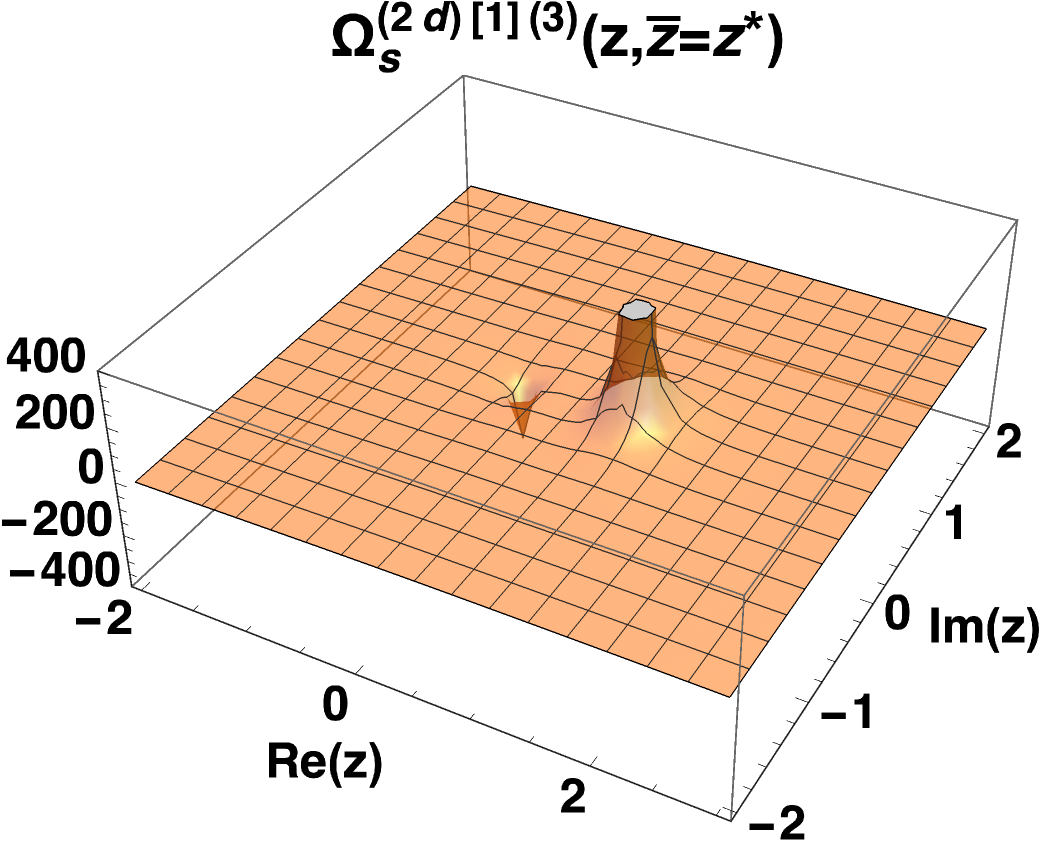}
\includegraphics[width=0.32\textwidth]{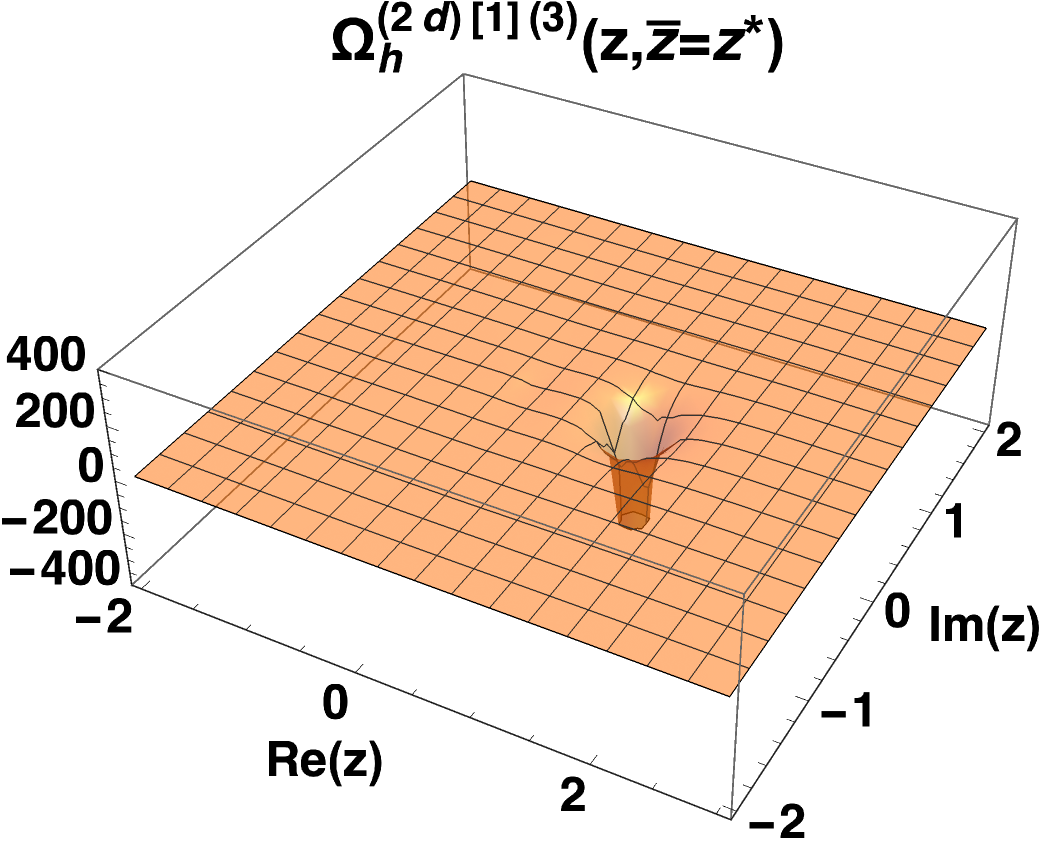}
\includegraphics[width=0.32\textwidth]{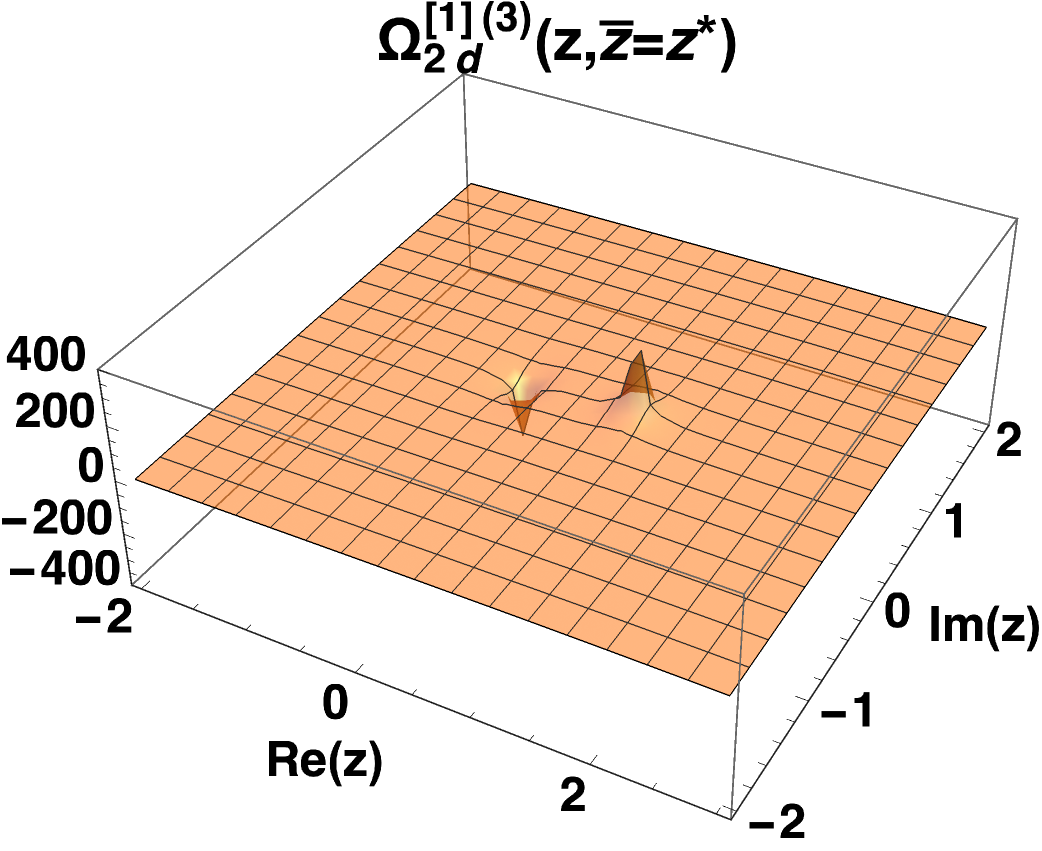} \\[0.3cm]
\includegraphics[width=0.32\textwidth]{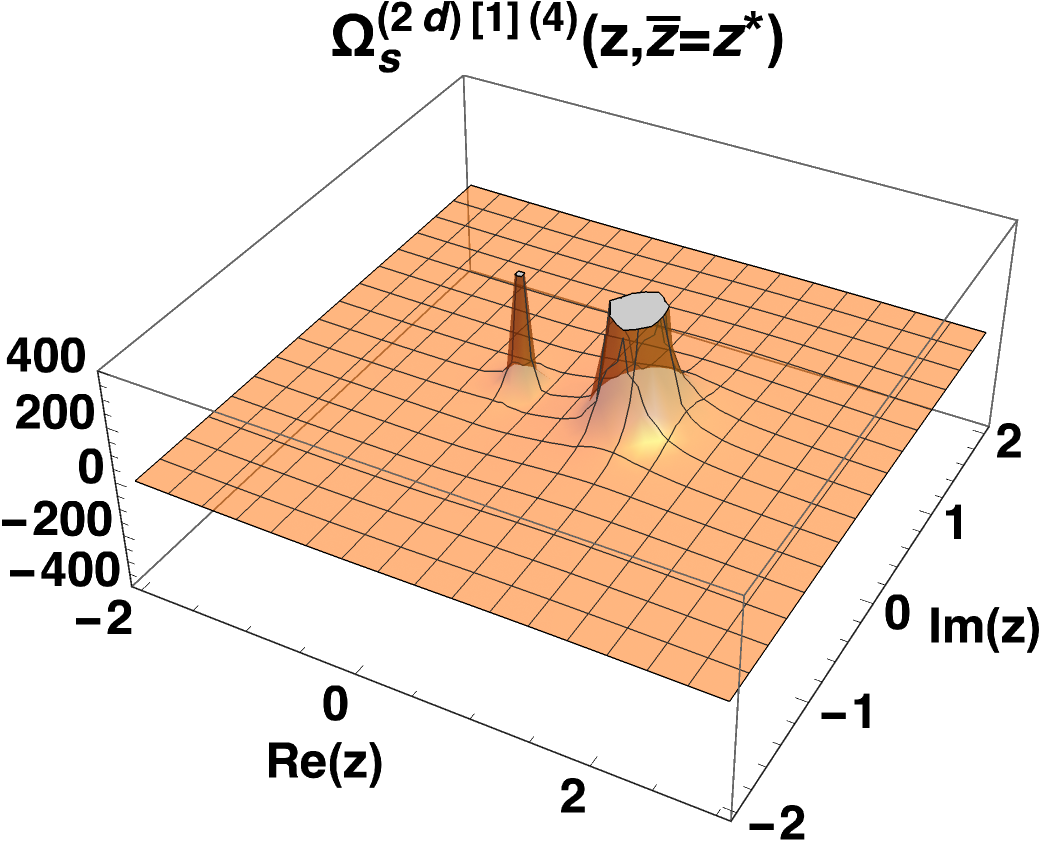}
\includegraphics[width=0.32\textwidth]{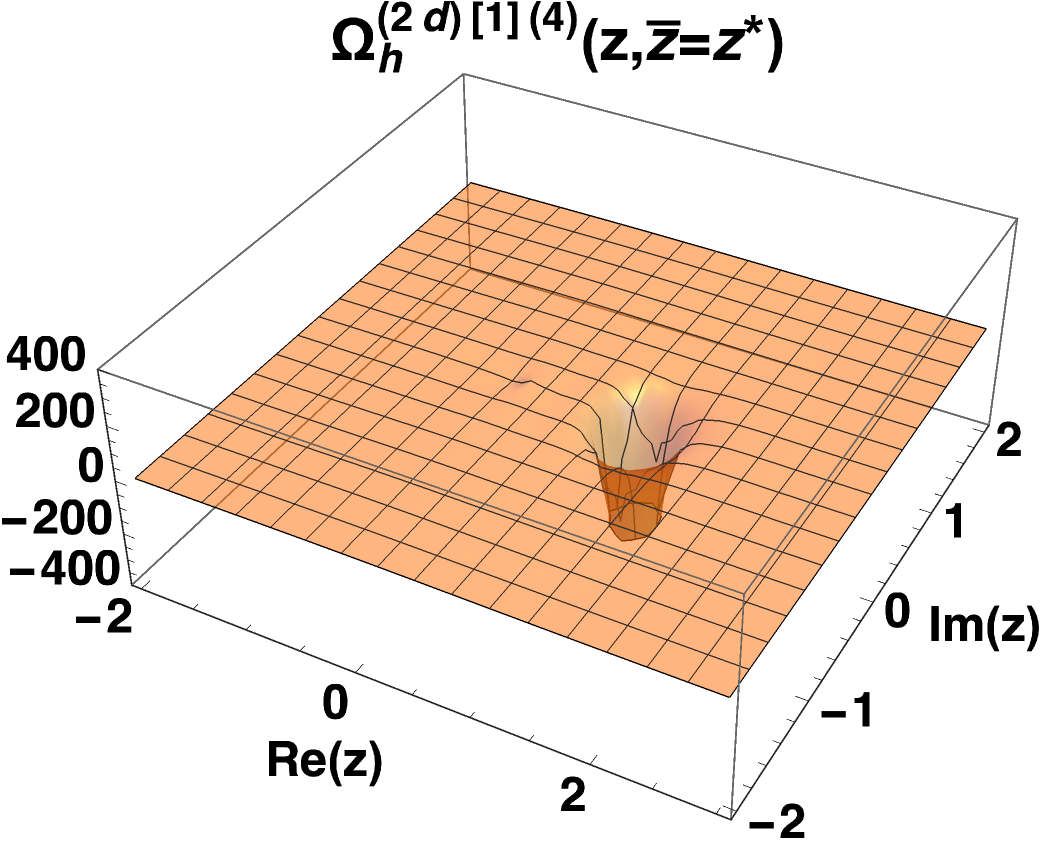}
\includegraphics[width=0.32\textwidth]{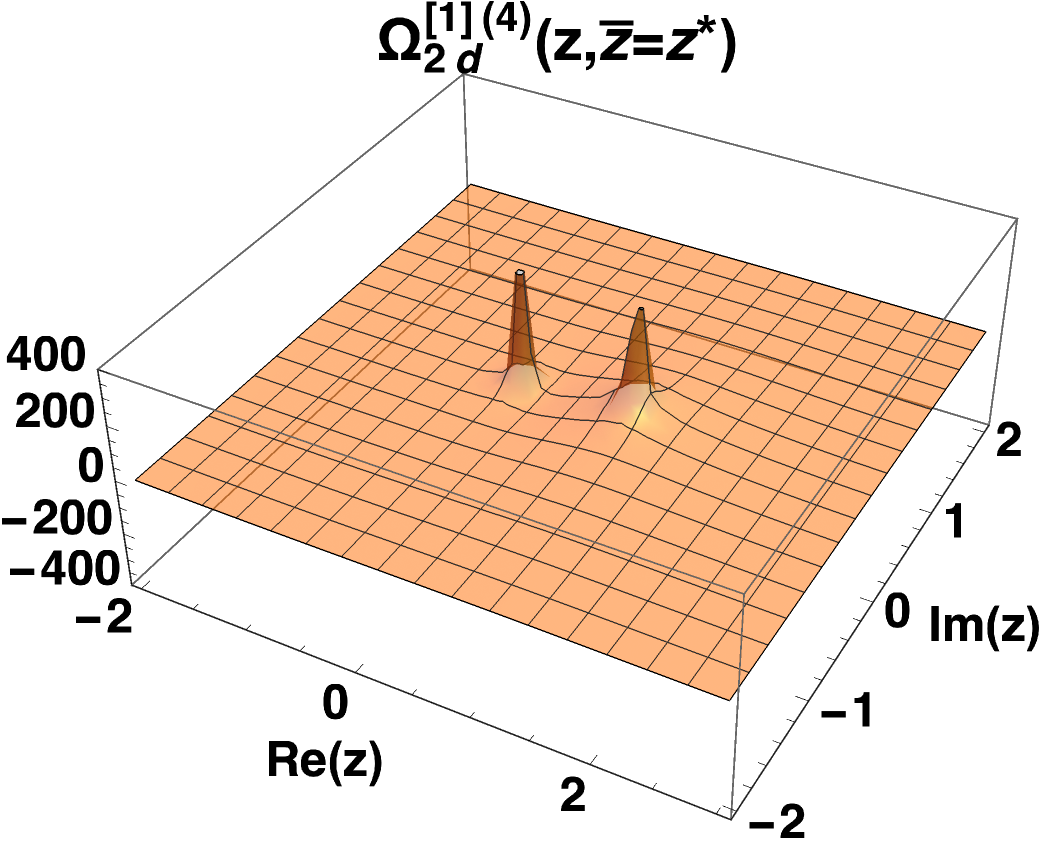} \\[0.3cm]
\includegraphics[width=0.32\textwidth]{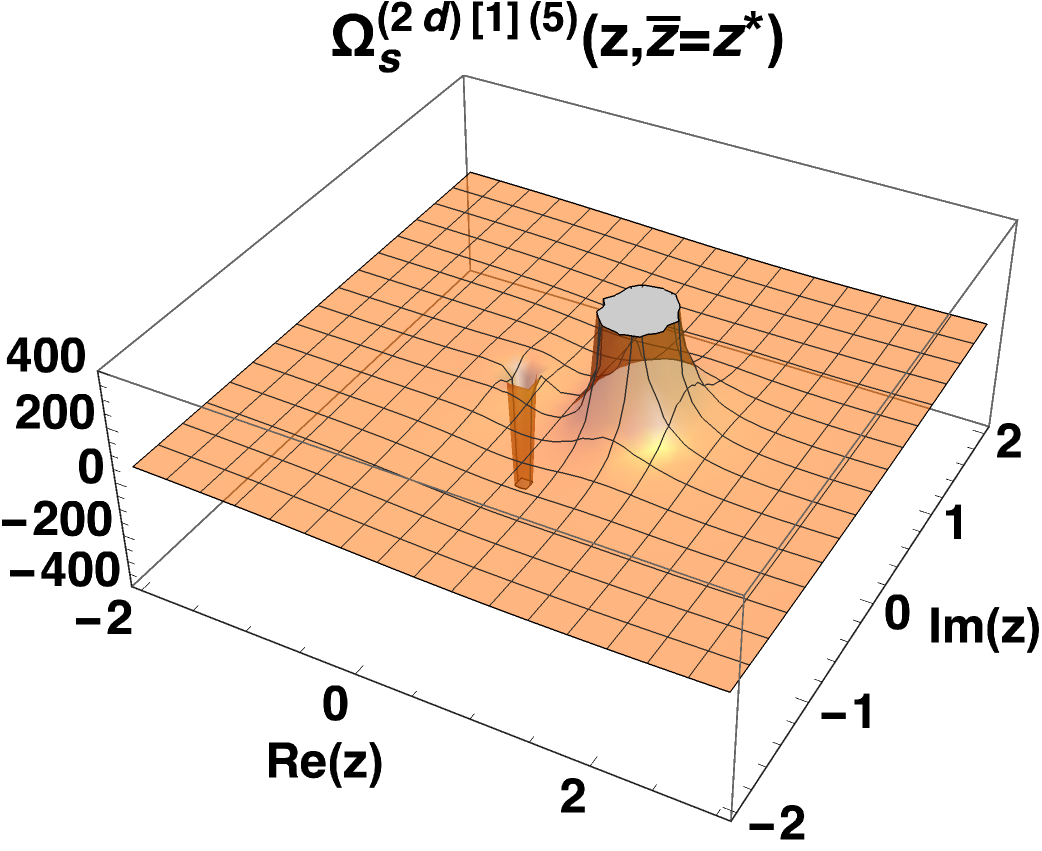}
\includegraphics[width=0.32\textwidth]{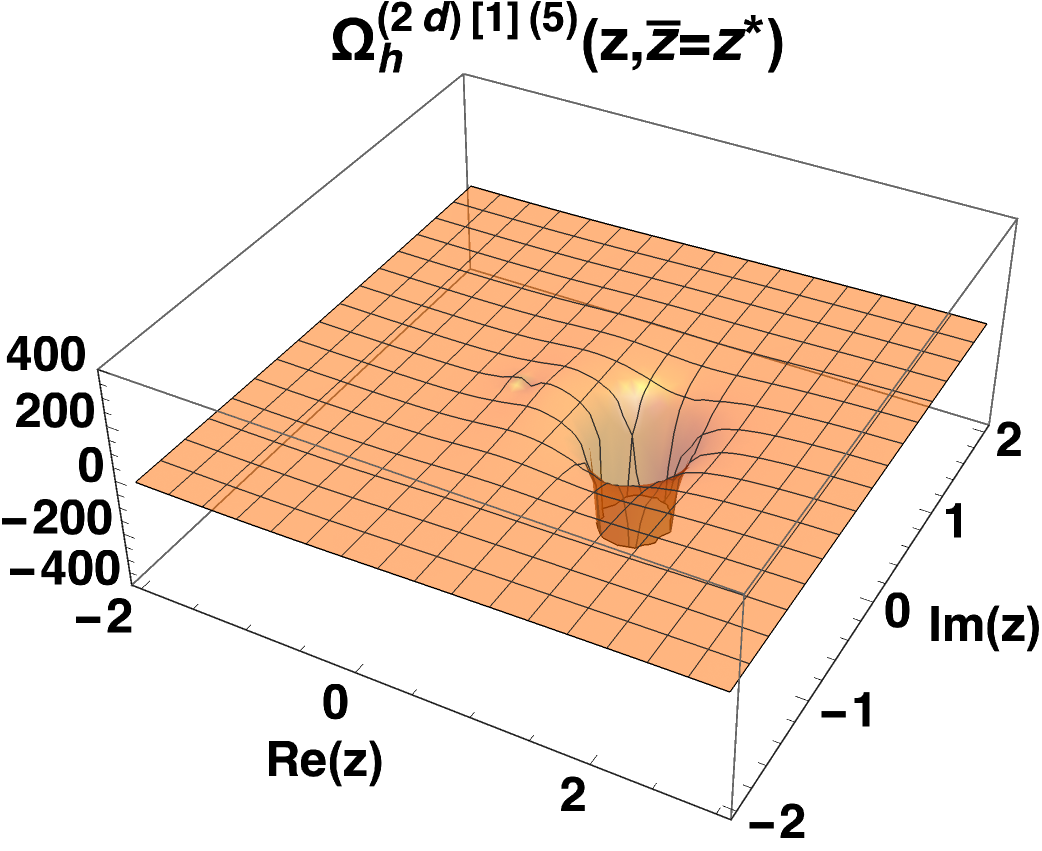}
\includegraphics[width=0.32\textwidth]{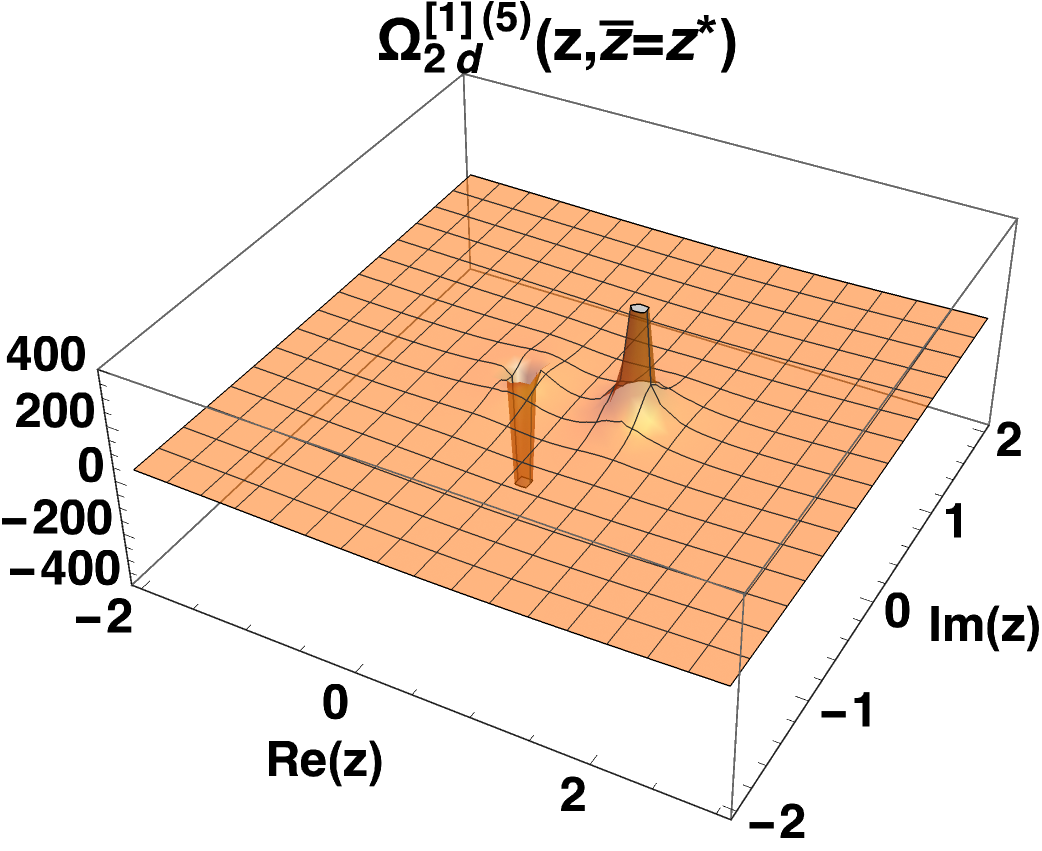}
\caption{Soft, hard and full wavefunction in the complex plane 
	${\rm Re}(z)$, ${\rm Im}(z)$. Here we plot the singlet component 
	of the wavefunction. The soft and the full wavefunction exhibit 
	singularities at $z = 0$ and $z = \infty$, due to the $z \leftrightarrow 1/z$
	symmetry, (the latter is not visible in the plots). In addition, there is a 
	singularity at $z = 1$, which appears also in the hard part of the 
	wavefunction. Notice that the singularities at $z = 1$ partly cancel 
	between the soft and hard wavefunctions, such that the full 
	wavefunction exhibits a peak near $z=1$ which is markedly 
	smaller relative to the two separate contributions.}
	\label{Wave1}
\end{figure}
\begin{figure}[!ht]
\centering
\includegraphics[width=0.32\textwidth]{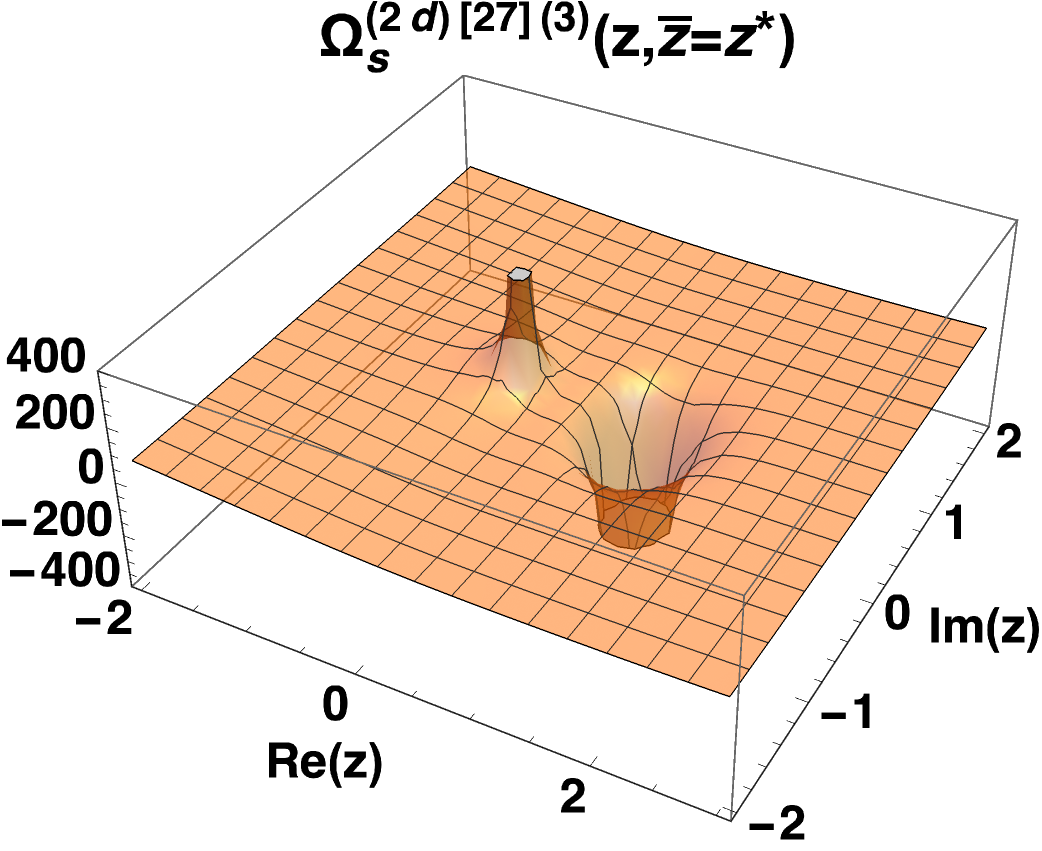}
\includegraphics[width=0.32\textwidth]{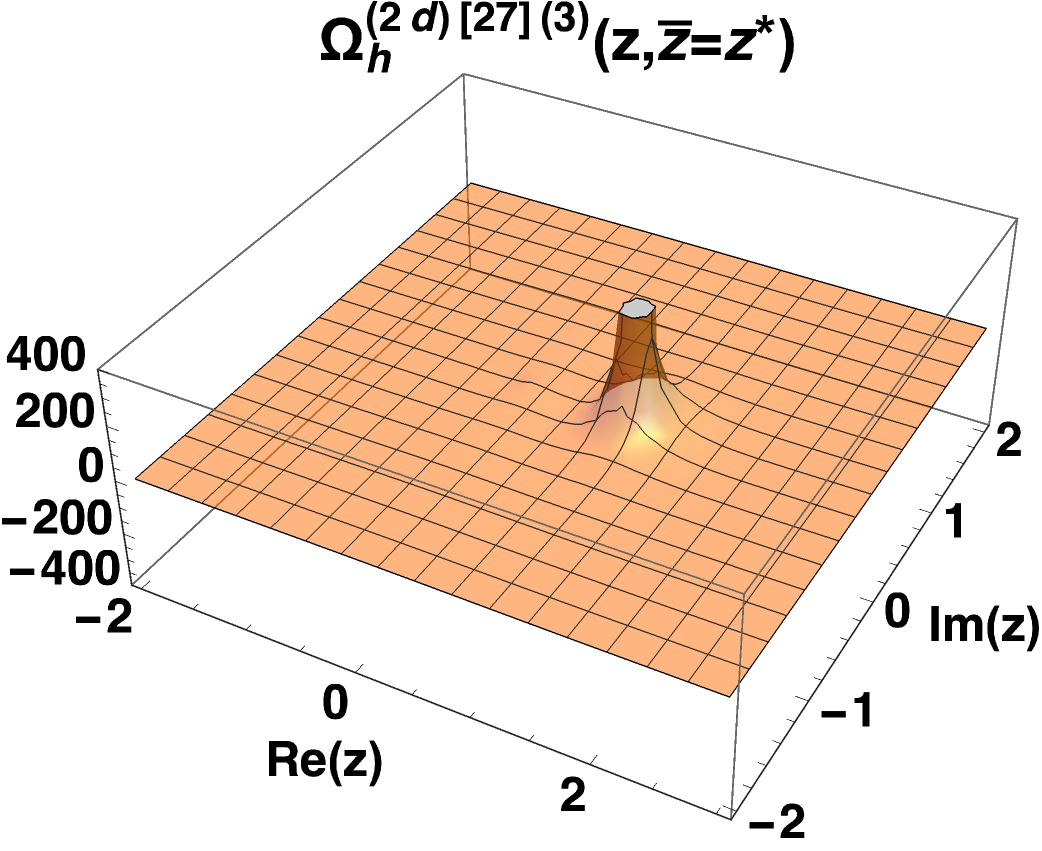}
\includegraphics[width=0.32\textwidth]{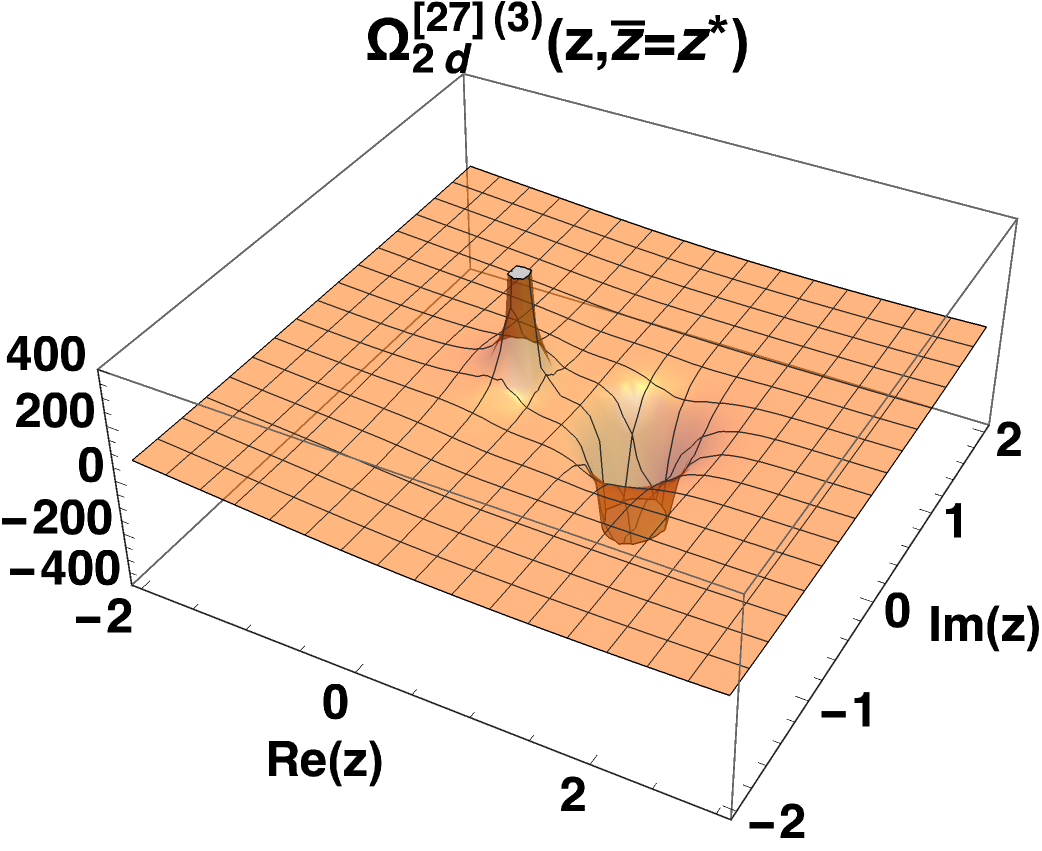} \\[0.3cm]
\includegraphics[width=0.32\textwidth]{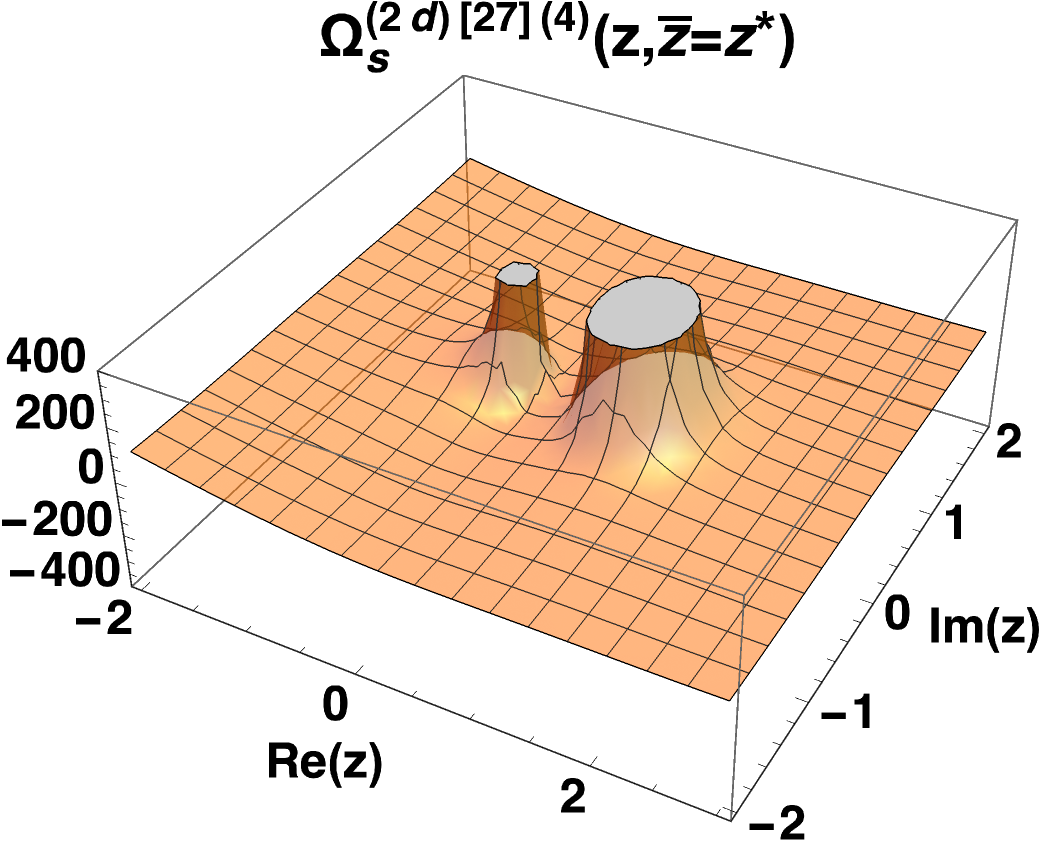}
\includegraphics[width=0.32\textwidth]{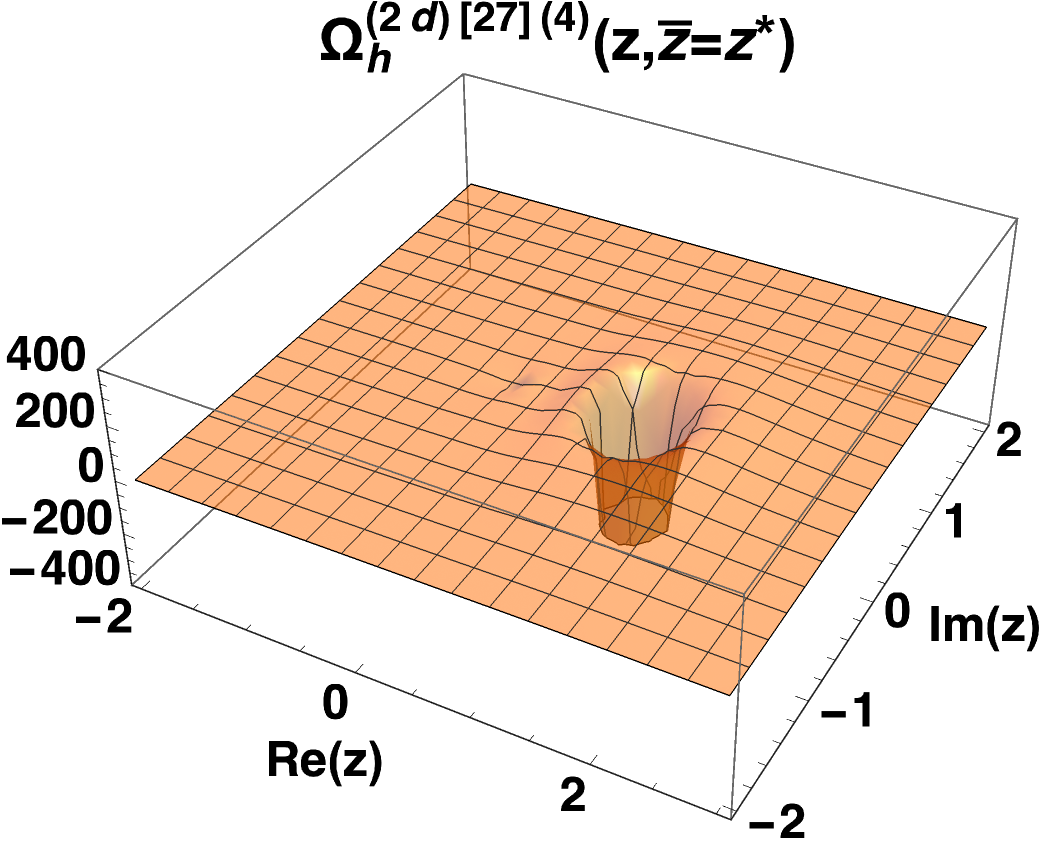}
\includegraphics[width=0.32\textwidth]{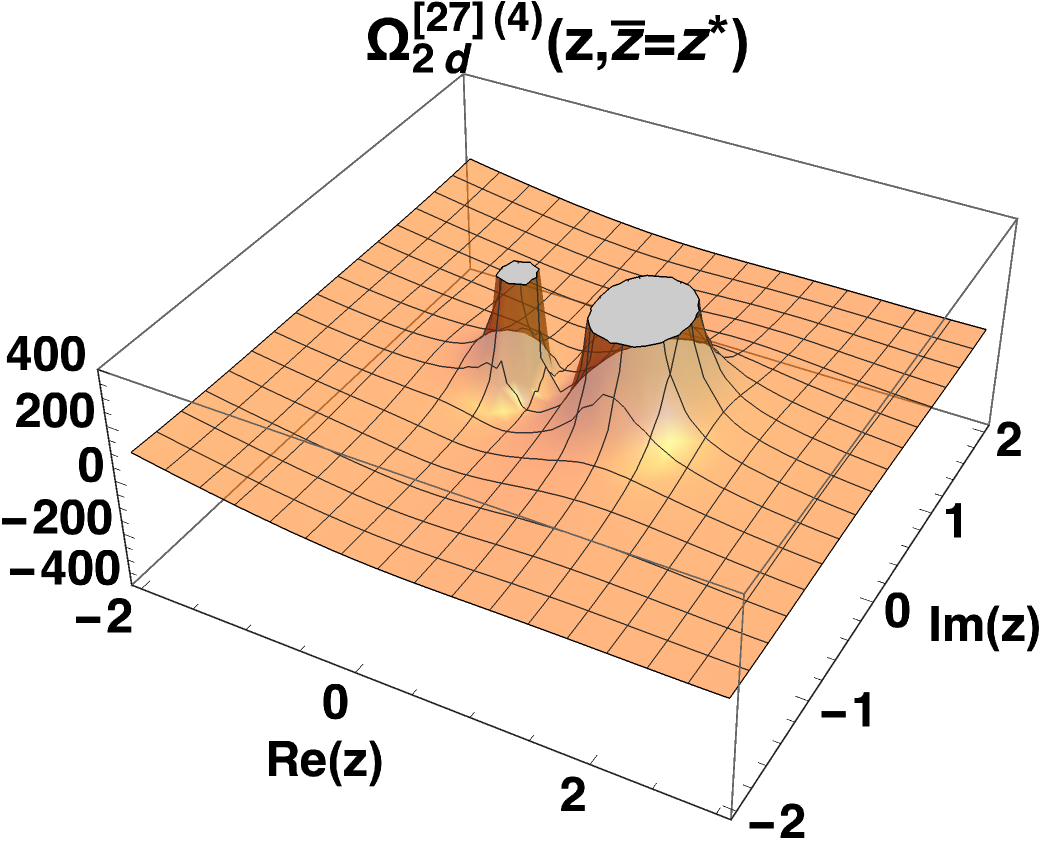} \\[0.3cm]
\includegraphics[width=0.32\textwidth]{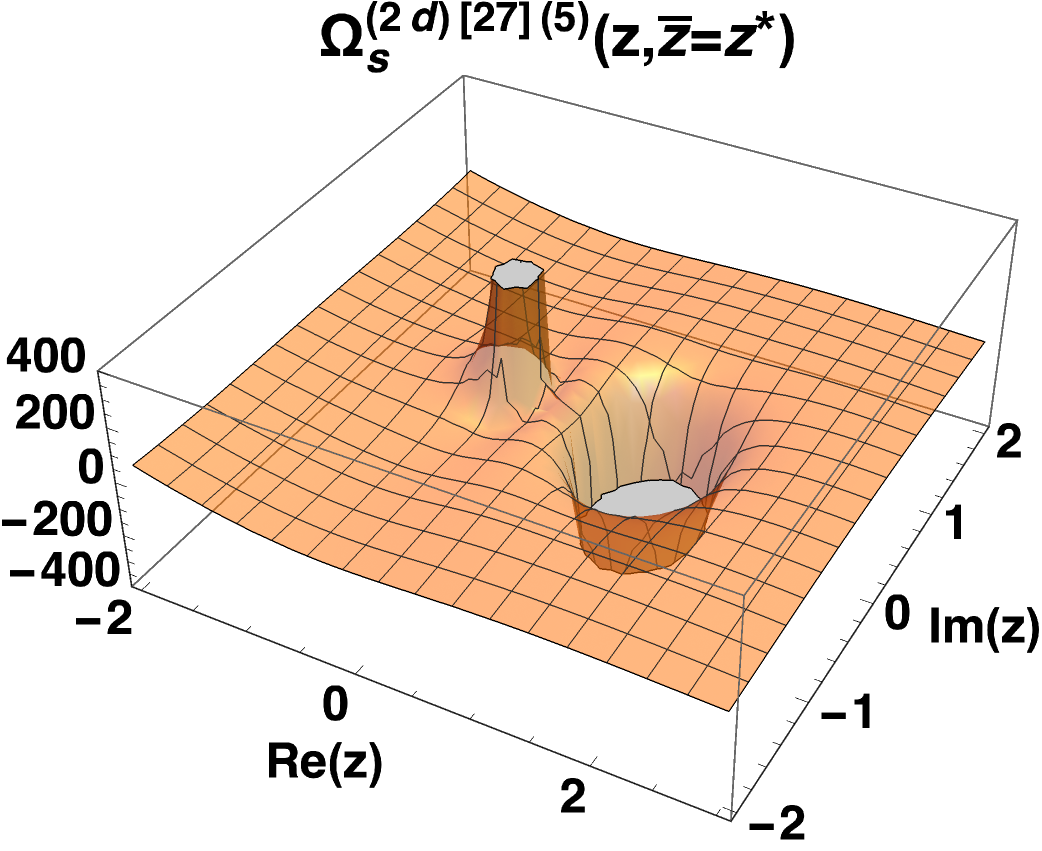}
\includegraphics[width=0.32\textwidth]{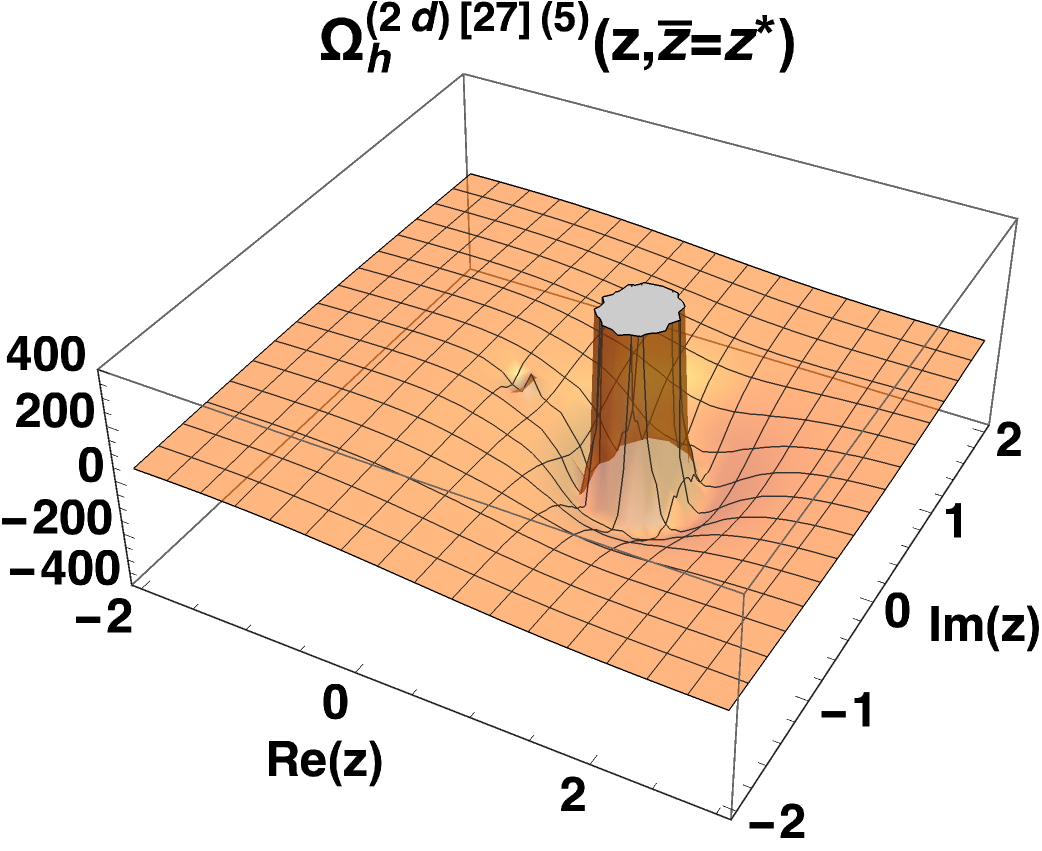}
\includegraphics[width=0.32\textwidth]{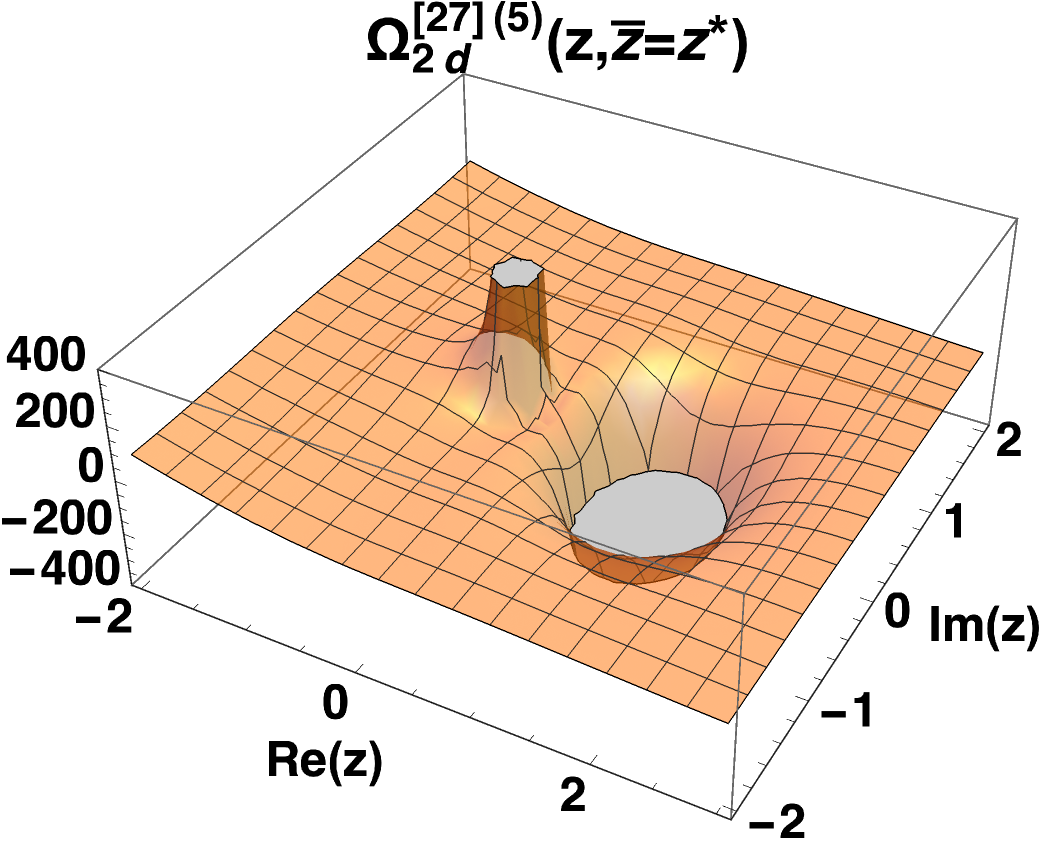}
\caption{Soft, hard and full wavefunction in the complex plane 
	${\rm Re}(z)$, ${\rm Im}(z)$. Here we plot the component 
	corresponding to the $27$ colour representation. }
\label{Wave27}
\end{figure}

As an example, in figures~\ref{Wave1} and~\ref{Wave27} we 
plot the coefficients $\Omega_{\rm s}^{(\rm 2d)(\ell)}$, 
$\Omega_{\rm h}^{(\rm 2d)(\ell)}$, $\Omega_{\rm 2d}^{(\ell)}$, 
at third, fourth and fifth order in perturbation theory. 
In these plots we fix $N_c=3$ and consider specific colour 
representations, namely the singlet and the 27 representation, 
such that the Casimir operator in the $t$-channel evaluates to  
\begin{align}
\label{Tt2num}
\begin{array}{ll}
	\text{singlet}:& {\mathbf T}_{t}^2 \, {\cal M}^{[1]} = 0, \\
	\text{27 representation}:\quad\qquad& 
	{\mathbf T}_{t}^2 \, {\cal M}^{[27]}  
	= 2(N_c+1)\, {\cal M}^{[27]}=8\, {\cal M}^{[27]}.
\end{array}
\end{align}
We plot the wavefunction in the complex $z$ plane, 
for $\bar z = z^*$. We observe that the soft 
and full wavefunctions exhibit peaks at $z = 0$; these 
are associated with the soft limit. Of course, by the 
$z \leftrightarrow 1/z$ symmetry discussed in 
sections~\ref{sec:wf2d} and~\ref{sec:asalphabet}, 
there is an identical singularity at $z = \infty$ (which 
is not visible in the patch of the complex plane shown 
in the plot). In the way we separated between the soft 
and hard components, the two-dimensional hard 
wavefunction is strictly zero in these soft limits (see 
the discussion following eq.~(\ref{eq:Iregdef})). 

All components of the wavefunction have 
singularities at $z = 1$. The $z=1$ singularity 
represents rather different physics, where both 
Reggeons are \emph{hard}, namely $k^2, (p-k)^2\gg p^2$.
It is interesting to note that 
the singularity at $z = 1$ is always of opposite sign 
between the soft and hard wavefunction, such that 
these contributions cancel to a large extent in the 
full wavefunction. This observation allows us to conclude 
already that the soft approximation, although convenient 
for calculation purposes, does not provide a good numerical 
approximation for the full wavefunction away 
from the soft limit. 

Focusing now on the full wavefunction, the singular 
behaviour near $z = 0$ and $z = 1$ at $\ell$ loop order 
can be described respectively by the leading logarithms 
in the two limits, $\sim c_\ell \log^\ell (z \bar z)$
and $\sim c'_\ell \log^\ell\big[1/(1-z)^2(1-\bar z)^2\big]$,
where both the magnitude and the sign of the coefficients 
$c_\ell$ and $c_{\ell}'$ depends 
on the colour representation considered. Concerning 
the limit $z = 0$, the asymptotic behaviour is entirely 
determined by the soft wavefunction, given  that 
$\Omega_{\rm h}^{\rm 2d}(0,0) = 0$. We obtain 
the coefficients $c_{\ell}$ expanding the soft function in 
\eqn{eq:wffullsoftresummed_SVHPL} (compare 
eqs.~(\ref{eq:wffullsoftresummed}) and 
(\ref{Well-1-ansatz-sym})). Taking into account 
\eqn{OmegaEven}, we find 
\be\label{asymptotic0}
\Wfl{\ell}(z,\bar z)|_{z \to 0} \simeq c_\ell \, \log^{\ell}(z \bar z),
\quad 
c_\ell= \frac{\Ctwo^{\ell}}{2^{\ell}}.
\ee
Given that $\Ctwo = C_A = 3$ for the singlet, while $\Ctwo = -C_A -2 = -5$
for the 27 representation, this explains the sign-oscillating behaviour of the
wavefunction near $z = \bar z = 0$ for the singlet, and the constant sign of the 
27 representation, which can be seen also in figures~\ref{Wave1} and~\ref{Wave27}. 

Determining the coefficients $c_{\ell}'$ is less trivial, given that near 
$z = 1$ also $\Omega_{\rm h}^{\rm 2d}$ contributes. An analysis 
of the asymptotic behaviour up to the 14th order allows us to 
deduce the pattern and extrapolate. We find:
\be\label{asymptotic1}
\Wfl{\ell}(z,\bar z)|_{z \to 1} \simeq c'_{\ell} \, \log^{\ell} 
\bigg[\frac{1}{(1-\bar z)(1-\bar z)^2}\bigg],
\quad 
c'_{\ell}= \frac{(-1)^{\ell}}{{\ell}!} \frac{\left(\Tt\right)^{\ell}}{2^{2\ell}} 
\frac{\Gamma\left( 2+\ell-\frac{2C_A}{\Tt}\right)}{\Gamma\left(2-\frac{2C_A}{\Tt}\right)}.
\ee
Once again, we see that the series has alternating or constant
signs depending on the color representation. Specifically, it is 
sign alternating for the 27 representation, and it has constant 
sign for the singlet. Notice that both asymptotic expansion of 
the wavefunction near $z=0$ and $z = 1$ can be summed using
\eqn{OmegaEven}. We obtain
\begin{align} \label{asymptotic-resum0}
\Wf(p,k)|_{z \to 0} &= \frac{\as}{\pi}  \, 
(z \bar z)^{\frac{\as}{2 \pi} \, L \, \Ctwo}, \\
\label{asymptotic-resum1}
\Wf(p,k)|_{z \to 1} &=  \frac{\as }{\pi}  \,
\left\{
\begin{array}{ll}
	{}_1F_1\left(2 - \frac{2C_A}{\Tt}, 1, -\frac{\as }{4\pi}\, L \, \Tt 
	\log\left( \frac{1}{(1-z)^2(1-\bar z)^2}\right)\right), & \mbox{if } \Tt \neq 0,  \\
	{}_0F_1\left(1,\frac{\as }{2\pi}\, L \, C_A 
	\log\left( \frac{1}{(1-z)^2(1-\bar z)^2}\right)\right), & \mbox{if } \Tt = 0.
\end{array}
\right.
\end{align}
where ${}_0F_1$ and ${}_1F_1$ are the confluent and Kummer's 
confluent hypergeometric function. These resummed expressions 
are valid only in the leading logarithmic approximation in $z\zb$ 
and $(1-z)(1-\zb)$, respectively. The generalization of (\ref{asymptotic-resum0}) 
to include subleading logarithms of $z\zb$ has been given in 
(\ref{eq:wffullsoftresummed}), while a closed form generalization 
of (\ref{asymptotic-resum1}) is yet unknown. 

\subsection{Convergence of the loop expansion of the infrared-renormalized amplitude}

\begin{figure}[htb]
\centering
	\includegraphics[width=0.7\textwidth]{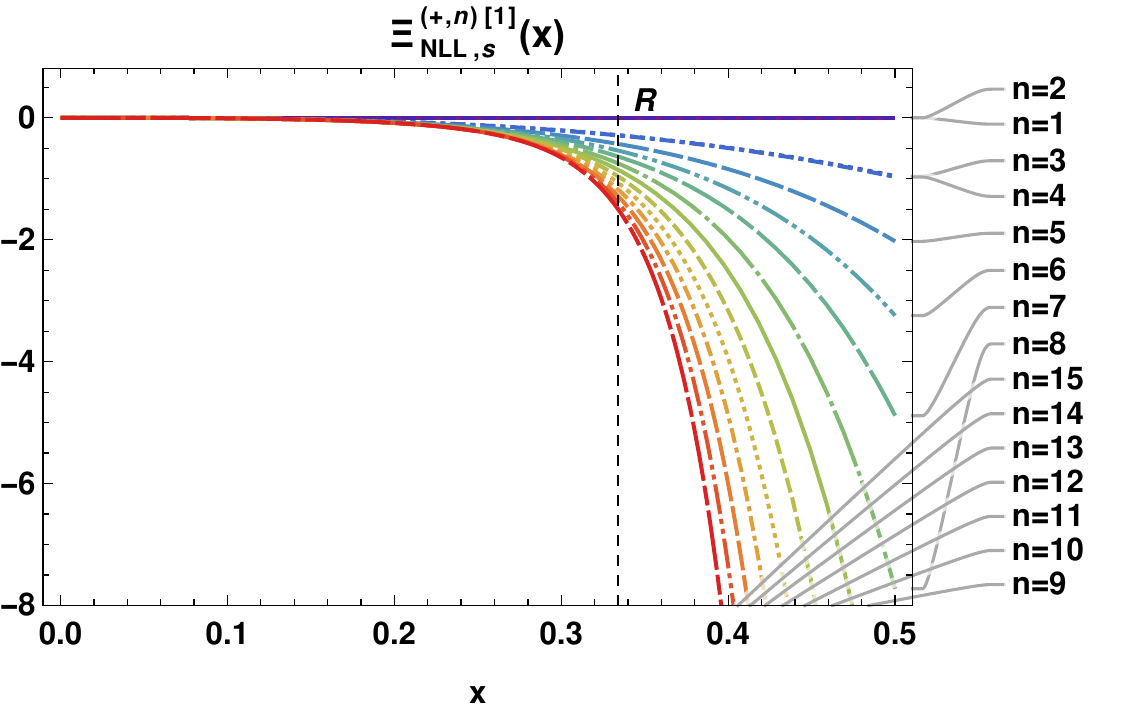}  \\[0.3cm]
	\includegraphics[width=0.7\textwidth]{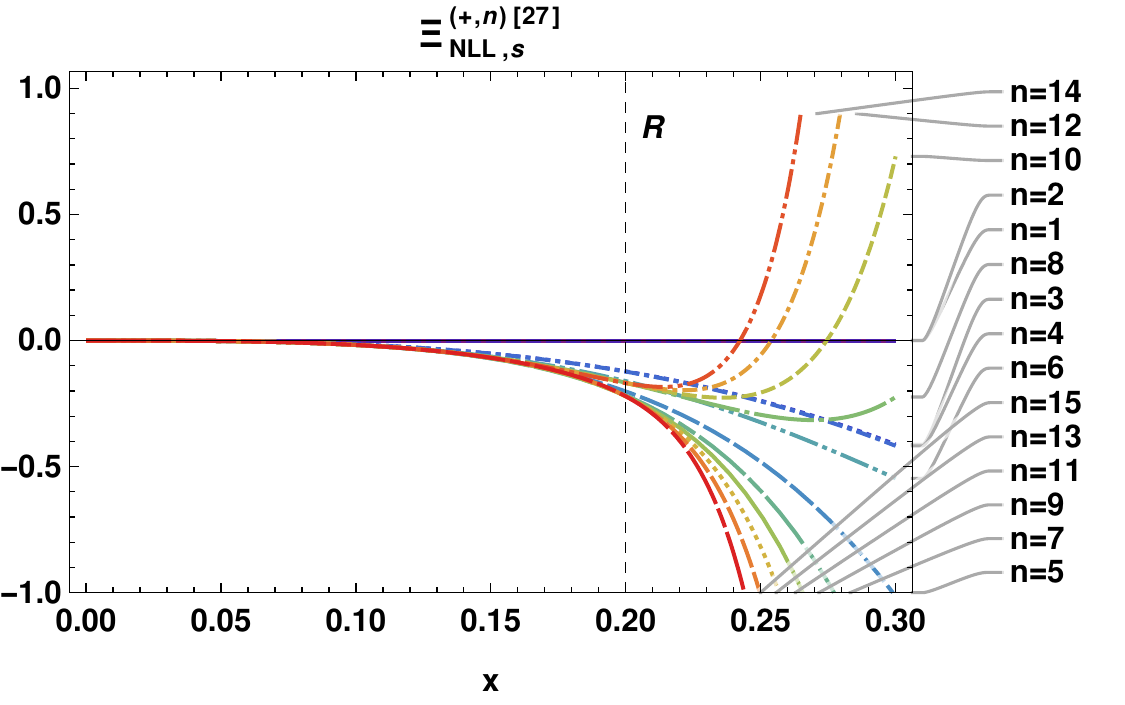}
	\caption{Partial sums of the \emph{soft} component of the 
	infrared-renormalized amplitude  coefficients $\Xi_{\rm NLL,s}^{(+,\ell)}$, 
	up to 15th order, for the singlet (upper plot) and 27 colour representation 
	(lower plot). The dashed vertical line represents the radius of 
	convergence, $R$, determined from the resummed expression.}
\label{Radius-Soft}
\end{figure}
\begin{figure}[htb]
\centering
	\includegraphics[width=0.7\textwidth]{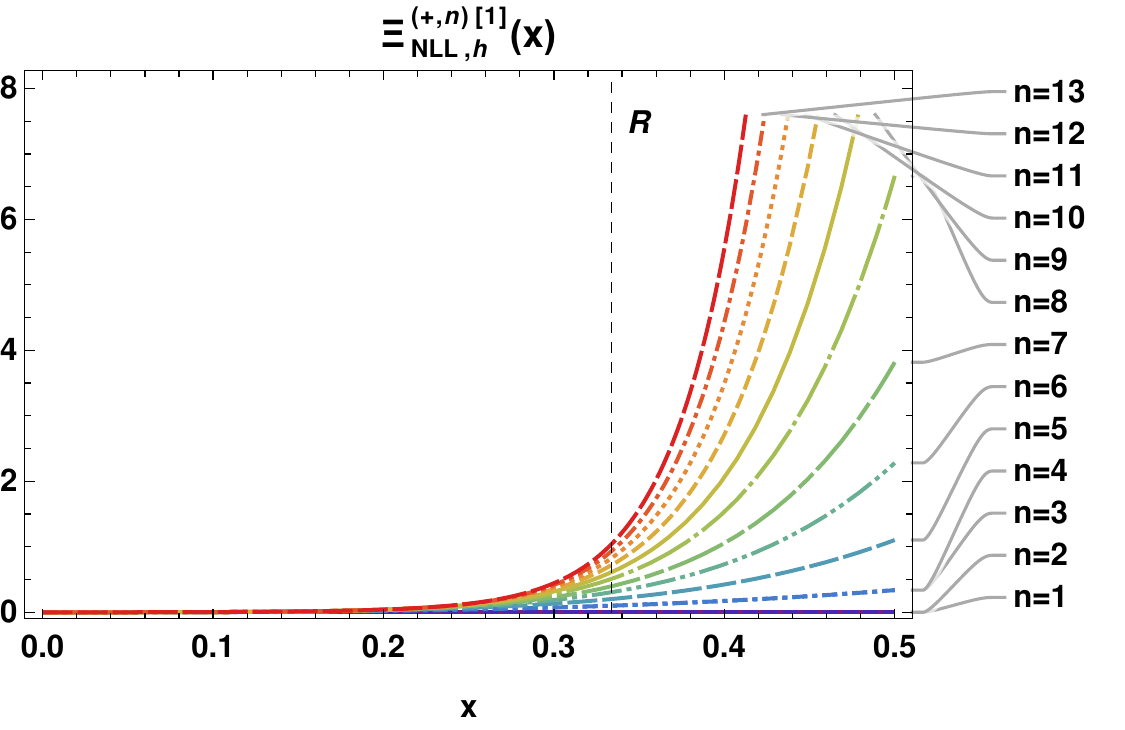}  \\[0.3cm]
	\includegraphics[width=0.7\textwidth]{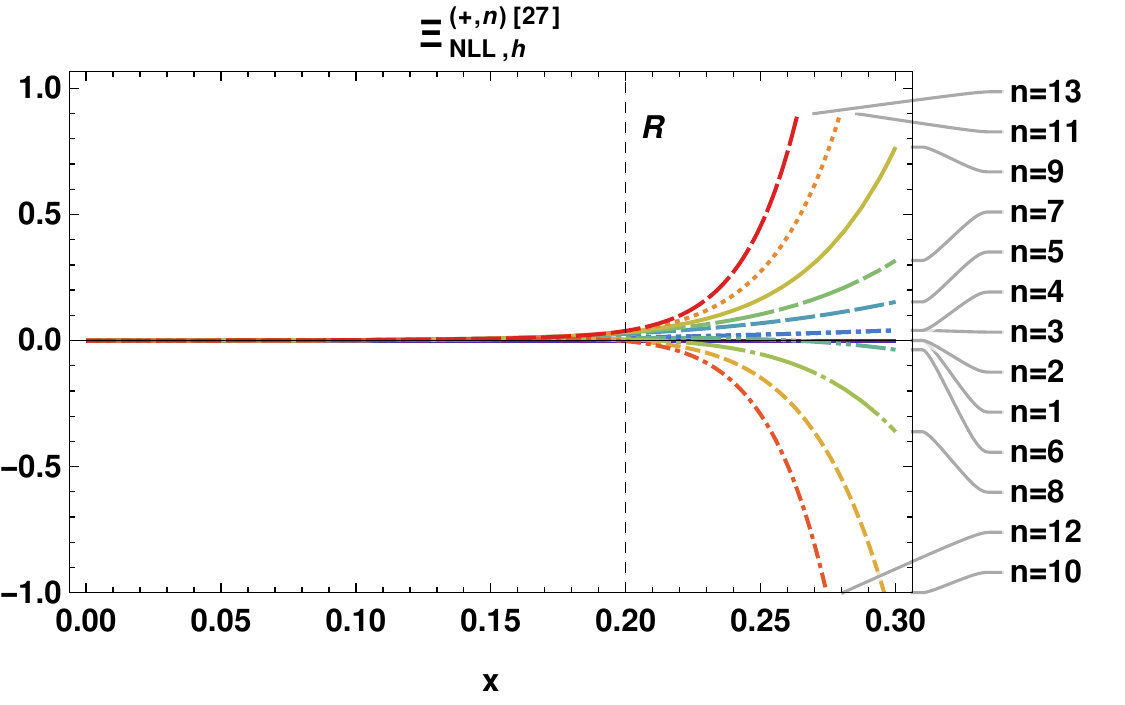}
	\caption{Partial sums of the \emph{hard} component of the 
	infrared-renormalized amplitude  
	coefficients $\Xi_{\rm NLL,h}^{(+,\ell)}$, up to 13th order, for the 
	singlet (upper plot) and 27 colour representation (lower plot). The 
	dashed vertical line represents the radius of convergence, $R$, determined 
	by the pole closest to $x = 0$, using the method of Pad\'{e} approximants.}
\label{Radius-Hard}
\end{figure}
\begin{figure}[bht]
\centering
\includegraphics[width=0.7\textwidth]{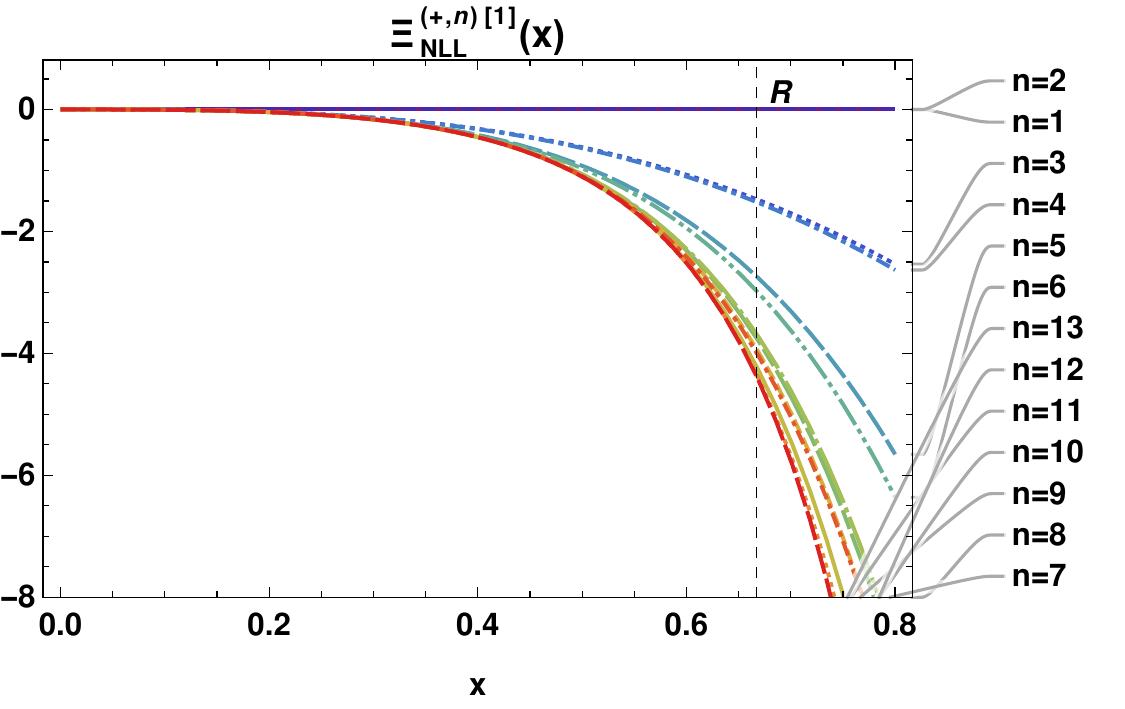}  \\[0.3cm]
\includegraphics[width=0.7\textwidth]{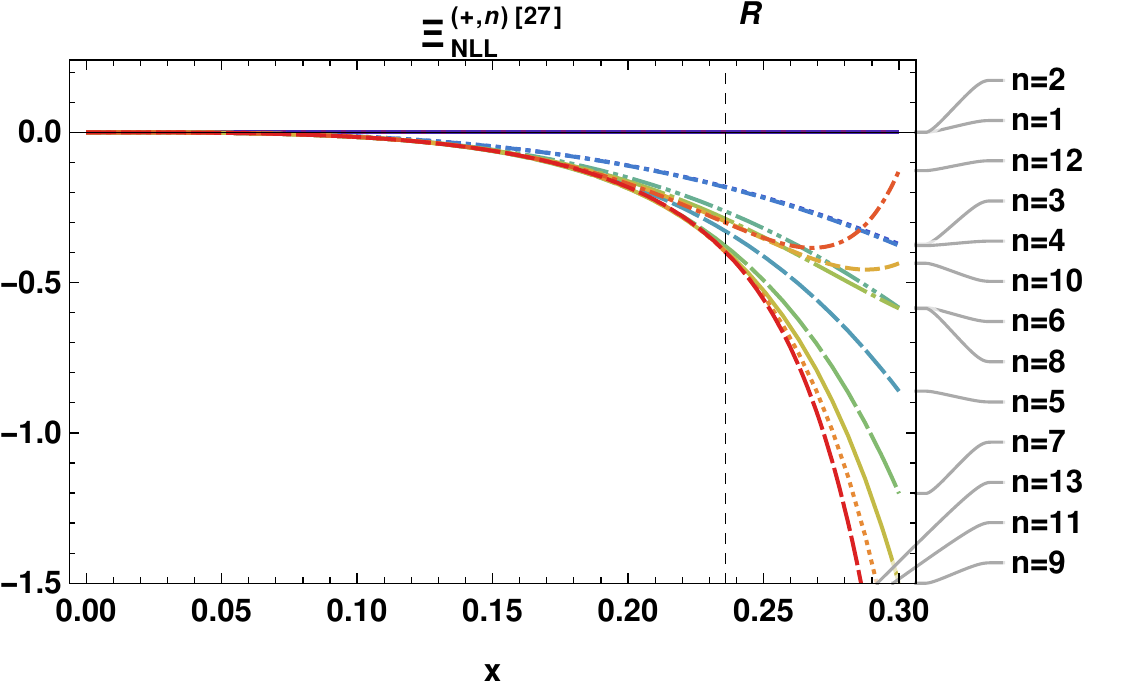}
\caption{Partial sums of the infrared-renormalized amplitude  coefficients 
	$\Xi_{\rm NLL}^{(+,\ell)}$, up to 13th order, for the singlet 
	(upper plot) and the 27 colour representation (lower plot). The 
	dashed vertical line represents the radius of convergence, $R$, 
	determined by the  pole closest to $x = 0$, using the method of 
	Pad\'{e} approximants.}
\label{Radius-Full}
\end{figure}
Having computed finite contributions to the imaginary part of 
the amplitude to high loop orders we are in a position to 
investigate a very interesting theoretical question, namely 
the convergence properties of the perturbative expansion. Of 
course, this is done here at a fixed-logarithmic accuracy, 
namely considering the amplitude as a function of 
$x \equiv L\, \as/\pi$.
The high-energy limit adds an interesting twist to 
the question of convergence, since within the a priori 
``perturbative regime'' where $\alpha_s(\mu^2)$ is small 
(recall that $\mu^2$ is naturally determined by the momentum 
transfer $-t$, and we assume $s\gg-t\gg\Lambda_{\rm QCD}^2$) 
high-energy logarithms \hbox{$L \sim \log |s/t|$} can be 
arbitrarily large, but then the effective expansion parameter 
$x \equiv L\, \as/\pi$ becomes large. Thus, while there is no 
obvious reason why perturbation theory should break down, the 
question arises whether we can extend the validity of the 
calculation to large values of the expansion parameter $x$. 
In \cite{Caron-Huot:2017zfo} we studied the infrared-divergent 
part of the amplitude in detail, and proved that these corrections 
exponentiate in terms of the soft anomalous dimension. We 
determined the latter to all orders in perturbation theory, 
and shown that it is an entire function, having \emph{an 
	infinite radius of convergence} in $x$. 

We are now in a position to study the convergence of 
the infrared-renormalized amplitude ${\cal H}_{\rm NLL}^{(+)}$, 
which we determined analytically 
to the 13th order in section \ref{amplitude}. For 
convenience, we introduce the amplitude $\Xi$ and 
its coefficients $\Xi^{(\ell)}$, defined through
\begin{align} \label{MhatEvenB}
	\begin{split}
		{\cal H}_{\rm NLL}^{(+)}  & 
		= \frac{i \pi}{L} \, \Xi_{\rm NLL}^{(+)} \, \Tsu \Mtree\,  \\
		&  = \frac{1}{L} \sum_{\ell=1}^\infty  x^\ell  {\cal H}_{\rm NLL}^{(+,\ell)}
		= \frac{i \pi}{L} \sum_{\ell=1}^\infty x^\ell \, \Xi_{\rm NLL}^{(+,\ell)} \, \Tsu \Mtree\, , 
	\end{split}
\end{align}
such that 
\beq  \label{MhatEvenC}
{\cal H}_{\rm NLL}^{(+,\ell)}
=  i \pi \, \Xi_{\rm NLL}^{(+,\ell)} \, \Tsu \Mtree\, ,
\eeq
{\it cf.} \eqns{MhatEven}{ReducedAmpNLL2}. In the following 
we will use equivalent definitions also for the soft and hard 
parts of the infrared-renormalized amplitude coefficients. 

Numerical expressions for the coefficients of the 
infrared-renormalized amplitude
$\Xi_{\rm NLL}^{(+,\ell)}$ up to thirteen loops 
can be obtained starting from the analytic expressions 
given in the ancillary files\footnote{The same result is 
provided explicitly in the main text in 
\eqnss{eq:HExpanded_1}{eq:HExpanded_8}, 
up to eight loops.}, using the relations in 
\eqn{Tt2num}, and converting the multiple zeta 
values there into decimal numbers. We arrive at
\beqa \label{Xi1} \nn
\Xi_{\rm NLL}^{(+)[1]} &=& -4.959 \, x^3 
- 0.2283 \, x^4 - 9.230 \, x^5 - 2.690 \, x^6  - 13.13 \, x^7 + 1.696 \, x^8 \\
&&\hspace*{5pt}-\, 20.44 \,  x^9 + 16.54\,  x^{10} - 35.99 \, x^{11} + 46.06 \, x^{12} 
- 74.05 \, x^{13} + \ord(x^{14}), \\[0.2cm] \nn
\label{Xi27}
\Xi_{\rm NLL}^{(+)[27]} &=& -13.77 \, x^3 
- 0.6342 \, x^4 - 199.2 \, x^5 + 381.1 \, x^6 - 2826 \, x^7  + 9380 \,  x^8 \\ 
&&\hspace*{5pt} -\, 46488 \, x^9  + 180393 \, x^{10}  - 797524 \, x^{11} 
+ 3.239 \times 10^{6} \,  x^{12} - 1.374 \times 10^{7} \, x^{13} \nonumber 
\\&&\hspace*{5pt}+ \,\ord(x^{14}). 
\eeqa
We consider also the soft and hard contribution to the 
infrared-renormalized amplitude  ${\cal H}_{\rm NLL}^{(+)}$, 
defined by the two terms in \eqn{Hsplit}.
Defining the soft $\Xi_{\rm NLL,s}^{(+)}$ and hard 
$\Xi_{\rm NLL,h}^{(+)}$, in analogy to \eqns{MhatEvenB}{MhatEvenC}, 
we can easily obtain a numerical expression for the singlet and 27 
colour representation, as in \eqns{Xi1}{Xi27}:  
\beqa \label{Xis1} \nn
\Xi_{\rm NLL,s}^{(+)[1]} &=&  -7.663\, x^3 
- 0.2283 \, x^4 - 33.73 \, x^5 - 78.04 \, x^6 - 210.0 \, x^7 - 726.9 \, x^8 \\
&&\hspace*{5pt}-\, 2023 \, x^9 - 6237 \,  x^{10} - 18605 \, x^{11} - 55822 \, x^{12} 
- 167566  \, x^{13} + \ord(x^{14}), \\[0.2cm] \nn
\label{Xis27}
\Xi_{\rm NLL,s}^{(+)[27]} &=&  -15.28\, x^3 
- 0.6342 \, x^4  - 245.7\, x^5 + 641.8 \, x^6 - 4445 \, x^7 + 19735  \, x^8 \\ \nn
&&\hspace*{5pt}-\, 103863  \,  x^9 + 507855   \,  x^{10} - 2.566 \times 10^{6}  \, x^{11}  
+ 1.277   \times 10^{7} \, x^{12} \\
&&\hspace*{10pt} -\, 6.398  \times 10^{7} \, x^{13}  + \ord(x^{14}), \,\,\,\,
\eeqa
for the soft part of the infrared-renormalized amplitude, and 
\beqa \label{Xih1} \nn
\Xi_{\rm NLL,h}^{(+)[1]} &=& 2.705\, x^3 + 24.50 \, x^5 + 75.34 \, x^6 
+ 196.9  \, x^7 + 728.6 \, x^8 + 2003 \, x^9 \\
&&\hspace*{5pt}+\,  6254 \, x^{10} + 18570 \, x^{11} + 55869  \, x^{12} 
+ 167492 \, x^{13} + \ord(x^{14}), \\[0.2cm] \nn
\label{Xih27}
\Xi_{\rm NLL,h}^{(+)[27]} &=& 1.503 \, x^3 + 46.45 \, x^5 - 260.6 \, x^6 
+ 1619 \, x^7 - 10356 \, x^8 + 57375 \,  x^9  \\ \nn
&&\hspace*{5pt}-\,  327462 \,  x^{10} + 1.768 \times 10^{6}  \, x^{11} 
- 9.527 \times 10^{6} \, x^{12} \\
&&\hspace*{10pt}+\, 5.024 \times 10^{7} \, x^{13} + \ord(x^{14}),
\eeqa
for its hard part. We plot the partial 
sums of the soft, hard and full infrared-renormalized amplitude 
as a function of $x$ respectively in 
figures~\ref{Radius-Soft},~\ref{Radius-Hard}, 
and~\ref{Radius-Full}.

Considering the all-order resummed 
expression for the soft component of the infrared-renormalized 
amplitude  in \eqn{getH4}, we can immediately  conclude that it 
exhibits a finite radius of convergence. The radius of convergence 
can be identified as the position of the pole closest to the origin 
in the complex $x$ plane, which we denote in what follows $R$. 
Inspecting \eqn{getH4}, and in particular the explicit expression 
for $\hat \Delta_{\rm NLL}^{(+)}$ in \eqn{DeltaResum2}, we see 
that the soft part of the infrared-renormalized amplitude has poles 
when the argument of the gamma functions in the numerator equals 
zero or negative integers. In general poles appear for both positive 
and negative $x$: this is an important point we shall return to below.
The pole closest to the origin is determined by $1- \Ctwo\, x = 0$, 
which in turn determines the radius of the convergence of the soft 
part of the infrared-renormalized amplitude to be $R_{ s}= 1/\Ctwo$ 
(the subscript ``s'' refers to the soft part of the infrared-renormalized 
amplitude).  This corresponds to $R_{s} = 1/3 \simeq 0.333$ for the 
colour singlet infrared-renormalized amplitude, 
and $R_{s} = -1/5 = -0.2$ for one in the 27 representation. 
The qualitative picture of convergence of the partial sums as a function 
of $\ell$ for any $x<R_s$, and divergence beyond that point, can indeed 
be confirmed upon inspecting figure~\ref{Radius-Soft}.

For the hard contribution to the infrared-renormalized amplitude, 
and thus also for the complete one, we do not have an 
all-order expression. Nevertheless, information on the radius 
of convergence can be extracted from the perturbative expansion
by constructing Pad\'{e} approximants of the infrared-renormalized 
amplitude. More specifically, we may use the partial sum of the 
infrared-renormalized amplitude at any order $\ell$ to construct 
a rational function of~$x$, which reproduces the partial sum upon 
expansion. Here 
we choose to use Pad\'e approximants of the form\footnote{There 
	is of course some freedom of in choosing the degrees of the 
	polynomials in the numerator and the denominator. After some 
	experimentation we found that Pad\'{e} approximants with 
	second-order denominators yield stable predictions for the 
	position of the first pole already at relatively low orders, 
	and hence we use this form as the default choice for the 
	analysis presented here. Qualitatively, the results are 
	the same using different Pad\'{e} approximants.}:
\be\label{padeDef}
\Xi_{\rm NLL}^{(+)}|_{\text{Pad\'{e}},\ell} = 
\frac{\sum_{n = 3}^{\ell-2} \, a_n \, x^n}{1+ b_1 \,x + b_2 \,x^2}. 
\ee
With this definition the Pad\'{e} approximant has two poles
at $x_{\pm} = \big(-b_1 \pm \sqrt{b_1^2 - 4 b_2}\big)/(2 b_2)$.
The pole closest to the origin provides a prediction for the 
radius of convergence of the series: $R = {\rm min}\{ x_{-}, x_+ \}$.
Of course, this prediction is expected to be reliable only upon 
considering sufficiently high orders, where the series approaches 
its asymptotic regime. The stability of the deduced value for the 
radius of convergence with respect to the order $\ell$ provides 
an indication of whether the asymptotic regime is reached.

Before describing the results a further comment is due regarding 
the sign of $R$. Strictly speaking, the radius of convergence 
would be the absolute value of $R$. Here, however, we are interested 
in keeping track also of the sign of the nearest pole, indicating 
whether the series is (asymptotically) of constant signs, or 
oscillating. We shall see that both scenarios are realised.

We test this method on the soft part of the infrared-renormalized 
amplitude, for which the all-order result is known, as discussed 
above. The results for the singlet and the 27 representation for 
$N_c=3$ are shown in Table \ref{tableXminSoft}. We see that 
in both cases the pole closest to $x = 0$ ($x^{[1]}_{\rm s,-} 
= 0.333$ for the singlet and $x^{[27]}_{\rm s,+} =  -0.200$ 
for the 27 representation) approximates very well the exact 
radius of convergence, $R_s = 1/\Ctwo$.
\begin{table}[ht]
\begin{center}
{\small
	\begin{tabular}{|l|l|l|l|l|}
		\hline
		$\ell$&                  singlet            &           27           \\ \hline \hline
		10 &    $x^{[1]}_{\rm s,-} = 0.335, \quad x^{[1]}_{\rm s,+} =  -0.703$     & $x^{[27]}_{\rm s,-} = 0.472, \quad x^{[27]}_{\rm s,+} =  -0.200$   \\ \hline
		11 &   $x^{[1]}_{\rm s,-} = 0.333, \quad x^{[1]}_{\rm s,+} =  -1.053$ &                    $x^{[27]}_{\rm s,-} = 0.446, \quad x^{[27]}_{\rm s,+} =  -0.200$\\  \hline
		12 &   $x^{[1]}_{\rm s,-} = 0.334, \quad x^{[1]}_{\rm s,+} =  -1.866$  &     $x^{[27]}_{\rm s,-} = 0.428, \quad x^{[27]}_{\rm s,+} =  -0.200$               \\  \hline
		13 & $x^{[1]}_{\rm s,-} = 0.333, \quad x^{[1]}_{\rm s,+} =   3.911$&   			 $x^{[27]}_{\rm s,-} = 0.419, \quad x^{[27]}_{\rm s,+} =  -0.200$   \\  \hline
	\end{tabular}}
\end{center}
\caption{Table summarising the values of $x=\frac{\alpha_s}{\pi}L$ 
	at the poles of the Pad\'e Approximants in eq.~(\ref{padeDef}), 
	considering the \emph{soft} component of the infrared-renormalized amplitude (indicated by the subscript ${\rm s}$) at orders $\ell=10$ 
	through 13 for the singlet and the 27 representation.}
\label{tableXminSoft}
\end{table}

We thus proceed and apply the same method to the hard component of
the infrared-renormalized amplitude. The results are summarised in table~\ref{tableXminHard}.
\begin{table}[ht]
\begin{center}
{\small
	\begin{tabular}{|l|l|l|l|l|}
		\hline
		$\ell$&                  singlet            &           27           \\ \hline\hline
		10 &    $x^{[1]}_{\rm h,-} = 0.333, \quad x^{[1]}_{\rm h,+} =  -0.753$     & $x^{[27]}_{\rm h,-} = 0.822, \quad x^{[27]}_{\rm h,+} =  -0.176$   \\ \hline
		11 &   $x^{[1]}_{\rm h,-} = 0.332, \quad x^{[1]}_{\rm h,+} =  -0.856$ &                    $x^{[27]}_{\rm h,-} = 0.096, \quad x^{[27]}_{\rm h,+} =  -0.179$\\  \hline
		12 &   $x^{[1]}_{\rm h,-} = 0.333, \quad x^{[1]}_{\rm h,+} =  -1.258$  &     $x^{[27]}_{\rm h,-} = -4.392, \quad x^{[27]}_{\rm h,+} =  -0.186$         \\  \hline
		13 & $x^{[1]}_{\rm h,-} = 0.333, \quad x^{[1]}_{\rm h,+} =  -1.244$&   			 $x^{[27]}_{\rm h,-} =-0.02, \quad x^{[27]}_{\rm h,+} = -0.185$   \\  \hline
\end{tabular}}
\end{center}
\caption{Table summarising the values of $x=\frac{\alpha_s}{\pi}L$ 
	at the poles of the Pad\'e Approximants in eq.~(\ref{padeDef}), 
	considering the \emph{hard} component of the infrared-renormalized amplitude  (indicated by the subscript ${\rm h}$) at orders $\ell=10$ 
	through 13 for the singlet and the 27 representation.}
\label{tableXminHard}
\end{table}
We observe that for the singlet there is a highly stable nearest 
pole at $x^{[1]}_{\rm h,-} = 0.333$. For the 27 representation, 
in turn, the stable pole at $x^{[27]}_{\rm h,+} \simeq -0.19$ is 
not always the one closest to the origin, due to the wide fluctuations 
of $x^{[27]}_{\rm h,-}$. Finally, for the complete infrared-renormalized amplitude we 
summarise the results in table~\ref{tableXminFull}. Here we find 
highly stable results: $x^{[1]}_{+} \simeq  -0.66$  and 
$x^{[27]}_{+} \simeq -0.24$.

We conclude that Pad\'{e} approximants based on partial sums 
of order $\ell=10$ through $13$, yield fairly stable predictions 
for the poles. Naturally, ones still finds some fluctuations, which 
can be attributed to subasymptotic  effects, but an overall 
consistent picture emerges, and we can deduce an approximate 
radius of convergence in each case from the position of the poles.
\begin{table}[ht]
\begin{center}
{\small
\begin{tabular}{|l|l|l|l|l|}
	\hline
	$\ell$&                  singlet            &           27           \\ \hline\hline
	10 &    $x^{[1]}_{-} = 1.092, \quad x^{[1]}_{+} =  -0.624$     & $x^{[27]}_{-} = 0.393, \quad x^{[27]}_{+} =  -0.236$   \\ \hline
	11 &   $x^{[1]}_{-} = 1.266, \quad x^{[1]}_{+} =  -0.666$ &                    $x^{[27]}_{-} = 0.437, \quad x^{[27]}_{+} =  -0.237$\\  \hline
	12 &   $x^{[1]}_{-} = 1.311, \quad x^{[1]}_{+} =  -0.661$  &     $x^{[27]}_{-} = 0.367, \quad x^{[27]}_{+} =  -0.238$                \\  \hline
	13 & $x^{[1]}_{-} = 1.466, \quad x^{[1]}_{+} = -0.669$&   			 $x^{[27]}_{-} =0.461, \quad x^{[27]}_{+} = -0.239$            \\  \hline
\end{tabular}}
\end{center}
\caption{Table summarising the values of $x=\frac{\alpha_s}{\pi}L$ 
at the poles of the Pad\'e Approximants in eq.~(\ref{padeDef}), 
considering the \emph{full amplitude} at orders $\ell=10$ through 
13 for the singlet and the 27 representation.}
\label{tableXminFull}
\end{table}

The final results of this analysis are summarised in table 
\ref{tableXmin}, where we compare the results for the soft 
part of the infrared-renormalized amplitude, deduced from the resummed result 
(which are highly consistent with the Pad\'e approach), with 
those for the hard component and complete infrared-renormalized amplitude, which 
are both based solely on the Pad\'e analysis. In the table we 
also provide an interpretation of the radius of convergence 
for the full infrared-renormalized amplitude in terms of the analytic dependence 
on the colour factors $C_1$ and $C_2$; this will be explained 
below.
\begin{table}[ht]
\begin{center}
{\small
	\begin{tabular}{|l||r|r|r|r|} \hline
	&  \multicolumn{2}{c|}{$x=\frac{\as}{\pi}L$ expansion: convergence radius  $R$}  \\ \hline
	Representation &                  singlet            &           27           \\ \hline \hline
	Colour factors   &     $C_1=6$, \quad$C_2=3$    &   $C_1=-2$,\quad $C_2=-5$ \\ \hline \hline
	\shortstack{soft contribution to ${\cal H}_{\rm NLL}^{(+)}$}   &    $1/{C_2}=1/3 $  &   $1/{C_2}=-1/5$   \\ 
	\shortstack{hard contribution ${\cal H}_{\rm NLL}^{(+)}$} &                  $\sim 0.333$  &            $\sim-0.19$ \\ 
	Full ${\cal H}_{\rm NLL}^{(+)}$    &  $-2/{C_2}\simeq -0.666$  &   $1/\big({C_2}- \frac{3}{8}{C_1}\big) \simeq -0.235$  
	\\  \hline
\end{tabular}}
\end{center}
\caption{Summary table for the radius of convergence $R$ of the 
	expansion of the infrared-renormalized amplitude in powers 
	of $x=\frac{\alpha_s}{\pi}L$, determined by identifying the pole 
	closest to $x= 0$ using Pad\'e approximants. We use the 
	shorthand notation $C_1 = \Cone$ and $C_2= \Ctwo$.}
\label{tableXmin}
\end{table}

The numerical results in the table indicate that the radius 
of convergence of the full infrared-renormalized amplitude is 
larger compared to both its soft and hard components. Indeed, 
better convergence is clearly observed  looking at successive 
orders in the full hard function in figure~\ref{Radius-Full} 
compared to its soft and hard components in figures \ref{Radius-Soft} 
and \ref{Radius-Hard}, respectively. The interpretation is clear: 
the pole that limits the convergence of the soft component of the 
infrared-renormalized amplitude in the resummed 
expression,~\eqn{getH4}, exactly cancels against a similar 
divergence in the hard component, hence the 
similar values of $R$ for the soft and hard components in 
table~\ref{tableXmin}. Upon cancelling the leading divergence, 
a subleading pole is exposed, which becomes the dominant 
obstruction for convergence of the full amplitude. 
This is of course another indication that the separation 
of the \emph{finite} ${\cal O}(\epsilon^0)$ terms between 
the soft and hard regimes is arbitrary; we have already 
seen that the soft wavefunction cannot approximate the 
full one away from the soft limit in figures \ref{Wave1} 
and~\ref{Wave27}.  

Even more interesting is the observation that the \emph{sign} 
of the first pole, $R$, which indicates whether the series is 
asymptotically sign-oscillating ($R<0$) or of constant signs 
($R>0$), is negative for the full infrared-renormalized amplitude, 
while it may be either positive or negative for the separate soft 
and hard components, as can be seen in table~\ref{tableXmin}. 
Upon resumming the perturbative expansion of the full 
infrared-renormalized amplitude, one expects a smooth 
extrapolation to high energies when taking the centre-of-mass 
energy large compared to the momentum transfer, 
$s\gg-t\gg\Lambda_{\rm QCD}^2$. Given that
the expansion parameter, $x=\frac{\alpha_s(-t)}{\pi} 
\log\frac{s}{-t}$, gets large (and positive) in this limit, 
smooth extrapolation (of the resummed expression) to high 
energies can only be consistent with a finite radius of 
convergence if the series is sign-oscillating, or put in 
stronger terms: if \emph{all} the singularities of the resummed 
infrared-renormalized amplitude are locate away from the 
positive real axis of $x$. In the example of the soft part of the 
hard function, singularities appear on the real axis at both 
positive and negative values. We expect that this would not 
happen for the full hardfunction. In other words, the 
singularities present in the resummed soft part
of the hard function at positive $x$ \emph{must} all cancel 
against similar divergences in the resummed hard part of the 
hard function. This explains the observations above regarding 
the radius of convergence of the full hard function versus its 
soft and hard components, but it applies more generally, also 
to poles further away from the origin.
\begin{figure}[htb]
\centering
\includegraphics[width=0.75\textwidth]{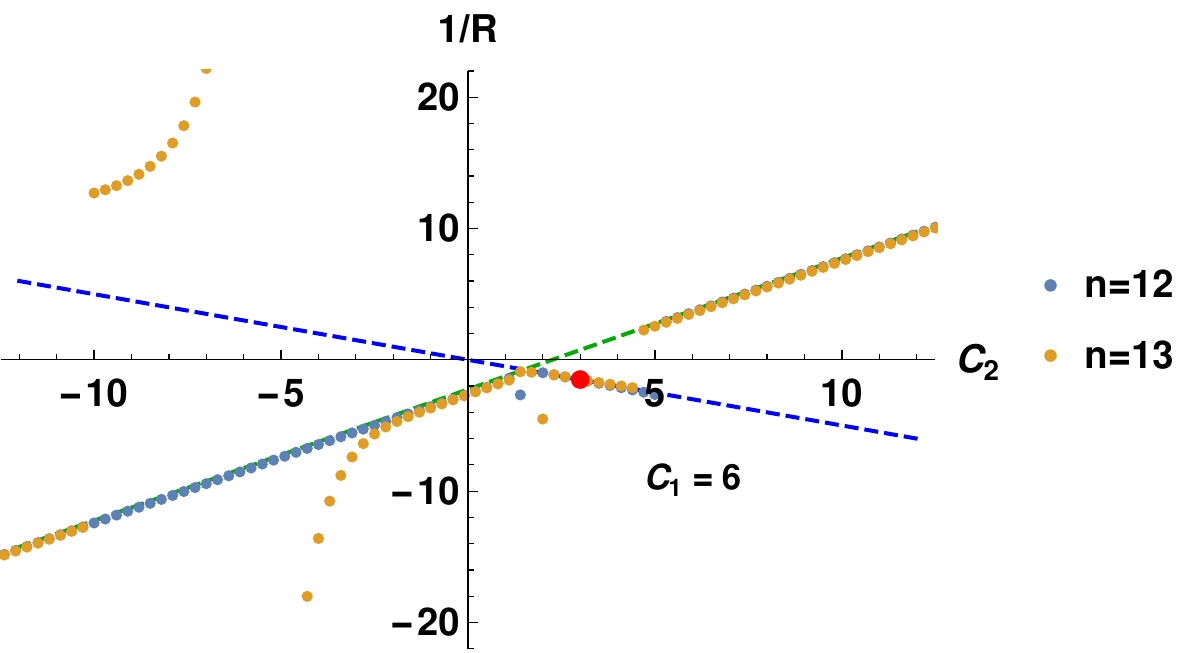}   \\[0.5cm]
\includegraphics[width=0.75\textwidth]{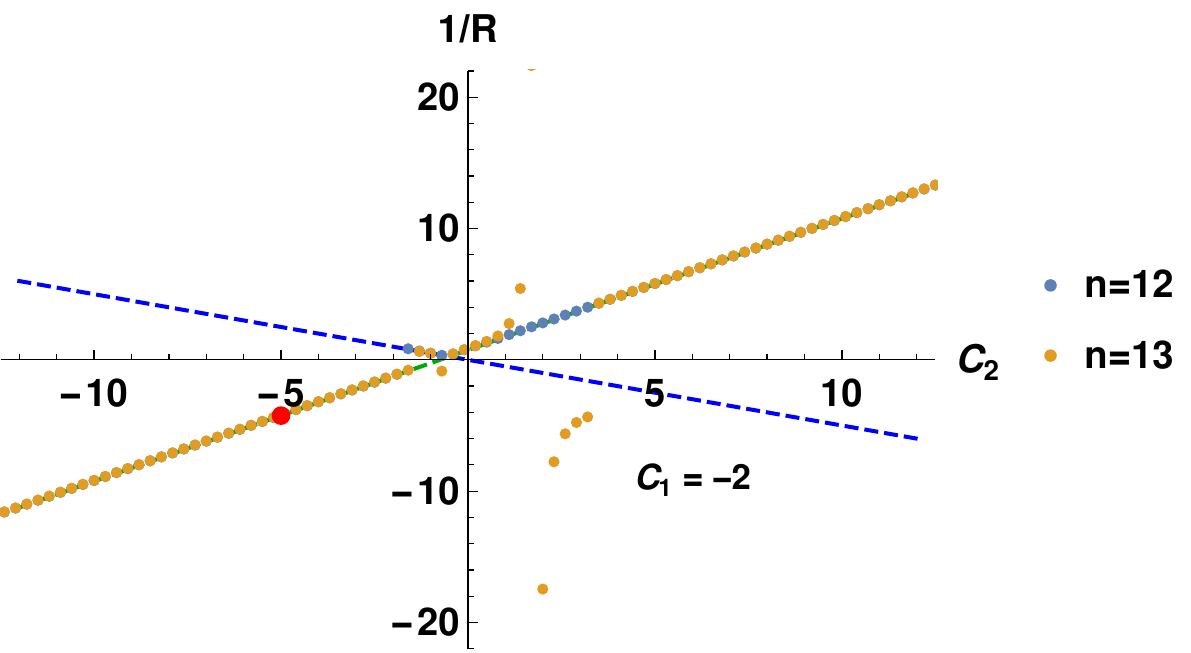}
\caption{The radius of convergence of the full infrared-renormalized amplitude as 
	a function of the colour operators. In these plots the dots 
	represent the value of $1/R$, for the corresponding value of 
	$C_1=\Cone$ and $C_2=\Ctwo$, based on the Pad\'e approximant 
	analysis for $\ell=11$ and $\ell=12$, as indicated in the plots. 
	We superimpose two linear lines, which determine the dependence 
	of $1/R$ on the colour operators, as summarised by 
	eq.~(\ref{Rfull_exact}).}
\label{RR}
\end{figure}

To complete the analysis of the radius of convergence in the 
full hard function we would now like to interpret the numerical 
values of $R$ obtained in the Pad\'e-based analysis in terms of 
the colour structures $C_1=\Cone$ and $C_2=\Ctwo$\footnote{The 
	hard function is provided in \eqns{eq:HExpanded_1}{eq:HExpanded_8}
	in terms of $C_A$ and $C_2$, but we find it more convenient for 
	this analysis to express it as a function of the color
	operators $C_1$ and $C_2$.}. We start by 
recalling that for the soft part of the hard function, 
$R_{ s}= 1/C_2$  depends on $C_2$ only. The Pad\'e-based 
analysis of the 27 colour representation indicates that 
this is not so for the full hard function. To obtain an 
analytic expression it proves useful to depart from the 
actual values of the colour factors corresponding to 
physically-relevant representations, and simply repeat 
the Pad\'e approximant analysis for a range of values 
of $C_2$ for a fixed $C_1$. To this end we plot in 
figure~\ref{RR} the numerical values of $R$ emerging 
from Pad\'e approximants, as a function of $C_2$, for 
fixed values of $C_1$ (we pick $C_1=6$ and $C_1=-2$, 
corresponding to the singlet and the 27 representation, 
for easy reference). More precisely, we display in figure 
\ref{RR} the value of $1/R$ rather than $R$ itself, which 
makes it easy to recognise the exact linear behaviour. 
Based on this analysis we deduce the radius of 
convergence of the full amplitude to be:
\begin{align}
\begin{split}
	\hspace*{-30pt}R=\min \left\{x_a,x_b \right\}\qquad
	\text{with}&\qquad x_a = \frac{1}{C_2 - \frac{3}{8}C_1}
	= \frac{1}{\Ctwo - \frac{3}{8}\Cone}, \\
	&\qquad x_b = - \frac{2}{C_2}=- \frac{2}{\Ctwo}.
	\label{Rfull_exact}
\end{split}
\end{align}
Returning to the physically-relevant representations, 
$x_b$ end up being closest to the origin ($|x_b|<|x_a|$) 
for the singlet representation where $C_1^{[1]} = 6$, 
$C_2^{[1]} = 3$. One then obtains from eq.~(\ref{Rfull_exact}) 
a radius of convergence of $R=x^{[1]}_b \simeq -0.6667$, in 
accordance with the result in table \ref{tableXmin}.
In turn, $x_a$ gives the pole closest to the origin for 
the 27 representation, namely for $C_1^{[27]} = -2$, 
$C_2^{[27]} = -5$, where one obtains from 
eq.~(\ref{Rfull_exact}) $x^{[27]}_a \simeq -0.235$, 
again, in accordance with table \ref{tableXmin}.


\section{Conclusions}  \label{conclusion}

In this paper we completed the perturbative calculation of $2\to 2$ 
partonic amplitudes at the next-to-leading logarithmic accuracy in 
the Regge limit to high loop orders. We focused on the previously-unknown
even-signature terms, corresponding to the imaginary part of the 
amplitude, which vanishes at the leading-logarithmic accuracy.
Building upon our previous work in ref.~\cite{Caron-Huot:2017zfo} 
where we determined the infrared singularities, we now computed 
the finite corrections to the hard amplitude~${\cal H}$, which remain 
after stripping off, or renormalizing, these singularities.  
We believe that these results --- the soft anomalous dimension 
and hard functions --- together exhaust the physical information 
contained in these partonic amplitudes.

Our results are based on the well-established BFKL evolution 
equation in momentum space. Since the even amplitude vanishes 
at the leading-logarithmic order, only the leading-order BFKL 
evolution kernel was needed in our calculation, and the final 
formulae apply equally to quark and gluon amplitudes.
We exploited the fact, observed in~\cite{Caron-Huot:2017zfo}, 
that the two-Reggeon wavefunction is finite. While it is unknown 
how to diagonalise the BFKL Hamiltonian for arbitrary colour 
structures beyond the planar limit, we were able to solve the 
BFKL equation iteratively, treating complementary regions 
using two different approaches. The first relies on the soft 
approximation keeping the dimensional regularization parameter 
finite -- the same method we used in ref.~\cite{Caron-Huot:2017zfo} 
to determine the singularities of the amplitude -- while the second 
relies instead on a computation in exactly two transverse 
dimensions, which captures general hard momentum 
configurations where both Reggeons carry momenta 
of the order of the total momentum transfer $p^2=-t$.
As shown in eqs.~(\ref{getH2})~and~(\ref{eq:redampSplit}), 
each separated part of the BFKL-motivated reduced amplitude 
needs only be calculated to order ${\cal O}(\epsilon^0)$, and 
by carefully recombining them we obtained the renormalized 
amplitude in eq.~(\ref{eq:HExpanded}). The result passes 
several consistency checks and agrees with a direct 
computation in dimensional regularization, which we 
performed through five loops.

The central new computation in this paper is the iterative 
solution of the BFKL equation in two dimensions, leading 
to a simple algorithm to compute the two-Reggeon 
wavefunction to any order, presented in section \ref{2d-bfkl}.
The result lives inside a very rigid space of functions:
the $\ell$-loop wavefunction is a linear combination of 
weight-$\ell$ single-valued harmonic polylogarithms 
(SVHPLs) of $z$ and $\zb$ with rational coefficients.
The algorithm is formulated as an operation on SVHPLs, 
and it works by producing differential equations in the 
holomorphic variable~$z$ that can be directly integrated 
in terms of HPLs of $z$, to which we subsequently apply 
the single-value map to recover the actual wavefunction 
in terms of SVHPLs of $z$ and $\zb$. 

The hard contribution to the infrared-renormalized amplitude 
${\cal H}$ computed using the two-dimensional method admits
a rather complex structure, and its resummation goes beyond 
the scope of the present paper. This is to be contrasted with 
the soft contribution, which we could resum to all orders
in terms of gamma functions, eq.~(\ref{ReducedAmpNLLresum2B}), 
which includes singular as well as finite corrections. The 
number-theoretical content of the hard contribution is interesting: 
by construction it is restricted to single-valued multi zeta values 
(see eqs.~(\ref{reduced_hard_results}) and (\ref{eq:m11g533})).
The presence of multi zeta values -- which make their first 
appearance at weight 11 involving a single-valued 
version~\cite{Brown:2013gia,Schnetz:2013hqa} of 
$\zeta_{5,3,3}$  -- precludes resummation in terms 
of gamma functions, so the resummed result would 
clearly be of different nature to that of the soft 
contribution.

Having obtained explicit analytic expressions for both 
the two-Reggeon wavefunction and the infrared-renormalized 
amplitude ${\cal H}$ to high loop order, it is straightforward to 
study the results numerically. In section~\ref{numerics} we 
examine a couple of aspects, first considering the wavefunction 
and then the infrared-renormalized amplitude. The wavefunction 
manifests highly regular behaviour as a function of the Reggeon 
kinematics variables, except for three specific limits. Two of these 
correspond to the soft limits, $z,\zb\to 0$ and $z,\zb\to \infty$, while 
the third $z,\zb \to 1$ corresponds to the limit of large internal 
momentum. The former are described analytically by the soft 
wavefunction in eq.~(\ref{eq:wffullsoftresummed}), and by 
definition the hard wavefunction vanishes there, while a peak 
at large momentum is present in both the soft and hard 
wavefunctions. Interestingly, there is a significant -- but 
incomplete -- cancellation between these two leading to a 
more modest peak in the full wavefunction.  While this 
phenomenon does not affect the validity of our results,
it would be interesting to independently predict this limit 
of the wavefunction (extending eq.~(\ref{asymptotic-resum1}))
which could help find simpler numerical approximations.

Considering the infrared-renormalized amplitude we focused 
on one interesting problem, namely the convergence of the 
perturbative expansion. We find that the ${\cal O}(\epsilon^0)$ 
infrared-renormalized amplitude has a finite radius of 
convergence in the expansion parameter $x=L\alpha_s/\pi$. 
For the soft contribution, where we have a resummed analytic 
expression, eq.~(\ref{ReducedAmpNLLresum2}), this radius of 
convergence can readily be identified as the first pole of a 
gamma function, generating asymptotic behaviour 
$\sim (x(C_A-\Tt))^{\ell}$ at high orders, $\ell\to \infty$.
The soft contribution is however not physically meaningful 
on its own, and the complete infrared-renormalized amplitude
features a larger radius of convergence, as shown in figure 
\ref{Radius-Full} (compare with figures \ref{Radius-Soft} and 
\ref{Radius-Hard} for the separate soft and hard components).
Estimating the convergence radius using Pad\'e approximants 
for different colour channels, we deduced an empirical formula 
for the radius of convergence $R$ of the full amplitude in terms 
of $C_A$ and $\Tt$, eq.~(\ref{Rfull_exact}) above.
Interestingly, the pole closest to the origin is always on the 
negative real axis, leading to an asymptotic behaviour of 
alternating signs. This matches our physical expectation that 
the resummed expression should smoothly extrapolate to high 
energies, corresponding to large positive values of $x$, and 
is similar to what was observed previously for non-global 
logarithms in ref.~\cite{Larkoski:2016zzc}. It remains for 
future work to understand the true high-energy 
(large $x=L\alpha_s/\pi$) behaviour.

Let us conclude with a brief summary of the state-of-the-art 
knowledge of partonic $2\to 2$ scattering amplitudes in the 
Regge limit. With the completion of this work these amplitudes 
are known in full to NLL accuracy. The signature odd part, 
corresponding to the exchange of a single Reggeized gluon 
was already known, and is given by a \emph{Regge pole} 
(\ref{Mreal}) with two-loop corrections to the trajectory 
$\alpha^{(2)}_g(p^2)$, and suitable impact factors (the former, 
in particular, was calculated in 
\cite{Fadin:1995xg,Fadin:1996tb,Fadin:1995km,Blumlein:1998ib};
it can also be extracted from two-loop calculations
of $2\to 2$ scattering amplitudes~\cite{DelDuca:2001gu}).
The signature even part, corresponding to a pair of Reggeized 
gluons, which generate a \emph{Regge cut}, was determined 
here. The next frontier is therefore NNLL accuracy. In the 
signature-odd sector the first step was taken in 
ref.~\cite{Caron-Huot:2017fxr}, where  the non-linear 
Balitsky-JIMWLK equation was used to compute the Regge 
cut contribution generated through the evolution of three 
Reggeized gluons and their mixing with one Reggeon 
through three loops. It is very interesting, and indeed -- 
using the techniques we developed in the present paper 
-- technically feasible, to compute higher-loop corrections 
in this tower of logarithms.  NNLL corrections in the 
signature-even sector are in turn simpler and can be deduced 
from  linear BFKL evolution with a NLO 
kernel~\cite{Fadin:1995xg,Fadin:1996tb,Fadin:1995km,Blumlein:1998ib}, 
supplemented by suitable impact factors.  At N$^3$LL one 
expects new phenomena such as the mixing of two and 
four Reggeon states, which can again be computed 
using the Balitsky-JIMWLK equation.

Finally, beyond their immediate relevance to the the 
study of the high-energy limit, the results in this paper 
can be used to check future multi-loop calculations, 
and ultimately serve as ``boundary data'' in a bootstrap 
programme in which amplitudes are deduced using 
knowledge of the space of functions, analytic properties, 
symmetries and special kinematic limits. Such a 
programme was highly successful in the context of 
${\cal N}=4$ supersymmetric Yang-Mills theory, see 
e.g.~\cite{Caron-Huot:2016owq,Dixon:2016nkn}, but 
also, more recently in the context of the singularity 
structure of gauge theories including 
QCD~\cite{Almelid:2017qju}. In both cases, 
the high-energy limit served as crucial input.


\vspace{20pt}
\acknowledgments

We would like to thank Lorenzo Magnea, 
Jenni Smillie and Claude Duhr for useful discussions. 
We would also like to thank Gudrun Heinrich and Stephan
Jahn for helpful feedback on the numerical evaluations of 
integrals with pySecDec. 
EG's research is supported by the STFC Consolidated Grant 
`Particle Physics at the Higgs Centre', JR's research was 
supported by the Walter Nimmo and Walter Scott PhD studentship,
LV research is supported by Fellini - Fellowship for Innovation 
at INFN, funded by the European Union's Horizon 2020 research 
programme under the Marie Sk\l{}odowska-Curie Cofund Action, 
grant agreement no. 754496. 
EG and JR would like to thank the McGill Physics department for 
their kind hospitality. EG, SCH and LV would like to thank the GGI, 
Florence, for support during the programme `Amplitudes in the LHC 
era'  in Autumn 2018. EG thanks the Simons Foundation for support 
as a Simons Visiting Scientist in GGI in Autumn 2018 and the CERN 
theory department for hospitality as a Scientific Associate in 2019. 
This research is also supported by the National Science 
and Engineering Council of Canada, the Canada Research Chair 
program, and the Fonds de Recherche du Qu\'ebec - Nature 
et Technologies.



\appendix

\section{Harmonic polylogarithms} \label{app:hpls} 

Harmonic polylogarithms (\ac{hpls}) \cite{Remiddi:1999ew} extend the 
natural logarithm $\log z$ with $z \in \mathbb{C}$ to nested integrals. 
Similarly to the well-known polylogarithms $\Li_n(z)$ they are defined 
recursively namely
\beq
\label{eq:hpldef} H_{0,\sigma}(z) = \int_0^z \dd t 
\frac{H_\sigma(t)}{t} \qquad \text{and} \qquad H_{1,\sigma}(z) 
= \int_0^z \dd t \frac{H_\sigma(t)}{1-t}
\eeq
where $\sigma$ is a ``word'' of any length made from the 
letters\footnote{The full alphabet of \ac{hpls} includes the 
letter $-1$. In the present work however we only encounter 
integrals corresponding to the letters $0$ and $1$.} $\{0,1\}$. 
The number of indices of a \ac{hpl} $H(z)$ is called the 
\emph{weight} of the function. By means of 
eq.~\eqref{eq:hpldef} it corresponds to the number 
of nested integrals. The recursion is closed by 
the weight-1 identities
\beq
\label{eq:hpl1def}
H_{0}(z) = \log z \qquad \text{and} \qquad H_{1}(z) = -\log (1-z).
\eeq

\ac{hpls} form a shuffle algebra and thus obey shuffle 
product identities
\beq
\label{eq:hplshuffle} 
H_{\rho}(z) H_{\sigma}(z) 
= \sum_{\tau \in \rho \shuffle \sigma} H_\tau(z)
\eeq
where $\rho \shuffle \sigma$ denotes the shuffle of the 
words $\rho$ and $\sigma$.

The indices of a \ac{hpl} may be shortened by means 
of a collapsed notation; one replaces strings of zeros 
followed by a one according to
\beq
\underbrace{0,0,\dots,0}_{n \text{ zeros}},1 
\longrightarrow n+1 \label{eq:collapsedn}
\eeq
for example $H_{0,1,0,0,1,1}(z) \to H_{2,3,1}(z)$. In the 
collapsed notation the number of indices is referred to 
as the \emph{depth} of the function (while their sum 
now equals the weight).

Depending on the context it may be useful to view 
the \ac{hpls} as nested sums. One commonly used 
definition is
\beq
\label{eq:hplseriesdef}
H_\sigma(z) = \sum_{j=1}^\infty z^j Z_j(\sigma)
\eeq
with
\beq
\label{eq:zsumdef}
Z_j(a,\sigma) = \frac{1}{j^a} \sum_{i=2}^j Z_{i-1}(\sigma) 
\qquad \text{and} \qquad Z_j(1) = 1/j
\eeq
where we assume the collapsed notation. Note that the 
aforementioned depth is equal to the number of nested 
sums.

The Taylor series of \ac{hpls}, defined by eq.~\eqref{eq:hpldef}, 
whose rightmost index is non-zero, is given by 
eq.~\eqref{eq:hplseriesdef} with \eqref{eq:zsumdef}. 
Trailing zeros in the indices of a \ac{hpl} point to 
logarithmic divergences at $z=0$. The $\log z = H_0(z)$ 
terms can be exposed using the shuffle algebra; one 
considers
\beq
H_\sigma(z) H_0(z) = H_{\sigma,0}(z) + \ldots + H_{0,\sigma}(z)
\eeq
and solves for $H_{\sigma,0}(z)$. This procedure 
can be applied recursivly until all trailing zeros are 
removed. Hence, \ac{hpls} can always be written 
as a series in $z$ and $\log z$.

For arguments between 0 and 1 \ac{hpls} yield 
real values. They show branch cuts on the real 
axis where $z \in [1,\infty)$ and are thus 
multi-valued functions.

\section{Single-valued harmonic polylogarithms}
\label{app:svhpls} Single-valued harmonic polylogarithms 
(\ac{svhpls}) \cite{Brown:2004ugm} are the class of all 
branch cut-free, single-valued, combinations of \ac{hpls}. 
Their construction is somewhat involved and we will only 
provide a short summary here. Further details can be 
found in e.g.~refs.~\cite{Pennington:2012zj,Dixon:2012yy,DelDuca:2013lma}. 

\ac{svhpls} are functions of a complex variable $z$ and 
its complex conjugate $\zb$. They correspond to the 
linear combinations of $H_\sigma(z) H_{\sigma'}(\zb)$ 
that solve 
\beq
\label{eq:ddzL} \ddz \El_{0,\sigma}(z,\zb) 
= \frac{\El_\sigma(z,\zb)}{z} \qquad \text{and} 
\qquad \ddz \El_{1,\sigma}(z,\zb) 
= \frac{\El_\sigma(z,\zb)}{1-z}
\eeq
and obey the boundary conditions \cite{Pennington:2012zj}
\beq
\El_\varnothing(z,\zb) = 1, \qquad \El_{\vec 0_n}(z,\zb) 
= \log^n (z \zb) / n! \qquad \text{and} \qquad 
\lim_{z \to 0} \El_{\sigma \neq \vec 0_n}(z,\zb) = 0.
\eeq

For the explicit construction one typically defines two 
alphabets $\{x_0,x_1\}$ and $\{y_0,y_1\}$ and the 
corresponding sets of all words $X^*$ and $Y^*$ 
formed from the respective alphabet. The letters 
of the former alphabet directly translate to $\{0,1\}$ 
when they appear as the indices of a (SV)HPL. 
The letters $y_0,y_1$ are related to $x_0,x_1$ via
\begin{align}
	y_0 &= x_0 \label{eq:y0x0} \\
	\tilde Z(y_0,y_1) y_1 \tilde Z(y_0,y_1)^{-1} 
	&= Z(x_0,x_1)^{-1} x_1 Z(x_0,x_1) \label{eq:y1x1}
\end{align}
where $Z$ is the so-called Drinfeld associator. 
It is defined as the generating series
\beq
Z(x_0,x_1) = \sum_{\sigma \in X^*} H_\sigma(1) 
\sigma \qquad \text{and} \qquad \tilde Z(y_0,y_1) 
= \sum_{\sigma \in Y^*} H_{\phi(\sigma)}(1) 
\tilde \sigma \label{eq:drinfeld}
\eeq
where the ``tilde'' operation reverses words and 
$\phi$ maps $y_i \to x_i$. The values of the \ac{hpls} 
at $z = 1$ in the definition \eqref{eq:drinfeld} are 
regularised by the shuffle algebra. Eq.~\eqref{eq:y1x1} 
can be solved iteratively for $y_1$. 

The \ac{svhpls} can then be extracted from the 
product of another two generating series
\beq
\sum_{\sigma \in X^*} \El_\sigma(z,\zb) = L_X(z) \tilde L_Y(\zb)
\eeq
where 
\beq
L_X(z) = \sum_{\sigma \in X^*} H_\sigma(z) \sigma 
\qquad \text{and} \qquad \tilde L_Y(\zb) 
= \sum_{\sigma \in Y^*} H_{\phi(\sigma)}(\zb) \tilde \sigma 
\eeq
with ``tilde'' and $\phi$ defined below eq.\eqref{eq:drinfeld}.

\ac{svhpls} obey the same shuffle product as \ac{hpls} 
\eqref{eq:hplshuffle}, namely
\beq
\label{svhplshuffle}
\El_{\rho}(z,\zb) \El_{\sigma}(z,\zb) 
= \sum_{\tau \in \rho \shuffle \sigma} \El_\tau(z,\zb).
\eeq

\subsection{Holomorphic part and single-value map}
\label{app:holomorphicpart} \ac{svhpls} are uniquely fixed 
by their holomorphic part (i.e.~their functional dependence 
on~$z$) and the requirement of single-valuedness. We 
define the holomorphic part of a function $\psi(z,\zb)$ 
as the limit
\beq
\label{eq:holopartdef}
\psi^{(h)}(z) = \psi(z,0) \big|_{\log \zb \to 0}.
\eeq
For a given linear combination of \ac{svhpls} taking this 
limit simply amounts to replacing $\El_\sigma(z,\zb) 
\to H_\sigma(z)$.

The dependence on $\zb$ is reconstructed by the 
single-value map 
\beq
\label{eq:sdef}
\mathbf{s} \left( \psi^{(h)}(z) \right) = \psi(z,\zb)
\eeq
which is discussed in detail in 
refs.~\cite{Brown:2013gia,DelDuca:2016lad}. 
Again, we restrict ourselves here to stating the 
(obvious) replacement rule $H_\sigma(z) \to 
\El_\sigma(z,\zb)$ which generates the 
corresponding single-valued expression 
from a linear combination of \ac{hpls} of 
$z$. As the action of the Hamiltonian 
$\Hitd$ \eqref{eq:hi2d} removes constant 
terms from the wavefunction prior to 
integration we shall not discuss this 
aspect in the context of 
eqs.~\eqref{eq:holopartdef} and 
\eqref{eq:sdef} here. The interested 
reader is referred to the above 
references.

\subsection{Variable transformations}
\label{app:svhplvariables} \ac{svhpls} obey 
relations under certain variable transformations. 
For the most part they are, in some sense, the 
\emph{same} relations that apply to \ac{hpls} due 
to the single-value map discussed above in 
appendix~\ref{app:holomorphicpart}. While the 
latter are much better documented (for an 
overview we recommend ref.~\cite{Maitre:2005uu}) 
we struggled to find a comprehensive list for 
\ac{svhpls} which motivated this appendix.

In section~\ref{sec:asalphabet} we transform 
$z \to 1/z$ and $z \leftrightarrow \zb$ to account 
for the symmetries of the two-dimensional 
wavefunction. In addition, we consider 
$z \to 1-z$ in section~\ref{sec:method2} to 
facilitate the ``last integration''. Let us 
discuss the latter transformation in 
detail.

At the level of \ac{hpls} it is straightforward 
to find relations under $z \to 1-z$. Effectively, 
the transformation moves the lower limit of the 
integral definition \eqref{eq:hpldef} from zero 
to one. Consider the weight-$w$ \ac{hpls} 
with argument $1-z$
\begin{align}
H_{0,a_2,\dots,a_w}(1-z) &= \int_0^{1-z} \frac{\dd t}{t} H_{a_2,\dots,a_w}(t) \nn \\
&= \int_0^1 \frac{\dd t}{t} H_{a_2,\dots,a_w}(t) - \int_{1-z}^1 \frac{\dd t}{t} H_{a_2,\dots,a_w}(t) \nn \\
&= H_{0,a_2,\dots,a_w}(1) - \int_0^z \frac{\dd t}{1-t} H_{a_2,\dots,a_w}(1-t) \label{eq:hpl0zerotoone}
\end{align}
and
\begin{align}
H_{1,a_2,\dots,a_w}(1-z) &= \int_0^{1-z} \frac{\dd t}{1-t} H_{a_2,\dots,a_w}(t) \nn \\
&= \int_0^1 \frac{\dd t}{1-t} H_{a_2,\dots,a_w}(t) - \int_{1-z}^1 \frac{\dd t}{1-t} H_{a_2,\dots,a_w}(t) \nn \\
&= H_{1,a_2,\dots,a_w}(1) - \int_0^z \frac{\dd t}{t} H_{a_2,\dots,a_w}(1-t) \label{eq:hpl1zerotoone}
\end{align}
with
\beq
H_0(1-z) = -H_1(z) \quad \text{and} \quad H_1(1-z) = -H_0(z).
\eeq
Since the \ac{hpls} inside the integrals in 
eqs.~\eqref{eq:hpl0zerotoone} and \eqref{eq:hpl1zerotoone} 
are of weight $w-1$ this defines a recursive prescription of 
how to write any \ac{hpl} of $1-z$ in terms of \ac{hpls} of $z$. 

By means of the holomorphic part of \ac{svhpls} and the 
single-value map, see appendix~\ref{app:holomorphicpart}, 
these relations can be applied to \ac{svhpls}. However, it is 
also possible to solve the recursion and write the answer 
directly as a sum. We find
\beq
\El_{a_1,\dots,a_w}(1-z,1-\zb) = \sum_{j=0}^w (-1)^j 
\El_{\tilde a_1,\dots,\tilde a_j}(z,\zb) \El_{a_{j+1},\dots,a_w}(1,1)
\eeq
with the ``$\sim$'' operation swapping the indices $0 \leftrightarrow 1$.

Similarly, on can derive identities for the transformation 
$z \to 1/z$, $\zb \to 1/\zb$. Again, the recursion can be 
solved and the resulting formula is simply yet slightly 
awkward to write out. To do so we define $n_0(\sigma)$ 
($n_1(\sigma)$) to count the number zeros (ones) in the 
indices $\sigma$ and $\hat s_{1 \to 0+1}$ to split 
$\El_\sigma(z,\zb)$ into a sum of $2^{n_1(\sigma)}$ 
\ac{svhpls} according to the index rule $1 \to 0 + 1$. 
For example,
\begin{multline}
\hat s_{1 \to 0+1} \left[ \El_{1,0,0,1,0}(z,\zb) \right] = \\
\El_{0,0,0,0,0}(z,\zb) + \El_{0,0,0,1,0}(z,\zb) 
+ \El_{1,0,0,0,0}(z,\zb) + \El_{1,0,0,1,0}(z,\zb)
\end{multline}
Then
\beq
\El_{a_1,\dots,a_w} \left( \frac1z,\frac1\zb \right) 
= \sum_{j=0}^w (-1)^{n_0(a_1,\dots,a_j)} 
\hat s_{1 \to 0+1} \left[ \El_{a_1,\dots,a_j}(z,\zb) 
\right] \El_{a_{j+1},\dots,a_w}(\infty,\infty).
\eeq
The values of \ac{svhpls} at $z,\zb \to \infty$ are 
related to the values at $z,\zb = 1$ by yet another 
transformation: $z \to z/(z-1)$.
\beq
\El_{a_1,\dots,a_w} \left( \frac{z}{z-1},\frac{\zb}{\zb-1} 
\right) = (-1)^{n_1(a_1,\dots,a_w)} \hat s_{0 \to 0+1} 
\left[ \El_{a_1,\dots,a_w}(z,\zb) \right]
\eeq
with $\hat s_{0 \to 0+1}$ defined like $\hat s_{1 \to 0+1}$ 
\eqref{eq:sdef} but based on the index rule $0 \to 0 + 1$. 
This last step is not strictly necessary but it reduces the 
amount of data needed to apply these kinds of 
transformations to a list of \ac{svhpls} at $z,\zb = 1$.

Lastly, let us examine the transformation $z \leftrightarrow \zb$ 
and how to related an \ac{svhpl} $\El_\sigma(\zb,z)$ to 
(a sum of) \ac{svhpls} $\El_{\sigma_i'}(z,\zb)$. The easy 
yet computationally heavy way is to translate 
$\El_\sigma(z,\zb)$ to \ac{hpls}, swap 
$z \leftrightarrow \zb$, extract the holomorphic 
part by means of eq.~\eqref{eq:holopartdef} and 
finally apply $\mathbf{s}$ \eqref{eq:sdef}. For 
\ac{svhpls} of weight less or equal to five this 
might be adequate but at higher weights it 
becomes inefficient due to the large size of 
expressions that the translation to \ac{hpls} 
causes. Like in the above examples this step 
can be avoided altogether.

The procedure relies on knowing the functional 
dependence of $y_1$ on the $x_i$, 
\emph{cf.}~eq.~\eqref{eq:y1x1}. Consider the 
weight-$n$ \ac{svhpl} $\El_\sigma(z,\zb)$ with 
$\sigma = \sigma_1,\dots,\sigma_n$ and swap 
$z \leftrightarrow \zb$. Then
\beq
\El_\sigma(\zb,z) = \El_{\tilde \sigma}(z,\zb) 
+ \sum_{i=4}^{|\sigma|} \sum_{j=0}^{|\sigma|-i} 
y1(\sigma_j,\dots,\sigma_{i+j}) 
\El_{\underbrace{\scriptstyle\sigma_1,\dots,
\sigma_{j-1}}_{(A)},\underbrace{\scriptstyle\sigma_{i+j+1},
\dots,\sigma_n}_{(B)},1} \label{eq:ztozb}
\eeq
where the ``tilde'' map was defined below 
eq.~\eqref{eq:drinfeld} and $y1(\sigma)$ is 
the coefficient of the product of $x_0$ and $x_1$ 
corresponding to $\sigma$, e.g.\ if $\sigma 
= 1,1,0,1,0$ then $y1(\sigma)$ is the coefficient 
of $x_1 x_1 x_0 x_1 x_0$. The indices (A) in 
eq.~\eqref{eq:ztozb} only appear if $j - 1 \geq 1$ 
and likewise (B) if $i + j + 1 \leq n$.


\bibliography{main}
\bibliographystyle{JHEP}

\end{document}